\documentclass[a4paper, intlimits, 11pt,reqno]{amsart}

\usepackage{mathtools}

\newcommand{\eqdef}{\vcentcolon=}

\usepackage{multirow,amssymb}

\usepackage[T1]{fontenc}
\usepackage{float}

\usepackage{dsfont}
\usepackage{times}
\usepackage[english]{babel}
\usepackage[T1]{fontenc}
\usepackage{enumitem}
\usepackage{optidef}
\usepackage{graphicx}
\usepackage{amsmath,bm}
\usepackage{subcaption}
\usepackage{booktabs}
\usepackage{tabularx}
\usepackage{color}
\usepackage{eurosym}
\usepackage{hyperref}
\usepackage{natbib}
\usepackage{tablefootnote}
\usepackage{diagbox} 
\usepackage[normalem]{ulem}

\usepackage{amsthm, amssymb, amsfonts}
\usepackage{amsmath}

\usepackage{accents}

\def\bo{\mathbf}

\def\b{\textcolor{blue}}
\def\r{\textcolor{red}}
\newcommand\g[1]{ {\color{teal}#1} }

\usepackage{url}

\usepackage{breakurl}

\def\hmath$#1${\texorpdfstring{{\rmfamily\textit{#1}}}{#1}}

\newtheorem{theorem}{Theorem}[section]

\newtheorem{lemma}[theorem]{Lemma}
\newtheorem{proposition}[theorem]{Proposition}

\theoremstyle{definition}
\newtheorem{definition}{Definition}[section]

\theoremstyle{remark}
\newtheorem{remark}{Remark}[section] 
\newtheorem{example}[remark]{Example}

\numberwithin{equation}{section}

\usepackage{tikz}
\usepackage{tkz-graph}
\usetikzlibrary{calc,positioning, shapes, fit,arrows,arrows.meta,decorations.pathreplacing,calligraphy}

\def\E{\mathbb{E}}

\newcommand{\beqa}{\begin{eqnarray}}
\newcommand{\eeqa}{\end{eqnarray}}
\def\bal{\begin{aligned}}
\def\eal{\end{aligned}}
\def\bll#1{
\beqa\label{#1}\bal
}\def\lel{\eal\eeqa}
\newcommand{\beql}[1]{\bll{#1}}
\newcommand{\eeql}{\lel}

\hypersetup{
 colorlinks  = true, %
 urlcolor   = black, %
 linkcolor  = black, %
 citecolor  = black %
}
\makeatletter
\renewcommand\@makefnmark{\hbox{\@textsuperscript{\normalfont\color{blue}\@thefnmark}}}
\renewcommand\@makefntext[1]{%
 \parindent 1em\noindent
      \hb@xt@1.8em{%
        \hss\@textsuperscript{\normalfont\@thefnmark}}#1}
\makeatother

\usepackage[hmargin=1.4in,vmargin=1.2in]{geometry}

\newcommand\IM{\mathrm{IM}}
\newcommand\DF{\mathrm{DF}}
\newcommand\FVA{\mathrm{FVA}}
\newcommand\CVA{\mathrm{CVA}}
\newcommand\MVA{\mathrm{MVA}}
\newcommand\KVA{\mathrm{KVA}}
\newcommand\XVA{\mathrm{XVA}}
\newcommand\FTP{\mathrm{FTP}}

\newcommand\CC{\mathrm{CC}}
\newcommand\AC{\mathrm{AC}}
\newcommand\CM{\mathrm{CM}}
\newcommand\MC{\mathrm{MC}}
\newcommand\LC{\mathrm{LC}}

\usepackage{hhline}
\usepackage[flushleft]{threeparttable}
\usepackage{makecell}
\usepackage{soul}
\usepackage{booktabs}

\def\sp{,\,}
\def\proof{\noindent \emph{\textbf{Proof.} $\, $}} 
\def\finproof {\hfill $\Box$ \vskip 5 pt }\def\finproof{\rule{4pt}{6pt}}\def\finproof{\ensuremath{\square}}
\def\finenv{}
\def\theQ{a_E(0)}\def\theQ{Q}
\def\qqq{\;\;\;\;\;\;\;\;\;}

\newcommand\TerminalTime{T}

\makeatletter
\newcommand\@notni[2]{\mathrel{\rotatebox[y=#1]{180}{$#2\notin$}}}
\newcommand\notni{
\mathchoice
  {\@notni{0.57ex}\displaystyle}
  {\@notni{0.57ex}\textstyle}
  {\@notni{0.39ex}\scriptstyle}
  {\@notni{0.26ex}\scriptscriptstyle}
}
\makeatother

\def\anE{E}

\def\CMset{A}
\def\CLset{B}
\def\OTCset{O}
\def\theD{\mathrm{D}}
\def\theE{\mathrm{E}}
\def\thec{\mathrm{c}}
\def\CM{\mathrm{CM}}

\def\emptyset{\varnothing}

\def\brho{\boldsymbol\rho}

\def\dloss{\mathcal{C}}

\begin{document}

\title{
Resolving a Clearing Member's Default
\\[1ex] {\small A Radner Equilibrium Approach}}
\author{Dorinel Bastide\textsuperscript{\lowercase {a}, \lowercase {b}},
St\'ephane Cr\'epey\textsuperscript{\lowercase {c}}, Samuel Drapeau\textsuperscript{\lowercase {d}}, Mekonnen Tadese\textsuperscript{\lowercase {e}}\\\today}

\maketitle

\begin{abstract}
For vanilla derivatives, which form the core of investment banks' hedging portfolios, central clearing via central counterparties (CCPs) has become the prevailing standard. A fundamental role of a CCP is to ensure an efficient and effective resolution process in the event of a clearing member's default. Upon such a default, the CCP is tasked with hedging and subsequently auctioning or liquidating the defaulted positions.
While the counterparty credit risk associated with auctioning has been examined in prior studies from a valuation adjustments (XVA) perspective, this work focuses on evaluating the costs associated with hedging or liquidation. This is achieved by contrasting pre- and post-default market equilibria through a Radner equilibrium framework applied to portfolio allocation and price discovery in both scenarios.
We establish the unique existence of Radner equilibria and provide both analytical and numerical solutions within elliptically distributed market settings. These insights equip CCPs with a rational basis for determining whether to hedge, auction, or liquidate defaulted portfolios in specific markets. Moreover, clearing members can leverage these findings to conduct what-if analyses, addressing inquiries from senior management and regulatory bodies. This study also underscores the advantages of central clearing over bilateral trading from a default resolution perspective.
  \vspace{5pt}

 
 \vspace{5pt}
\end{abstract}

{\let\thefootnote\relax\footnotetext{Python notebooks reproducing the results of this paper are available  
on \url{https://github.com/mekonnenta/CCP-Radner.git}}}

{\let\thefootnote\relax\footnotetext{\textsuperscript{a} \textit{BNP Paribas Stress Testing Methodologies \& Models. This article expresses the author's opinions and does not represent the position or opinions of BNP Paribas or its members.} dorinel.2.bastide@bnpparibas.com }}
{\let\thefootnote\relax\footnotetext{\textsuperscript{b} \textit{Laboratoire de Mathématiques et Modélisation d'Evry (LaMME), Université d'Evry/Université Paris-Saclay CNRS UMR 8071.} }}

{\let\thefootnote\relax\footnotetext{\textsuperscript{c} \textit{Laboratoire de Probabilités, Statistique et Modélisation (LPSM), Sorbonne Université et Université
Paris Cit\'e, CNRS UMR 8001.} stephane.crepey@lpsm.paris, \url{https://perso.lpsm.paris/~crepey}. 
The research of S. Cr\'epey benefited from the support of the Chair Stress Test, Risk Management and Financial Steering, led by the French Ecole Polytechnique and its Foundation and sponsored by BNP Paribas.}}

{\let\thefootnote\relax\footnotetext{\textsuperscript{d} \textit{Shanghai Jiao Tong University, Shanghai, China.} sdrapeau@saif.sjtu.edu.cn, \url{http://www.samuel-drapeau.info}.}}

{\let\thefootnote\relax\footnotetext{\textsuperscript{e} 
\textit{Sorbonne Université / LPSM and Centre de Math\'ematiques Appliqu\'ees (CMAP), Ecole Polytechnique, France.} demeke@lpsm.paris. 
The research of M. Tadese has been co-funded by the program PAUSE (Collège de France) and the Chair Stress Test, Risk Management and Financial Steering (\'Ecole Polytechnique and BNP Paribas).}}

{\let\thefootnote\relax\footnotetext{\textit{Acknowledgments:} We thank
Yannick Armenti, head of front office risk Europe derivatives execution and clearing at BNP Paribas securities services, 
and Mohamed Selmi,
head of market and liquidity risk at LCH SA,
for useful discussions.
}
}

  \section{Introduction}\label{s:intro}
   
Exchanges serve as platforms where financial actors can find counterparties for their transactions.
In the context of derivatives, the exchange is supported by a central counterparty (CCP).
A CCP is a financial institution that acts as an intermediary in derivatives and securities markets, standing between two trading parties to reduce counterparty credit risk.
When two parties engage in a transaction, the CCP becomes the buyer to the seller and the seller to the buyer, effectively guaranteeing the trade's completion even if one party defaults.
This mechanism ensures that the failure of one market participant does not directly impact others.
The CCP manages this risk by netting positions across multiple trades, requiring members to post collateral (margin) and maintaining a default fund composed of contributions from its clearing members.
The primary challenge for a CCP is to effectively handle the default of a clearing member by liquidating the member's portfolio within a few days of the default.
In the event of a default, the CCP follows a specific procedure (see {Section} \ref{s:RegFmwk}):
first, it hedges and auctions large portions of the portfolio to the remaining clearing members;
second, it sells the remaining part of the portfolio on the exchange.
Any shortfall arising during this fire sale procedure is covered sequentially by the collateral of the defaulted member, the private fund of the CCP (skin in the game), and finally the default fund.

From this simplified description, it is evident that the impact of fire selling the portfolio in a short amount of time through auction and liquidation should be reflected in the amount of collateral required from participants and in their contributions to the default fund.
Several studies have addressed the auctioning aspect \citep*{oleschak2019, ferera2020,BastideCrepeyDrapeauTadese21a}, as well as the liquidation process \citep*{ContAvellaneda13,VicenteCerezettiDefariaIwashitaPereira15}.
While the auctioning impact is typically derived endogenously from a game-theoretical perspective, the liquidation impact has mainly been examined from an exogenous financial engineering perspective, considering liquidation scheduling.

Using a one-period Radner equilibrium approach, we measure the equilibrium price movement between pre- and post-default resulting from the sudden fire sale of a defaulted member's portfolio.
From a mathematical perspective we provide existence and uniqueness for the specific objective measures of the participants in the CCP in the general case.
We further derive explicit solutions for such equilibrium in an elliptical market case which allows us to compare quantitatively the impact of liquidation on the market.
This perspective is however complicated by the following factors:
first, a single CCP can clear on different exchanges, while multiple CCPs can operate on the same exchange;
second, although a CCP is theoretically market-neutral by not holding any positions, in practice, it may engage in proprietary trading during fire sale situations by hedging the portfolio of the defaulted member.
We compare different scenarios in our study: where the CCP liquidates the portfolio on the defaulter's exchange, on an external exchange, or engages in proprietary trading by hedging the position.
To measure the impact of these different liquidation procedures, we assess the resulting price impact from the equilibrium in each scenario.

The use of competitive Walrasian equilibrium in economics to derive asset prices and optimal allocation has a long history, as surveyed in \citet*{magill1991} or \citet*{radner1982}.
In this context, we utilize methods developed in a dynamic setting by \citet*{cheridito2015} to derive equilibrium prices specific to CCPs.
Although the mathematical theory is well established, the study of equilibrium disruption resulting from a default is a novel approach.
We further provide precise quantification of this price impact by deriving explicit solutions to the equilibrium in a realistic context, using monetary risk measures that are established regulatory instruments to determine the required amount of collateral. 

 \subsection*{Outline}
Section \ref{sec:re} provides our Radner equilibrium market model.
Section \ref{sec:model} introduces the related comparative statics approach for the analysis of the market costs of hedging or liquidating a defaulted clearing member portfolio, either on the exchange of the CCP of the defaulter, or on an external exchange.
Sections \ref{sec:entropic} and \ref{sec:es} detail these costs in the case of entropic and expected shortfall risk measures. 
Section \ref{s:landc} analyzes the additional impact of counterparty credit risk, based on
 XVA specifications detailed in Section \ref{app:xva}.
 Section \ref{app:proof} proves some of the results of Section \ref{sec:re}.  
 Regulatory guidelines related to CCP default management can be found in Section \ref{s:RegFmwk}. Section \ref{app:examples} provides
examples of default resolution strategies where the CCP operates outside its own exchange.


 \subsection*{Standing notation}
Given vectors $x\in \mathbb{R}^m$ and $y\in \mathbb{R}^n$ (understood as column matrices), $x^\top $ is the transpose of $x$ and $(x,y)$ is the vector of $\mathbb{R}^{m+n}$ formed by stacking  $x$ above $y$. We denote by
$\mathcal{N}_n(\mu , \Gamma )$, the $n$-variate Gaussian distribution with mean  $\mu $ and covariance matrix $\Gamma$, and by $\phi$ and $\Phi$, the standard univariate Gaussian probability density and cumulative density functions; 
by $ \mathcal{E}_n(\mu, \Gamma,\psi)$, the $n$-variate elliptical distribution with mean  $\mu $, covariance matrix $\Gamma$, and characteristic generator function $\psi$, by $ \mathcal{T}_n(\mu, \Gamma,\nu)$, the $n$-variate Student $t$-distribution of degree of freedom $\nu$ with mean  $\mu $ and covariance matrix $\Gamma$, and by $t_{\nu}$ and $ T_\nu$, the standard univariate Student $t$ probability density and cumulative density functions of degree of freedom $\nu$.
 Throughout the paper, $(\Omega, \mathcal{A}, \mathbb{P})$ denotes a fixed probability space, with expectation, variance and covariance operators $\E  $,  $\mathbb{V}\mathrm{ar}$ and $\mathbb{C}\mathrm{ov}$;  $\mathcal{L}^0$ and $\mathcal{L}^1$  respectively denote the space of all the measurable and  integrable random variables (identified in the $\mathbb{P}$ almost sure sense), $\mathcal{X}$ is a  %
 linear subspace of $\mathcal{L}^1$ containing the constants.
 Bold letters refer to the solution of a Radner equilibrium.

For a function $f\colon \mathbb{R}^m \to \mathbb{R}$, its  directional derivative at $x$ in the direction of $y$ is defined as 
   \beql{defi:dirder}
       \mathcal{D}_y f(x) =\lim_{\epsilon \searrow 0} \frac{ f(x+\epsilon y)- f(x)}{\epsilon};
   \eeql
for $f$ convex,  a point $y \in \mathbb{R}^m$ is said to be a subgradient of $f$ at $x$,  denoted $y\in\partial f(x  )$, if 
    \begin{equation}\label{defi:subgrad}
        f( z)\geq f (x ) + y^\top  (z -x ), \quad 
        z \in \mathbb{R}^m;
    \end{equation}
the convex conjugate $f^\ast $ of $f$ is defined as 
\begin{equation}\label{defi:convconj}
    f^\ast (y) = \sup\left\{y^\top  x  - f (x ) ;\,  x\in\mathbb{R}^m \right\}, \quad y\in \mathbb{R}^m.
\end{equation}
    Let $f_i\colon \mathbb{R}^m \to \mathbb{R}$ be convex functions, for $i$ in a finite set $E$.
    The inf-convolution $f$ of the $f_i$ is defined as 
    \begin{equation}\label{defi:infconv}
  f(x)=\inf \left\{ \sum_{i\in \anE}f_i(x_i);\, \sum_{i\in \anE}x_i =x  \right\}, \quad x \in \mathbb{R}^m.
\end{equation}

\section{Radner Equilibrium Market Model}\label{sec:re}

We consider a one-period model of a financial exchange, with trading participants indexed by a finite set $\anE$.
A unitary position in each of the assets traded on the exchange pays a vector of random payoffs $P \in \mathbb{R}^m$ (with $m\ge 1$) at the terminal time $\TerminalTime$.
We assume that each participant $i$ in $\anE$ is endowed with a real valued random receivable $R_i$ (i.e.~$R_i\ge 0$ means a cash flow promised to the participant $i$).
For each $i\in \anE$, the participant $i$ hedges\footnote{unhedged market risk can generate significant regulatory capital requirements, rendering the contractual commitments non viable for the trading participant \citep*[Section MAR23, pp. 64 and 93]{BCBS2019}.} $R_i$ by entering a portfolio $\bo{q}_i  \in  \mathbb{R}^m $ of traded assets.  
The corresponding 
market loss of member $i$  is
\beql{e:lnp}
    -R_i + {\bo{q}_i^\top}  (\bo{p} - P ),
\eeql
 where $\bo{p} \in   \mathbb{R}^m $ is the vector of  prices of the traded assets at initial time:
see Table \ref{t:nbotation}.

For the monetary valuation of the risk of the participant $i$, we consider a normalized law invariant risk measure $\rho_i$ as per Definition \ref{def:riskdef}.  
\begin{definition}\label{def:riskdef}
  A function $\rho: \mathcal{X} \to \mathbb{R}$ is called a risk measure if it is
  {\rm\hfill\break $\bullet$  monotonous:} $\rho(X)\leq \rho(Y)$ for all $X$ and $Y$ in $\mathcal{X}$ such that $X\leq Y$;
  {\rm\hfill\break $\bullet$  convex:} $\rho(\lambda X+(1-\lambda)Y)\leq \lambda \rho(X)+(1-\lambda)\rho(Y)$ for all $X$ and $Y$ in $\mathcal{X}$, and $0\leq \lambda \leq 1$;
  {\rm\hfill\break $\bullet$  translation invariant:} $\rho(X-m)=\rho(X)-m$ for all $X$ in $\mathcal{X}$ and $m$ in $\mathbb{R}$;
    
  \noindent Such a risk measure is called normalized if $\rho(0)=0$ and law invariant if $\rho(X)=\rho(Y)$ whenever $X$ and $Y$ have the same law.
\end{definition}
We want to determine the portfolios $\bo{q}_i$ and the prices $\bo{p} $ endogenously as a Radner equilibrium driven by the offer and demand of all the participants to the exchange:

\begin{definition}\label{defi:radner}
A matrix of positions $(\bo{q}_i)_{i\in \anE }$ and a price vector $\bo{p} \in   \mathbb{R}^m $ form a Radner equilibrium on $\anE$ if 
{\rm\hfill\break $\bullet$  (optimality condition relative to each market participant $i\in \anE$)}
  \begin{equation}\label{eq:mogeneral}
    \rho_i(-R_i +\bo{q}_i^\top(\bo{p} - P ) ) \leq\rho_i( -R_i +q_i^\top  (\bo{p} - P ) )\sp q_i \in  \mathbb{R}^m ,
  \end{equation}
  \vspace{-0.7cm}
{\rm\hfill\break  $\bullet$ (zero clearing condition)}
  \begin{equation}\label{eq:zclearing}
    \sum_{i\in \anE} \bo{q}_i = 0.\ \finenv
  \end{equation} 
\end{definition}

\begin{table}[htp]  
    \begin{centering}
        {
            \begin{tabular}{@{}cll@{}} 
            \toprule
            $\bo{p} $  &    & equilibrium prices of the traded (hedging) assets \\ 
                        $\bo{q}_i $  &{}    & equilibrium positions of participant $i$ in the traded (hedging) assets \\ 
            $P$ & &   a vector of random payoff at terminal time $T$ \\
            $R_i$  & & receivable to be hedged by the participant $i\in E$  \\
            $\rho_i$  &{} & risk measure of the participant $i$ \\ 
                \bottomrule
            \end{tabular}
        }
        \caption{Main notation relative to an exchange $\anE$.}
    \label{t:nbotation}
    \end{centering}
\end{table}

\begin{remark}\label{rem:Ri} Since $R_i$ is assumed to be exogenously given, the price of the corresponding receivable to the participant $i$ is not part of the equilibrium. 
We say nothing on this price (assumed exogenously given and in fact implicitly part of  $R_i$ itself in our setup) in the paper, nor on the way it could be impacted  (in our setup it is simply not) by the instant default of a participant to the exchange.
\end{remark}

\subsection{Generic Results}

For each trading participant $i$ in $\anE$, we consider the convex function $r_i\colon \mathbb{R}^m \to \mathbb{R}$ defined as
\begin{equation*}
 r_i(q_i)=\rho_i(-R_i-q_i^\top   P )\sp q_i\in\mathbb{R}^m.
\end{equation*} 
By translation equivariance of $\rho_i$, the member $i$ optimality condition \eqref{eq:zclearing} can be rewritten as 
\begin{equation*}
    r_i(q_i) \geq r_i(\bo{q}_i) + (-\bo{p})^\top (q_i-\bo{q}_i)\sp q_i \in \mathbb{R}^m,
\end{equation*}
i.e., by  \eqref{defi:subgrad}, 
\begin{equation}\label{eq:membopt}
  -\bo{p} \in   \partial r_i(\bo{q}_i).
\end{equation}
By \citet[Theorem $23.5$]{rockafellar1970}, this is in turn equivalent to 
\begin{equation}\label{eq:membopt2}
     r_i(\bo{q}_i) = -{\bo{q}_i^\top}   \bo{p}   -r^\ast_i (-\bo{p} ), 
  \end{equation}
where $r^\ast_i$ is the convex conjugate  \eqref{defi:convconj} of $r_i$.
Note that
\beql{eq:membopt3} 
& -{q}_i^\top  {p}  -r^\ast_i (- {p}) \le r_i( {q}_i)\sp  q_i ,p \in  \mathbb{R}^m ,\\
&\qqq \mbox{with equality if and only if }-   p   \in \partial r_i( {q}_i),
\eeql
by \citet[Theorem $23.5$]{rockafellar1970}.

Lemma \ref{lem:radnerequivalence} and Theorems \ref{thm:largeloss}-\ref{thm:uniqueness} below are variants, for a single period model but with unbounded $(R_i,P)$ (as we want to endorse elliptical factor models later in the paper), of \citet*[Theorems 1 and 2]{cheridito2015}.
Radner equilibria admit the following dual characterization in terms of the inf-convolution \eqref{defi:infconv} $r$ of the $r_i$.

\begin{lemma}\label{lem:radnerequivalence}
  A matrix of positions $(\bo{q}_i)_{i\in \anE}$ and a price vector $\bo{p} \in \mathbb{R}^m$ form a Radner equilibrium on $E$
  if and only if
{ \rm\hfill\break \textbf{(i)}} $-\bo{p} \in  \partial r(0)$,
  {\rm\hfill\break \textbf{(ii)}} $r(0)=\sum_{i\in \anE}r_i(\bo{q}_i)$, and
  {\rm\hfill\break \textbf{(iii)}} $\sum_{i\in \anE}  \bo{q}_i=0$. 
  \end{lemma}
\proof See Section \ref{app:proof1}.\\ 

\noindent
Since the subgradient of a real valued convex function is non-empty, Lemma \ref{lem:radnerequivalence} implies that a Radner equilibrium exists if and only if the inf-convolution $r$ is attained at $0$.
It also implies that, whenever a Radner equilibrium exists, the optimal price is unique if and only if $r$ is differentiable at $0$.

\begin{theorem}\label{thm:largeloss}
  If $\rho_i$ is sensitive to large losses, i.e.\footnote{see \citet[Section $2.3$]{cheridito2015}.} $\lim_{\lambda \to \infty}\rho_i(\lambda L)=\infty$ for all $L\in \mathcal{X}$ such that $\mathbb{P}[L>0]>0$,  $i$ in $\anE$, then there exists a Radner equilibrium on $E$.
\end{theorem}

\proof
  Let $\theQ $  be the set of vector of positions $(q_i)_{i\in \anE}$ satisfying the zero clearing condition, i.e.
  \begin{equation*}
    \theQ =\left\{q\in \mathbb{R}^{m|E|}\colon q^\top   b^k=0, k=1,\dots,m\right\},
  \end{equation*}
  where $b^k$ is a vector in $\mathbb{R}^{m|E|}$ such that, for all $j=0,\dots,|E|-1$, the $k+jm$ entries of $b^k$ equal $1$ and all the other entries  of $b^k$ are  0 .
Note that $\theQ $ is a non-empty closed convex polyhedral subset of $\mathbb{R}^{m|E|}$.
Let $\mu = \E  [P]$. 
By \citet*[Theorem 3.3]{dalang1990},
$q_i^\top (  \mu- P )= 0$ almost surely holds or $\mathbb{P}[q_i^\top (  \mu- P )>0]>0$ holds, for any $ q_i$ in $\mathbb{R}^m$.
 The closed proper convex function 
  \begin{equation}\label{e:beta}
   \mathbb{R}^{m|E|} \ni q=(q_1,\dots, q_{|E|})\stackrel{\beta}{\mapsto} \sum_{i\in \anE}\rho_i(-R_i+q_i^\top ( \mu- P )) \in \mathbb{R}
  \end{equation}
 is such that
  \begin{equation*}
      \inf_{q \in \theQ  } \beta(q) = \inf_{\sum_{i\in E} q_i =0 }  \sum_{i\in \anE} r_i(q_i).
      \end{equation*}
In view of the comment preceding the statement of the theorem, it suffices to show that $\beta$ attains its minimum on $\theQ $.
  Let $B=\{q\in \mathbb{R}^{m|E|}\colon \beta(q)\leq \beta(0)\}$, with recession cone $0^+B =\{y \in \mathbb{R}^{m|E|} \colon b +\lambda y \in B,\forall \lambda \geq 0,\forall b \in B\}$ .
 Let $0^+\beta=\{y \in\mathbb{R}^{m|E|}\colon \beta 0^+(y) \leq 0\}$ denote  the recession cone of $\beta$, where $\beta 0^+$ is its recession function\footnote{The recession function $\beta 0^+$ is the map defined on $\mathbb{R}^{m|E|}$ as 
$ \beta 0^+(y) = \inf \left\{m \in \mathbb{R}\colon (y,m) \in 0^+\text{epi } \beta \right\},$  where $\text{epi } \beta = \left\{(q,n) \in \mathbb{R}^{m|E|}\times \mathbb{R} \colon n\geq  g(q) \right\}  $.}. 
  By \citet[Theorem $8.7$, page $70$]{rockafellar1970},  $0^+B=0^+\beta$.
  Since $B$ is a closed convex set containing the origin, \citet[Corollary $8.3.2$, page $64$]{rockafellar1970} yields
  \begin{equation}\label{e:0+B}
    0^+B=\{y\in \mathbb{R}^{m|E|}\colon \lambda y \in B \quad \forall \lambda>0\}.
  \end{equation}
Let $y = (y_1,\dots, y_{|E|}) \in \mathbb{R}^{m|E|}\setminus\{0\}$ (where each $y_i$ is in $\mathbb{R}^m$).
(i)  If $\mathbb{P}[y_i^\top (\mu- P )>0]>0$ holds for some $i\in \anE$, then, by the sensitivity to large losses condition on $\rho_i$,  $\beta(\lambda y )$ goes to infinity as $\lambda$ goes to infinity,
 which implies that $ y \notin 0^+B$.
(ii) If, instead, $y_i^\top  ( \mu- P )=0$ holds for all $i\in \anE$, then $\pm  y \in 0^+B$ hold by definitions \eqref{e:beta} of $\beta$, $B$ and \eqref{e:0+B} of $0^+B$. 
In particular, for any $y \in \mathbb{R}^{m|E|}\setminus\{0\}$ such that   $ y \in 0^+\beta$,  
we have
$- y \in 0^+\beta$.
  Hence, by \citet[Corollary $8.6.1$, page $69$]{rockafellar1970},  every direction of recession is a direction in which $\beta$ is constant. Following \citet[Theorem $27.3$, page $267$]{rockafellar1970}, in either case (i) or (ii), $\beta$ attains its minimum over $\theQ $.\ \finproof
 
\begin{remark}\label{rem:largeloss}
Let $\mathcal{X}$ be given as the Orlicz heart corresponding to the Young function $\theta\colon [0,\infty)\to [0,\infty)$ given by $\theta(t)=\exp(t-1)-\exp(-1)$, i.e.
  \begin{equation*}
    \mathcal{X}=\{L\in \mathcal{L}^0
    \colon \E  [\theta(c|L|)]<\infty\sp
    c>0\}\subseteq \mathcal{L}^1,
  \end{equation*}
  where $\mathcal{L}^0
  $ is the space of all real valued measurable random variables.
  An entropic risk measure of the form, for some $\varrho_i>0$,
  \begin{equation}\label{e:erm}
   \rho_i(L)=\frac{1}{\varrho_i}\ln(\E  [\exp(\varrho_i L)]), \quad L\in \mathcal{X},
  \end{equation}
is sensitive to large losses \citep[Section $2.3$]{cheridito2015}.\ \finenv
\end{remark}

Regarding the uniqueness of an optimal solution:
\begin{theorem}\label{thm:uniqueness}
  Let $((\bo{q}_i)_{i\in \anE}, \bo{p})$ be a Radner equilibrium on $\anE$.\\
  \textbf{(i)} If $r_i$ is differentiable at $\bo{q}_i$ for some $i$ in $\anE$, then $\bo{p} $ is unique.\\
  \textbf{(ii)} For any $i\in \anE$, if $r_i$ is differentiable and strictly convex on $\mathbb{R}^m$  then $\bo{q}_i$ is unique.
  \end{theorem}

\proof
  Let $((\bo{q}_i)_{i\in \anE}, \bo{p} )$ be optimal.
  If $r_i$ is differentiable at $\bo{q}_i$, then Lemma \ref{lem:radnerequivalence}(i) together with \citet[Theorems $23.2$, p. 216 and $25.2$, page $244$]{rockafellar1970} yield
  \begin{equation}\label{e:InfIneq}
   \mathcal{D}_xr_i(\bo{q}_i) %
   =x^\top  (-\bo{p})\leq \mathcal{D}_x r(0),
  \end{equation}
   where $\mathcal{D}_x r_i(\bo{q}_i) $ is the directional derivative \eqref{defi:dirder} of $r_i$ at $\bo{q}_i$ along $x$. 
  Take $\hat{q}_i = \bo{q}_i + \epsilon x$ and $\hat{q}_j=\bo{q}_j$ for all $j\neq i$.
By definition  \eqref{defi:infconv} of the inf-convolution,
  \begin{equation*}
    r(\epsilon x)\leq \sum_{j\in \anE}r_j(\hat{q}_j)=\sum_{j\neq i }r_j(\bo{q}_j)+r_i(\bo{q}_i+\epsilon x).
  \end{equation*}
  This together with Lemma \ref{lem:radnerequivalence}(ii) yields
  \begin{equation*}
    \mathcal{D}_x r(0)=\lim_{\epsilon \searrow 0} \frac{ r(\epsilon x)- r(0)}{\epsilon}\leq\lim_{\epsilon \searrow 0} \frac{ \sum_{j\in \anE}r_j(\hat{q}_j)-r(0) }{\epsilon}=\mathcal{D}_x r_i(\bo{q}_i) \leq \mathcal{D}_x r(0),
  \end{equation*}
  where the second inequality is due to \eqref{e:InfIneq}. 
  Hence $x\mapsto \mathcal{D}_x r(0)$ is linear.
  Thus, by \citet[Theorem $25.2$, page $244$]{rockafellar1970}, $r$ is  differentiable at  0 , i.e.~$\partial r(0)$ is a singleton, which, in view of Lemma  \eqref{lem:radnerequivalence} (ii)   implies (i).
  As for (ii), if $r_i$ is a strictly convex  and differentiable on $\mathbb{R}^m$, then it is closed and proper. 
  Following \citet[Corollary $26.3.1$, page $254$]{rockafellar1970},  $\partial r^\ast_i(-\bo{p})$ is a singleton and $\bo{q}_i=\nabla r^\ast_i(-\bo{p})$ is unique,  by \eqref{eq:membopt2}.\ \finproof

\begin{remark}\label{rem:differentiablity}
    Let $((\bo{q}_i)_{i\in \anE}, \bo{p} )$ be optimal.
    Following \citet[Proposition $2$]{cheridito2015}, if a risk measure $\rho_i$ is differentiable\footnote{
      $\rho_i$ is \emph{differentiable} at $L\in \mathcal{X}$ if there exist a random variable $W\in \mathcal{X}^\ast$ (the dual space of $\mathcal{X}$) such that (cf.~\eqref{defi:dirder})
      \begin{equation*}
        \lim_{\epsilon \searrow 0}\frac{\rho_i(L+\epsilon Y)-\rho_i(L)}{\epsilon}=\E  [YW] \quad  Y\in \mathcal{X}.
      \end{equation*}
      In this case, we write $W=\nabla\rho_i(L)$.
    } at $-R_i-{\bo{q}_i^\top}   P $ for some $i$ in $\anE$,  then $r_i$ is differentiable at $\bo{q}_i$, hence the optimal price $\bo{p} $ is unique.\ \finenv\end{remark}

 \subsection{Results Specific to Entropic or Expected Shortfall Risk Measures}
With explicit solutions and regulatory standards in view, from now on, $\rho_i$ is either an entropic or an expected shortfall risk measure.
In elliptical markets, entropic or coherent\footnote{e.g.~expected shortfall.} risk measures lead to analytical expressions for equilibria.
A summary of the notations that will be used hereafter is provided in Table \ref{t:notationentrop}, where further computations yield 
\beql{e:ER} \Gamma_i=\begin{bmatrix}
    \mathbb{V}\mathrm{ar}(R_i) & \mathrm{cov} _i\\
    \mathrm{cov}_i &\Gamma .
\end{bmatrix}.
\eeql
\begin{table}[H]  
    \begin{centering}
        {
            \begin{tabular}{@{}cll@{}} 
            \toprule
            \multicolumn{3}{c}{\emph{Entropic and expected shortfall case} }\\
		\cmidrule{1-3}
           $\mu $  &   & the vector $\E   [ P]    $ \\ 
           $\Gamma $ &    & the matrix $\mathbb{C}\mathrm{ov}( P  ) $ \\ 
            $\mu_i$ &{}    &  the vector  $\E  [(R_i, P)] $ \\ 
            $\mathrm{cov}_i$ &{}    &  the vector  $ \big(\mathbb{C}\mathrm{ov}(R_i,P_1), \dots,\mathbb{C}\mathrm{ov}(R_i,P_m) \big)$ \\ 
            $\mathrm{cov}$ &{}    &    $\sum_{i\in E} \mathrm{cov}_i$  \\ 
             $\Gamma_i$ &{}    &  the covariance matrix $\mathbb{C}\mathrm{ov}((R_i,P)) $, i.e.~of the vector  $(R_i, P)$  \\ 
             \cmidrule{1-3}
             \multicolumn{3}{c}{\emph{Specific to entropic  case} }\\
		\cmidrule{1-3}
            $\varrho_i $  &{} & risk-aversion parameter of an entropic risk measure of the participant $i $ \\
           $\varrho$  &{} & the number $(\sum_{i\in E} (1/\varrho_i) )^{-1}$  \\ 
           \cmidrule{1-3}
            \multicolumn{3}{c}{\emph{Specific to expected shortfall  case} }\\
		\cmidrule{1-3}
             $\alpha_i $  &  &  confidence level for an expected shortfall  risk measure of the participant $i$ \\
             $\mathbb{ES}_{\alpha_i}$ & & an expected shortfall risk measure with confidence level $\alpha_i$ \\
                \bottomrule
            \end{tabular}
        }
        \caption{Summary of the main notations in Propositions \ref{prop:entropy} and \ref{prop:es}.}
    \label{t:notationentrop}
    \end{centering}
\end{table}

We first consider the case of entropic $\rho_i$ and normally distributed $(R_i, P)$.

\begin{proposition}\label{prop:entropy}
  Let $\left(R_i,  P \right)\sim \mathcal{N}_{m+1}(\mu_i, \Gamma_i)$, $i$ in $\anE$, and $\Gamma$ be invertible\footnote{see Tables \ref{t:nbotation}-\ref{t:notationentrop}.}.
  If $\rho_i(L) =\frac{1}{\varrho}_i\ln(\E  [\exp(\varrho_i L)]$ for some $\varrho_i>0$, $i$ in $\anE$, then
  \begin{equation}\label{e:ps}
    \bo{q}_i =\Gamma^{-1}\left(\frac{\varrho}{\varrho_i} \mathrm{cov}-\mathrm{cov}_i\right)\sp i\in E, \quad \text{and} \quad \bo{p} =\mu - \varrho \mathrm{cov},
  \end{equation}
  where $\varrho = \left(\sum_{i\in \anE} \frac{1}{\varrho_i}\right)^{-1}$ and  $\mathrm{cov}=\sum_{i\in \anE}\mathrm{cov}_i$,  is a unique Radner equilibrium.
\end{proposition}

 \proof See Section \ref{app:proof2}.\\

We now turn to the case where each $\rho_i$ is an expected shortfall risk measure \citep[page $69$]{McNeilFreyEmbrechts2015}
\begin{equation}\label{e:ES}
    \rho_i(L)=\mathbb{ES}_{\alpha_i}(L)=\frac{1}{1-\alpha_i}\int_{\alpha_i}^1 q_u(L)du, \quad L \in \mathcal{X}=\mathcal{L}^1,
\end{equation}
for some $0\leq \alpha_i<1$, where $q_u(L)$ is the left $u$-quantile of $L$.
\begin{proposition}\label{prop:es}
  If $(R_i,  P )\sim \mathcal{N}_{m+1}(\mu_i, \Gamma_i)$ and  $\rho_i=\mathbb{ES}_{\alpha_i}$ for some $0\leq \alpha_i<1$, $i\in \anE$, then there exists a Radner equilibrium with a unique equilibrium price.
  If $\Gamma_i$ is further positive definite, $i\in E$, then the Radner equilibrium is unique.
\end{proposition}

 \proof See Section \ref{app:proof3}.\\

\begin{remark}\label{rem:spancov}
  If  $\mathbb{V}\mathrm{ar}(R_i)>0$, $\mathrm{cov}_i=0$, and $\Gamma$ is invertible, then 
  $z^\top  \Gamma_i z=z^2_1 \mathbb{V}\mathrm{ar}(R_i)+2z_1 \hat{z}^\top  \mathrm{cov}_i + \hat{z}^\top  \Gamma\hat{z}>0$ holds for any $z=(z_1,\dots,z_{m+1})\in \mathbb{R}^{m+1}\setminus\{0\}$, where $\hat{z}=(z_2,\dots,z_{m+1})$.
Hence $\Gamma_i$ is positive definite as assumed in the last part of Proposition \ref{prop:es}.

Instead, the positive definiteness of $\Gamma_i$ is not guaranteed when  $\mathbb{V}\mathrm{ar}(R_i)=\mathrm{cov}_i^\top\Gamma^{-1}\mathrm{cov}_i$, because $z^\top \Gamma_i z=0$ for $z =(-1,\Gamma^{-1}\mathrm{cov}_i)$.
This is for instance the case when $R_i$ is in the span of $P$, i.e. $R_i =  a_i^\top P+b_i$  for some constants $a_i \in \mathbb{R}^m$ and $b_i\in \mathbb{R}$, whence  $\mathrm{cov}_i = \Gamma a_i$ and $\mathbb{V}\mathrm{ar}(R_i)= a_i^\top \Gamma a_i = \mathrm{cov}_i^\top\Gamma^{-1}\mathrm{cov}_i$. \ \finenv
\end{remark}

\begin{remark}\label{rem:elliptical}
An $n$-variate  random vector $L$ has an elliptical distribution written as $ L \sim \mathcal{E}_n(\mu, \Gamma,\psi )$ if its characteristic function is expressed as 
  \begin{equation*}
     \E \Big[e^{{\rm i}\, z^\top L } \Big]  
    =\exp({\rm i}\ z^\top \mu )\psi\left(\frac{1}{2} z^\top   \Gamma z\right), \quad z \in \mathbb{R}^n,
  \end{equation*}
  for $\mu =\E  [L], \Gamma = \mathbb{C}\mathrm{ov}(L)$, and a function  $\psi \colon [0,\infty) \to \mathbb{R}$ such that $-\psi'(0)=-1$. 
 As is well known \citep*{landsman2003, McNeilFreyEmbrechts2015}, if $ L  \sim \mathcal{E}_n(\mu,\Gamma, \psi)$ (or, more specifically\footnote{ a Student $t$-distribution is elliptical \citep*{gaunt2021}.}, $ \mathcal{T}_n(\mu,\Gamma, \nu)$), 
 then  $a^\top   L \overset{d}{=}  a^\top  \mu + \sqrt{ a^\top   \Gamma\,  a } \;Z$, where $Z \sim \mathcal{E}_1(0,1, \psi)$ (specifically, $ \mathcal{T}_1(0,1, \nu)$). Hence, for any coherent risk measure $\rho$, 
  \begin{equation}\label{eq:eselliptical}
  \rho( a^\top    L ) =  a^\top  \mu +  \rho(Z)\sqrt{ a^\top   \Gamma  a } .
\end{equation}

Assuming $(R_i,  P )\sim \mathcal{E}_{m+1}(\mu_i,\Gamma_i, \psi)$ (e.g.  $ \mathcal{T}_{m+1}(\mu_i,\Gamma_i, \nu)$) with $\Gamma_i$ positive definite,  
  $i\in \anE$,  the above implies that  
\begin{equation}\label{eq:smalrgeneral}
  r_i(q_i)=-\E  [R_i]-q_i^\top  \mu+ \rho_i(Z)\sqrt{\mathbb{V}\mathrm{ar}(R_i)+2q_i ^\top \mathrm{cov}_i+q_i^\top \Gamma q_i}. 
\end{equation}
The proof of Proposition \ref{prop:es} in  Section \ref{app:proof3} thus works for any law invariant and coherent risk measure $\rho_i$ differentiable on the linear space spanned by the components of $(R_i,  P )$, $i$ in $\anE$.\ \finenv
\end{remark}

\section{The Comparative Statics Approach for Default Resolution Analysis}\label{sec:model}

\subsection{Methodology}\label{ss:methodo}

Let an index $d$ represent a clearing member of a CCP defaulting instantaneously at time 0\footnote{considering several instant defaulters would mainly mean replacing $\bo{q}_d$ by $\sum_d \bo{q}_d$ hereafter, 
see e.g.~Remark \ref{rem:lcnormal}. We refrain from doing so for parsimony of notation.}. 
We want to analyze and compare different close-out procedures, of the `hedging or not and liquidation or auctioning' types (one of the contributions of the paper, and not the least, being to clarify what exactly is meant by the latter), for the CCP portfolio of member $d$. Our methodology is based on a comparative statics analysis, whereby, for each envisioned default resolution strategy, we compute the sum of the associated risk incrementals of market participants, dubbed funds transfer price (FTP). Following an idea of Pareto optimality where numerous exchanges (and trading participants themselves in Section \ref{s:landc}) compete with one another, we then elect the strategy that minimises the funds transfer price.

{Until Section \ref{sec:es}, we describe the market cost (MC) component of the FTP, deferring to Section \ref{s:landc} the credit cost component calculation.} 
Along with auctions that will be considered in Section \ref{s:landc}, 
a default resolution strategy conducted by a CCP consists in two types of operations: liquidating or hedging. The CCP can execute these operations either on its own exchange or an external exchange. Liquidation involves the CCP selling the defaulted positions on an exchange, whereas hedging requires the CCP to retain those positions and act as a participant on an exchange to eliminate the market risk stemming from maintaining those positions. Performing these operations individually represents only the extreme ends of the default resolution strategies spectrum. In practice, a CCP would implement a combination of these operations on both its own and external exchanges: see \citet{BoE2018}, \citet{bis2020}, and Section \ref{s:RegFmwk}.

{Let $\theD$ be the exchange of the CCP itself, and $\anE$ label any exchange, for instance $\theD$ but without restriction to it.}
For each (pre-default) exchange $\anE$, we denote by $\anE'$ its advent in the wake of the instant default of $d$, depending on the settlement procedure implemented by the CCP.
If the CCP of an exchange $\theD$ faces the default of a clearing member $d$, then this CCP can envision different default resolution procedures, impacting possibly different exchanges $E'$ (starting with $\theD'$ itself), for the CCP portfolio  $\bo{q}_d$ of the defaulter  (in a pre-default equilibrium on $\theD$).  
 For each considered default resolution strategy, each of the impacted exchanges $E'$ (or their corresponding CCPs) would compute its corresponding market cost $\MC_E
$ as per \eqref{e:mastergen}
and communicate it to the CCP of $\theD$.  
The ensuing MC \eqref{e:mastergen} of the strategy  is the price that the markets  would charge to 
the CCP of $\theD$, should the latter choose this strategy for resolving  $\bo{q}_d$. 
The CCP of $\theD$  would then choose the most efficient strategy, i.e.~the one minimizing MC (or a broader FTP also encompassing a credit cost as the way detailed in Section \ref{s:landc}).

We assume that the different exchanges $\anE$ trade the same assets with terminal payoff $P$, possibly at different initial prices $\bo{p}$ (interpreted in this setup as ``time $0-$'', pre-default prices), reflecting different market equilibria.
Hedging procedures involve the CCP as a new trading participant, represented for this purpose by a new index $c$ (not involved in any exchange $\anE$). 
We use similar notation for $d$ and $c$ as for participants $i$ of $\anE$ in Section \ref{sec:re} (see Tables \ref{t:nbotation}-\ref{t:notationentrop}).  
Although other choices could be used without methodological change in what follows,
and due to non-suitable calibration data in this regard, we assume that any data other than  $\bo{p} $ and $\bo{q}_i$  in  Tables \ref{t:nbotation}-\ref{t:notationentrop} are not affected by the instant default of $d$.
The only exception is given in \eqref{eq:changerhogen} below for $R_c$, which represents the post-default receivable\footnote{see \eqref{eq:ccpendowment} below.} of the hedging CCP due to the portfolio of the defaulted member $d$ taken over by the hedging CCP, whereas the pre-default receivable of the CCP is zero (a CCP should not bear any position, except for the ones inherited from defaulted market participants during the close-out period of their portfolios).

The pair $((\bo{q}'_i)_{i\in E'}, \bo{p}')$ relative to any post-default exchange $E'$ involved in the settlement of the defaulted portfolio is derived using a Radner equilibrium in $E'$. Note that all the receivables and equilibrium portfolios and prices implicitly depend on the corresponding exchange.  
Regarding prices, we make this dependence explicit hereafter, denoting by  $\bo{p}^E$ a pre-default (``time $0-$'') equilibrium price on $E$ and  by $\bo{p}'^E$ a post-default (``time 0'') equilibrium price on $E'$.
Price moves generate a liquidity cost analyzed in Section \ref{ss:pi}. As illustrated in Sections \ref{ss:ftp} and \ref{app:examples}, this liquidity cost will be part of the market cost.

  \subsection{Liquidity Cost}\label{ss:pi}

We define $ \bo{q}_i=0,    i\in E'\setminus E ,$ and $ \Delta \bo{q}_i =   \bo{q}'_i  - \bo{q}_i ,  i\in E'\cup E$,
hence 
\beql{eq:changeqgen} 
\sum_{ i\in  \anE'\cap E  } \bo{q}_i + \sum_{i\in  \anE'  }  \Delta \bo{q}_i   
= \sum_{ i\in  \anE'  } \bo{q}_i+ \sum_{   \anE'   } \Delta \bo{q}_i  =    \sum_{i\in   \anE'   }   \bo{q}'_i. 
\eeql
If the CCP chooses to liquidate a portion $\bo{q}^l_d$ of
$\bo{q}_d$ and hedge the remaining $\bo{q}^h_d = \bo{q}_d -\bo{q}^l_d$, then the incremental positions of the participants to
any post-default exchange $\anE'$ can be split as  $\Delta\bo{q}_i= \Delta\bo{q}^l_i + \Delta\bo{q}^h_i$, where $\Delta\bo{q}^l_i$ and $\Delta\bo{q}^h_i$ are the increments implied by the liquidation and hedging legs of the strategy (see e.g.
Sections
\ref{rem:hybrid} and \ref{rem:hybrid2})--with always in particular \beql{e:dc}\Delta \bo{q}^l_c=0 ,\eeql
as a CCP does not take part as a participant to a liquidation.
Since the amount demanded should be equal to the amount supplied on both legs of the strategy, we have 
\begin{equation}\label{e:qlh}
    \sum_{i\in \anE'} \Delta \bo{q}^l_i = \bo{q}^l_d\quad \text{and} \quad  \sum_{i\in \anE'} \Delta \bo{q}^h_i = 0\text{, hence } \sum_{i\in\anE'} \Delta \bo{q}_i = \bo{q}^l_d.
\end{equation} 
The first consequence of a default resolution strategy is then a liquidity cost
\begin{equation}\label{eq:lccost}
  \LC=\sum_E \LC_E,
  \end{equation}
where \beql{e:lcenative}
\LC_E&=\sum_{i\in  \anE'\cap \anE } \bo{q}_i^\top  (\bo{p}^\anE -\bo{p}'^\anE) +\sum_{i\in \anE'}{(\Delta\bo{q}^l_i} )^\top (\bo{p}^\anE-\bo{p}'^\anE) 
\\&
=\sum_{i\in  \anE' } \underbrace{(\bo{q}_i + \Delta\bo{q}^l_i)^\top  (\bo{p}^\anE -\bo{p}'^\anE)}_{\LC_i} 
\eeql 
(as $\bo{q}_i =0,i \in \anE'\setminus \anE$) 
corresponds to margin payments (like in futures markets) by market participants at time 0 in response to the default settlement procedure of $d$, i.e. the price they have to pay for the transition from the pre-default to the post-default exchanges. 

As reflected in \eqref{e:lcenative}, it is only the contracts $\Delta\bo{q}^l_i$ involved in the liquidation leg of the strategy, which are old (``time $0-$'') contracts with the pre-default prices $\bo{p}^E$, that deserve margin payments, while the new (``time $0$'') contracts $\Delta\bo{q}^h_i$ involved in the hedging leg of the strategy are post-default contracts with the new prices $\bo{p}'^E$.
However, the following reformulation of $\LC_E$ in terms of the $\bo{q}_i + \Delta\bo{q}_i=\bo{q}'_i$  (instead of $\bo{q}_i + \Delta\bo{q}^l_i$ natively in \eqref{e:lcenative}) is possible:
\begin{lemma}\label{lem:lce} On each exchange $E$,
\beql{e:lce}
\LC_E =\sum_{i\in  \anE'\cap \anE } \bo{q}_i^\top  (\bo{p}^\anE -\bo{p}'^\anE) + \sum_{i\in \anE'}\Delta\bo{q}_i^\top (\bo{p}^\anE-\bo{p}'^\anE) =  \sum_{i\in  \anE'  } (\bo{q}'_i)^\top (\bo{p}^\anE -\bo{p}'^\anE)  .
\eeql 
\end{lemma}
\proof
By \eqref{e:qlh}, $\sum_{i \in \anE'} \Delta \bo{q}^h_i =0$. Hence \eqref{e:lcenative} yields 
\begin{equation*}
    \LC_E  
    = \sum_{i\in  \anE' } \big(\bo{q}_i + \Delta\bo{q}^l_i + \Delta\bo{q}^h_i \big )^\top  (\bo{p}^\anE -\bo{p}'^\anE),
\end{equation*}
where $\Delta\bo{q}^l_i + \Delta\bo{q}^h_i=\Delta \bo{q}_i$.\ \finproof

 \subsection{Market Cost}\label{ss:ftp}
Let 
\begin{equation}\label{eq:changerhogen}
    \Delta\brho_i =  
 \rho_i\big(-R_i +(\bo{q}'_i)^\top  (\bo{p}'^\anE - P )\big) - \rho_i\big(-\mathbf{1}_{i\neq c} R_i +\bo{q}_i^\top  (\bo{p}^\anE - P )\big)\sp  i\in E'. 
  \end{equation}   
Using \eqref{eq:changeqgen} and $ \Delta\bo{q}^l_i + \Delta\bo{q}^h_i=\Delta\bo{q}_i$, the post-default market loss of any trading participant $i\in\anE'$ is
\beql{e:PostDefTdgLoss}
& \underbrace{-R_i+\bo{q}_i^\top (\bo{p}^\anE-P)}_{\text{pre-default market loss}}+(\Delta\bo{q}^h_i)^\top (\bo{p}'^\anE-P)+ (\Delta\bo{q}^l_i )^\top(\bo{p}^\anE-P)\\
 =&  -R_i+\bo{q}_i^\top   (\bo{p}^\anE-P) + \Delta\bo{q}_i^\top (\bo{p}'^\anE-P)+
 \big(\Delta\bo{q}^l_i\big)^\top(\bo{p}^\anE-\bo{p}'^\anE) \\
= & -R_i+(\bo{q}'_i)^\top (\bo{p}'^\anE-P)+ \LC_i,
\eeql
for $\LC_i$ as per \eqref{e:lcenative}.
Hence, by translation equivariance of $\rho_i$, the post-default risk of participant $i$ is
\begin{equation*}
    \rho_i\big(-R_i +(\bo{q}'_i)^\top  (\bo{p}'^\anE - P )\big) +   \LC_i .
\end{equation*}
The risk incremental of participant $i$ is therefore  $ \LC_i +  \Delta\brho_i  $,  $i\in \anE'$.
Accordingly, we assess the market cost (MC) of a default resolution strategy by  
\beql{e:mastergen}
& \MC = \sum_{E}\MC_E,  \text{ where }
   \MC_E =\LC_E +\sum_{i\in  \anE' }   
{\Delta\brho_i}.
\eeql

\subsection{Examples of Default Resolution Strategies}\label{ss:fourcases}
The pre-default equilibria $((\bo{q}_i)_{i\in    E}, \bo{p}^\anE)$ involved in \eqref{e:mastergen}
are obtained by direct application of the results of Section \ref{sec:re}.
We now detail the corresponding post-default Radner equilibria  $((\bo{q}'_i)_{i\in    E '}, \bo{p}'^\anE)$  in different cases (without post-default new invited participants other than the CCP itself in the hedging cases, though; extra new invited participants will only be considered later in the paper).
The member optimality condition for the post-default market participant $i\in \anE'$ is always stated as
\begin{equation}\label{eq:mo}
    \rho_i\big(-R_i +(\bo{q}'_i)^\top (\bo{p}'^\anE - P )\big) \leq \rho_i\big(-R_i +q_i^\top  (\bo{p}'^\anE - P )\big), \; q_i \in \mathbb{R}^m.
\end{equation}
The clearing condition, instead,  depends on the considered default resolution strategy. 
Hereafter in this section, we examine strategies in which the CCP exclusively operates on its own exchange.
See Section \ref{app:examples} for further scenarios also involving external exchanges.

\subsubsection{The CCP fully liquidates on its own exchange} \label{sec:liqonE}
 
As a first default resolution alternative, the CCP may want to liquidate the defaulter's position $\bo{q}_d$  on its own exchange $\theD$.
Then  $\MC_\anE =0, \anE\neq \theD$, and $\sum_{i\in \theD' = \theD \setminus \{d\}} \Delta \bo{q}_i =\bo{q}_d.$
As $\sum_{i \in \theD\setminus\{d\}}\bo{q}_i + \bo{q}_d=0$,  we obtain a post-default equilibrium clearing condition 
\begin{equation}\label{e:Dq}
    \sum_{i\in\theD' =\theD \setminus \{d\}} \bo{q}'_i =0 
\end{equation} 
and
\begin{equation*}
   \LC = \LC_\theD = 0\sp  \MC = \MC_\theD =\sum_{ i \in \theD'=\theD \setminus \{d\}}\Delta\brho_i.
\end{equation*}

\subsubsection{The CCP fully  hedges on its own exchange} \label{ss:he}
If $\bo{q}_d$ is not instantaneously liquidated upon the default of member $d$ at time 0, then the CCP $c$ of $d$ endorses at time 0 the receivable 
\begin{equation}\label{eq:ccpendowment}
  R_\mathrm{c}=\bo{q}_d^\top   ( P -\bo{p}^\theD) ,
\end{equation} 
which it can hedge by holding  on its own exchange $\theD$ a portfolio $\Delta \bo{q}_\mathrm{c}$ minimizing some risk measure $\rho_\mathrm{c}$.
The corresponding member optimality condition \eqref{eq:mo}  for the CCP $c$, playing the role of a new post-default trading participant, is   
\begin{equation}\label{e:CCPrhoc}
\rho_\thec\big( \bo{q}_d ^\top  (\bo{p}^\theD-P)+( \bo{q}'_\thec )^\top (\bo{p}'^\theD-P) \big) \leq \rho_\thec\big(\bo{q}_d ^\top  (\bo{p}^\theD-P)+ q_c^\top (\bo{p}'^\theD-P)\big), \quad q_\thec \in \mathbb{R}^d.
\end{equation}
In this case, $\MC_\anE=0, \anE\neq \theD$, and $\sum_{i\in \theD' =(\theD\setminus\{d\}) \cup \{c\}  } \Delta \bo{q}_i  =0$ (as, in this hedging case, on the post-default market $\theD'$, the amount demanded must be equal to the amount supplied).
Since $\sum_{i\in  \theD\setminus\{d\} }  \bo{q}_i  =-\bo{q}_d$  and  $\Delta \bo{q}'_c  =\bo{q}'_c$, we obtain a post-default equilibrium clearing condition
\begin{equation}\label{eq:cchedgingonE}
  \sum_{i\in  \theD' = (\theD\setminus\{d\})\cup \{\thec\}}\bo{q}'_i =-\bo{q}_d 
\end{equation}
and
\begin{equation*}
    \MC=\MC_{\theD} = \underbrace{-\bo{q}_d ^\top  (\bo{p}^\theD-\bo{p}'^\theD)}_{\LC_\theD} \; + \sum_{i\in \theD'=(\theD\setminus\{d\})\cup\{c\}}\Delta\brho_i .
\end{equation*}    
\begin{remark}\label{rem:cchedging}  
  By change of variable $\bo{z}'_i =\Delta \bo{q}_i, i\in(\theD\setminus\{d\})\cup \{\thec\}$,  and $R'_i  = R_i+\bo{q}_i^\top P $, $i\in \theD\setminus\{d\})$ and  $R'_\thec=R_\thec $, the clearing condition \eqref{eq:cchedgingonE} relative to  the post-default equilibrium $((\bo{q}'_i)_{i\in    \theD'}, \bo{p}'^\theD)$ can be converted to a zero clearing condition as per Definition \ref{defi:radner} on $ \theD'$.\ \finenv
  \end{remark}
 
\subsubsection{The CCP fully  replicates on its own exchange} \label{rem:fullreplication}
By replication, we refer to a default resolution strategy whereby the CCP $c$ replicates the portfolio $\bo{q}_d$  (if not liquidated) that the CCP inherits from $d$ by mirroring position $\bo{q}'_c=-\bo{q}_d$ on its own exchange $\theD$. 
In this case $\MC_E =0, \anE\neq \theD$, and we have $\theD'=(\theD\setminus\{d\})\cup \{c\}, \Delta\bo{q}_c=-\bo{q}_d$ (in the replication case, the only admissible trading strategy for  $\thec$ as a post-default trading participant is $-\bo{q}_d$), $\Delta\brho_\thec =\rho_\thec \Big(\bo{q}_d^\top   (\bo{p}^\theD-P)- \bo{q}_d^\top   (\bo{p}'^\theD-P)\Big)= \bo{q}_d^\top \big(\bo{p}^\theD-\bo{p}'^\theD \big)$.
On the post-default exchange $\theD'$ where the hedge is implemented, the amount demanded must be equal to the amount supplied i.e. $ \sum_{i\in \theD'= (\theD \setminus \{d\}) \cup\{\thec\} }\Delta  {\bo{q}}_i  =0$, whence the post-default clearing condition  
\begin{equation}\label{eq:replicationonD}
  \sum_{i\in  \theD' = (\theD\setminus\{d\})\cup \{\thec\}}\bo{q}'_i = \underbrace{\sum_{i\in   \theD\setminus\{d\} }\bo{q}'_i}_{0} + \underbrace{\bo{q}'_\thec}_{-\bo{q}_d} = -\bo{q}_d.
\end{equation}
Therefore
\begin{equation*}
    \MC=\MC_\theD  = \underbrace{-\bo{q}_d^\top  (\bo{p}^\theD-\bo{p}'^\theD)}_{\LC=\LC_D} +\sum_{i\in \theD\setminus\{d\}} \Delta\brho_i + \underbrace{\Delta\brho_\thec}_{\bo{q}_d^\top (\bo{p}^\theD-\bo{p}'^\theD)} =\sum_{i\in \theD\setminus\{d\}} \Delta\brho_i.
\end{equation*}
The market cost is the same as in the liquidation case of \ref{sec:liqonE} (note that the embedded post-default Radner equilibria are the same), but its split between LC and  $ \sum_{i\in  \theD'  } \Delta\brho_i$ is different, see Table \ref{tab:SixCasesSummary}.
\begin{remark}\label{rem:ccreplication}
  As in Remark \ref{rem:cchedging}, by change of variable $\bo{z}'_i =\Delta \bo{q}_i, i\in(\theD\setminus\{d\})\cup \{\thec\}$,  and $R'_i  = R_i+\bo{q}_i^\top P $, $i\in \theD\setminus\{d\})$,  $R'_\thec=R_\thec $, the clearing condition \eqref{eq:replicationonD} relative to  the post-default equilibrium $((\bo{q}'_i)_{i\in    \theD'}, \bo{p}'^\theD)$ can be converted to a zero clearing condition as per Definition \ref{defi:radner} on $\theD'$.\ \finenv
  \end{remark}

\subsubsection{The CCP partially liquidates and hedges on its own exchange}\label{rem:hybrid}
The CCP can also liquidate a portion $\bo{q}_d^l$ of the defaulted position $\bo{q}_d$ and hedge the remaining $\bo{q}_d^h=\bo{q}_d-\bo{q}_d^l$ on its own exchange $\theD$.
The amount demanded should be equal to the amount supplied on each leg of the strategy, hence
$\sum_{i\in \theD\setminus\{d\}} \Delta \bo{q}^l_i = \bo{q}^l_d$ and $\sum_{i\in (\theD\setminus\{d\}) \cup \thec} \Delta \bo{q}^h_i = 0$, thus $\quad \sum_{i\in (\theD\setminus\{d\}) \cup \thec} \Delta \bo{q}_i =\bo{q}^l_d.$
As $\sum_{i\in \theD \setminus\{d\}}\bo{q}_i=-\bo{q}_d$, the ensuing post-default clearing condition on $\theD'$ is
\begin{equation}\label{eq:cchl}
    \sum_{i\in \theD' = (\theD \setminus\{d\}) \cup \{\thec\}} \bo{q}'_i = -\bo{q}^h_d,
\end{equation} 
We assume that both liquidation and hedging happen simultaneously at the same price $\bo{p}'^\theD$.
Hence each trading participant on the post-default market $\theD'$ has a single member optimality condition \eqref{eq:mo} (with, in particular,
$R_\thec = (\bo{q}^h_d )^\top (P-\bo{p}^\theD)$). 
Then
\begin{equation*}
    \MC=\MC_\theD  = \underbrace{-(\bo{q}^h_d)^\top  (\bo{p}^\theD-\bo{p}'^\theD)}_{\LC=\LC_D} \; +\sum_{i\in \theD' =(\theD\setminus\{d\})\cup \{\thec\}} \Delta\brho_i .
\end{equation*}

\begin{remark}\label{rem:ccrhconE}
    By change of variables $\bo{z}'_i= {\bo{q}}_i'+k_i\bo{q}^h_d$ and $ {R}'_i= {R}_i-k_i(\bo{q}^h_d)^\top  P$, for reals $k_i$ such that $\sum_{i\in  {\theD'}  } k_i =1,$  the clearing condition \eqref{eq:cchl} and the optimality conditions \eqref{eq:mo} relative to the post-default equilibrium  $((\bo{q}'_i)_{i\in    \theD'}, \bo{p}'^\theD)$ respectively become  $  \sum_{i\in  \theD'}\bo{z}'_i=0$   and
  \begin{equation*}
     {\rho}_i(- {R}'_i + (\bo{z}'_i)^\top  ( {\bo{p}}'^\theD-P)) \leq  {\rho}_i(- {R}'_i +z_i^\top  ( {\bo{p}}'^\theD-P))\sp  z_i\in\mathbb{R}^m.\ \finenv
  \end{equation*}
 \end{remark}
 
\begin{table}[htp]  
        \begin{centering}
           \begin{tabular}{@{}lcc@{}}
           \toprule
             & $\mathrm{LC}$ &   $\displaystyle 
             \sum_{i\in  \theD' } \Delta\brho_i$\\ 
                \midrule
                1.  & 0   & $\displaystyle\sum_{i\in \theD\setminus\{d\}}\Delta\brho_i$ \\     
                2.  & $-\bo{q}_d ^\top  (\bo{p}^\theD-\bo{p}'^\theD) $    &  $\displaystyle\sum_{i\in (\theD\setminus\{d\})\cup\{c\}}\Delta\brho_i$ \\ 
                3.  & $-\bo{q}_d ^\top  (\bo{p}^\theD-\bo{p}'^\theD)$   & $\displaystyle \sum_{i\in ( \theD\setminus\{d\} ) \cup \{\thec\}} \Delta\brho_i$     with $\Delta\brho_\thec = \bo{q}_d^\top \big(\bo{p}^\theD-\bo{p}'^\theD \big) $\\  
                4.  & $-(\bo{q}^h_d)^\top  (\bo{p}^\theD-\bo{p}'^\theD) $    &  $\displaystyle\sum_{i\in (\theD\setminus\{d\})\cup\{c\}}\Delta\brho_i$ \\ 
                \bottomrule
            \end{tabular}
            \caption{Decomposition of the  market costs in the four default resolution strategies  of Section \ref{ss:fourcases}.}
       \label{tab:SixCasesSummary}
   \end{centering}
\end{table}

In Sections \ref{sec:entropic}-\ref{sec:es}, we provide explicit or numerical solutions regarding the market cost of default resolutions on $\theD$, hence 
\begin{equation}\label{e:lmcD}
   \LC = \LC_\theD = \sum_{i\in \theD'} (\bo{q}'_i)^\top (\bo{p}^\theD-\bo{p'}^\theD), \quad \ \quad  \MC = \MC_\theD = \LC_\theD+\sum_{i\in\theD'}\Delta\brho_i.
\end{equation} 
By translation equivariance of the $\rho_i,$  \eqref{eq:changerhogen} yields
 \begin{equation*}
    \Delta\brho_i =  
        r_i(\bo{q}'_i) -\mathbf{1}_{i\neq c} r_i(\bo{q}_i) + (\bo{q}'_i)^\top \bo{p}'^\theD - \bo{q}_i^\top \bo{p}^\theD , \; i\in \theD'  .
\end{equation*}
A further computation based on \eqref{e:lmcD} then yields 
\begin{equation}\label{eq:marketcost}
\MC  =    \MC_\theD =\sum_{i\in \theD'}(\Delta\bo{q}_i)^\top  \bo{p}^\theD +   \sum_{i\in \theD' } 
  \big(r_i(\bo{q}'_i ) -\mathbf{1}_{i\neq c} r_i (\bo{q}_i)\big). 
\end{equation}
 
\section{Market Cost: the Case of Entropic Risk Measures}\label{sec:entropic}
Throughout this section, we assume that the risk preference of each trading participant $i$ in $\theD \cup \theD'$ is 
an entropic risk measure of the form 
\beql{rhoil}
 \rho_i(L) = \frac{1}{\varrho_i} \ln(\E  [\exp(\varrho_i L)]), \mbox{ for some }\varrho_i>0.
\eeql
We also assume that each ($R_i,P$) is jointly normal, so that  $r_i(q_i)$ is given by  \eqref{eq:entropicr}.
Building on these assumptions and Proposition \ref{prop:entropy}, which provides a unique closed-form solution for the Radner equilibrium, we use the equilibrium outcomes before and after default to express liquidation and market costs for default resolution scenarios involving either total liquidation or complete hedging on the CCP's exchange. We numerically demonstrate these calculations through a specific example, where both liquidation and market costs are derived and analyzed based on the given parameter configuration.

\subsection{Liquidation on $\theD$}\label{ss:entropicliqD}

\begin{proposition}\label{prop:lcnormal}
    Let $(R_i,  P )\sim \mathcal{N}_{m+1}(\mu_i, \Gamma_i)$,  $i\in \theD \cup \theD'$, and $\Gamma$ be invertible.
    If the CCP liquidates the defaulter position $\bo{q}_d$ on its own exchange $\theD$, then 
      \beql{eq:mcnormalentro}
    &  \LC_\theD=0 \quad \text{and}\\
     & \MC_\theD  =\frac{1}{2}\varrho'  \bo{q}_d^\top\Gamma \bo{q}_d-\varrho' \Big[ \Gamma^{-1}\sum_{j\in \theD'\setminus \theD} \Big(\frac{\varrho}{\varrho_j}\mathrm{cov}-\mathrm{cov}_j \Big)\Big]^\top    \Big( \mathrm{cov}' +\frac{1}{2}\Gamma \bo{q}_d \Big)  + \\
 &\qqq   \sum_{j\in \theD'\setminus\theD  }  \Big( \Gamma^{-1} ( \varrho\mathrm{cov}-\varrho_j\mathrm{cov}_j ) \Big)^\top  \Big(\mathrm{cov}_j+\frac{1}{2}\Gamma \bo{q}'_j \Big),
  \eeql
  where  $\varrho' =\left(\sum_{i\in   \theD'} \frac{1}{\varrho_i}\right)^{-1}$.
\end{proposition}
\proof 
By the clearing condition $\sum_{i\in \theD'}\bo{q}'_i =0$ (established like \eqref{e:Dq}), \eqref{e:lmcD} yields $\LC_\theD = \sum_{i\in \theD'} (\bo{q}'_i)^\top (\bo{p}^\theD-\bo{p'}^\theD)=0$.
Letting $\theD$ and $\theD'$ successively play
the role of $\anE$ in Proposition \ref{prop:entropy}, the pre-default and the post-default equilibrium are uniquely given by 
\begin{equation}\label{eq:prepositionentropy}
    \bo{q}_i =\Gamma^{-1}\Big(\frac{\varrho}{\varrho_i} \mathrm{cov}-\mathrm{cov}_i\Big)\sp i\in \theD;~ \bo{p}^\theD =\mu - \varrho \mathrm{cov},
  \end{equation}
  and 
  \begin{equation}\label{eq:postpositionentropy}
     \bo{q}'_i =  \Gamma^{-1}\Big(\frac{\varrho' }{\varrho_i} \mathrm{cov}' -\mathrm{cov}_i\Big), \quad i\in \theD';~ 
        \bo{p}'^\theD=\mu-\varrho'  \mathrm{cov}',
  \end{equation}
  where $\varrho = \Big(\sum_{i \in \theD} \frac{1}{\varrho_i}\Big)^{-1}$, $\mathrm{cov}=\sum_{i \in\theD}\mathrm{cov}_i$,  $\varrho' =\left(\sum_{i\in   \theD'} \frac{1}{\varrho_i}\right)^{-1}$,  $\mathrm{cov}' = \sum_{i\in  \theD'} \mathrm{cov}_i$.
  From \eqref{eq:entropicr}, we obtain
  \begin{equation}\label{eq:change}
      r_i(\bo{q}'_i\big) - r_i\big(\bo{q}_i) = -\Delta \bo{q}_i ^\top \mu + \varrho_i \Delta \bo{q}_i ^\top\Big[\mathrm{cov}_i + \Gamma\Big(\bo{q}_i + \frac{1}{2}\Delta\bo{q}_i\Big)\Big]\sp i\in  \theD' .
  \end{equation}
As also $\sum_{i\in \theD'}\Delta\bo{q}_i = \bo{q}_d $ holds in a liquidation setup and since $\bo{q}_i=0$ for each trading participant $i\in \theD'\setminus \theD$, \eqref{eq:marketcost} yields
  \begin{multline}\label{eq:mcsimpl}
      \MC_\theD = \bo{q}_d^\top  (\bo{p}^\theD - \mu ) + \sum_{i \in \theD' } \varrho_i (\Delta \bo{q}_i)^\top\Big[\mathrm{cov}_i + \Gamma\Big(\bo{q}_i + \frac{1}{2}\Delta\bo{q}_i\Big)\Big] \\
     = -\varrho \bo{q}_d^\top  \mathrm{cov}+ \sum_{i \in \theD' } \varrho_i (\Delta \bo{q}_i)^\top \Big[ \mathrm{cov}_i +\Gamma \Big(\bo{q}_i+\frac{1}{2} \Delta\bo{q}_i \Big)\Big],  \hspace{3cm}
  \end{multline}
by the second identity in \eqref{eq:prepositionentropy}. 

To compute $\Delta\bo{q}_i$ therein, note that
  \begin{multline*}
    \mathrm{cov}' =\mathrm{cov}-\mathrm{cov}_d +\sum_{j \in \theD' \setminus \theD}\mathrm{cov}_j 
    =\frac{\varrho}{\varrho_d} \mathrm{cov} + \Big(1-\frac{\varrho}{\varrho_d}\Big)\mathrm{cov}-\mathrm{cov}_d +\sum_{j \in \theD' \setminus \theD}\mathrm{cov}_j \\
    =\Gamma \bo{q}_d + \frac{\varrho}{\varrho' }\mathrm{cov}-\sum_{j\in \theD' \setminus \theD}\Big(\frac{\varrho}{\varrho_j}\mathrm{cov}-\mathrm{cov}_j\Big), \hspace{5.5cm}
  \end{multline*}
by the first identity in \eqref{eq:prepositionentropy} and the fact that  $\frac{1}{\varrho'} =\frac{1}{\varrho} -\frac{1}{\varrho_d} + \sum_{j\in \theD'\setminus\theD} \frac{1}{\varrho_j}$.
  This implies 
  \begin{equation}\label{eq:gammacov}
      \varrho' \mathrm{cov}' =\varrho' \Gamma \bo{q}_d + \varrho \mathrm{cov}-\varrho' \sum_{j \in \theD' \setminus \theD}\Big(\frac{\varrho}{\varrho_j}\mathrm{cov}-\mathrm{cov}_j\Big).
  \end{equation}
The definition of  $\Delta \bo{q}_i$,  \eqref{eq:prepositionentropy},  \eqref{eq:postpositionentropy}, and  \eqref{eq:gammacov} yield
  \begin{equation}\label{eq:changeq}
  \Delta \bo{q}_i = 
  \begin{cases}
      \frac{\varrho'}{\varrho_i} \bo{q}_d - \frac{\varrho'}{\varrho_i} \Gamma^{-1}\sum_{j \in \theD'\setminus \theD} \big(\frac{\varrho}{\varrho_j}\mathrm{cov}-\mathrm{cov}_j\big) &  i\in \theD\setminus \{d\}\\
       \frac{\varrho'}{\varrho_i} \bo{q}_d - \frac{\varrho'}{\varrho_i} \Gamma^{-1}\sum_{j \in \theD'\setminus \theD} \big(\frac{\varrho}{\varrho_j}\mathrm{cov}-\mathrm{cov}_j\big) + \Gamma^{-1} \big( \frac{\varrho}{\varrho_i}\mathrm{cov}-\mathrm{cov}_i \big) & i\in \theD'\setminus \theD.
  \end{cases}
  \end{equation}
As  $\sum_{i\in \theD\setminus \{d\}} \bo{q}_i = -\bo{q}_d$ and  $\sum_{i\in \theD'} \Delta \bo{q}_i =\bo{q}_d$, substituting  \eqref{eq:changeq} into \eqref{eq:mcsimpl} yields
\begin{multline*}
    \MC_\theD  =-\varrho \bo{q}_d^\top  \mathrm{cov}+  \Big[ \varrho' \bo{q}_d -\varrho' \Gamma^{-1} \sum_{j \in \theD'\setminus \theD} \Big(\frac{\varrho}{\varrho_j}\mathrm{cov}-\mathrm{cov}_j\Big)\Big]^\top \Big( \mathrm{cov}' -\frac{1}{2}\Gamma \bo{q}_d \Big)  + \\
    \sum_{j \in \theD'\setminus\theD} \Big(  \Gamma^{-1} ( \varrho\mathrm{cov}-\varrho_j\mathrm{cov}_j) \Big)^\top  \Big(\mathrm{cov}_j+\frac{1}{2}\Gamma \bo{q}'_j \Big), \hspace{3cm}
\end{multline*} 
whence the expression for $\MC_\theD$ in  \eqref{eq:mcnormalentro} .\ \finenv
  
\begin{remark}\label{rem:lcnormal}
To cope with the case of several instant defaulters $d$ at time 0, one just needs to replace $\bo{q}_d$, $ \mathrm{cov}_d $ and $\frac{\varrho}{\varrho_d}$ by $\sum_d \bo{q}_d$, $\sum_d \mathrm{cov}_d$ and $\sum_d \frac{\varrho}{\varrho_d}$ in Proposition \ref{prop:lcnormal} and its proof.\ \finenv
 \end{remark}

\begin{remark}
  If the CCP liquidates the defaulter's position among the surviving members, i.e.~for $\theD'\setminus\theD=\varnothing$ in the above, then Proposition \ref{prop:lcnormal} yields
  \begin{equation*}
    \MC_\theD = \frac{1}{2}\varrho'  \bo{q}_d^\top  \Gamma \bo{q}_d\geq  0.
 \end{equation*}
Using \eqref{eq:prepositionentropy}, \eqref{eq:change}, and \eqref{eq:gammacov}, we obtain  $\bo{p}'^\theD=\bo{p}^\theD -\varrho' \Gamma \bo{q}_d.$
Moreover, \eqref{eq:changeq} yields $\Delta \bo{q}_i =\frac{\varrho' }{\varrho_i}\bo{q}_d$, $i\in \theD\setminus \{d\}$.
In the case $m=1$ for simplicity, the reason why  $\MC\geq 0$  when the CCP  liquidates among the surviving member can thus be explained as follows.
If $\bo{q}_d > 0$, then the CCP replaces the defaulter's contract with each surviving member by selling at a ``fire sales'' price $\bo{p}'^\theD< \bo{p}^\theD$.
If $\bo{q}_d< 0$, then the CCP buys from each surviving member at a ``dear'' price $\bo{p}'^\theD>\bo{p}^\theD$.
In both cases, there is a market cost.\ \finenv\end{remark}

\begin{example}\label{ex:ccpliquidation}
    Let $\theD=\{1,\dots,15\}$, $d=\{15\}$, $\theD'\setminus D=\emptyset,$ 
    $\varrho_i=1$, $m=1$ and $(R_i, P) \sim \mathcal{N}_{2}(\mu_i, \Gamma_i)$, $i\in \theD$ .
    Suppose $\mathrm{cov}_i = c_i \sigma \sqrt{\mathbb{V}\mathrm{ar} (R_i)}$, where $\sigma^2=\mathbb{V}\mathrm{ar}(P)$, $c_i =(-1)^{i+1} 0.8$ (the correlation coefficient between $R_i$ and $P$), and $\mathbb{V}\mathrm{ar} (R_i)=0.09i^2$, $i \in \theD$.
    Fix $\sigma= 0.2$.
    The corresponding pre- and post-default optimal positions computed from \eqref{eq:prepositionentropy} and \eqref{eq:postpositionentropy} are given by Table \ref{tab:practexample}.
    \begin{table}[htp]  
        \begin{centering}
            \begin{tabular}{@{}lcccccccc@{}}
                \toprule
                $\CM_i$ &$1$  &$2$  &$3$     &$4$     &$5$   &$6$  &$7$  &$8$ \\
                \midrule
                $\mathrm{cov}_i$ &$0.05$   &$-0.10$ &$0.14$   &$-0.19$ &$0.24$   &$-0.29$   &$0.34$   &$-0.38$ \\
                $\bo{q}_i$ &$-0.56$   &$3.04$  &$-2.96$  &$5.44$  &$-5.36$   &$7.84$   &$-7.76$    &$10.24$\\
                $\bo{q}'_i$ &$-1.80$  &$1.80$  &$-4.20$  &$4.20$ &$-6.60$  &$6.60$   &$-9.00$  &$9.00$ \\
                \bottomrule
                \\
                \toprule
                $\CM_i$ &$9$ &$10$   &$11$  &$12$   &$13$ &$14$  &$15$  &{} \\
                \midrule
                $\mathrm{cov}_i$ &$ 0.43$ &$-0.48$   &$0.53$  &$-0.58$   &$ 0.62$   &$-0.67$  &$0.72$   & {}\\
                $\bo{q}_i$ &$-10.16$   &$12.64$   &$-12.56$  &$15.04$    &$-14.96$  &$17.44$    &$-17.36$ & \\ 
                $\bo{q}'_i$ &$-11.40$    &$11.40$    &$-13.80$   &$13.80$    &$-16.20$  &$16.20$  &{}  & {}\\
                \bottomrule
            \end{tabular}
            \caption{Pre- and post-default optimal positions of each clearing member $i$ ($\CM_i$) when the CCP liquidates $\bo{q}_d$ on  its own exchange in the entropic case.}
        \label{tab:practexample}
    \end{centering}
\end{table}
\noindent
Note that  each $\bo{q}_i$ or $\bo{q}'_i$ is positive (negative) provided
   $\mathrm{cov}_i$  is negative (positive) (here and again in Table \ref{tab:practexample3} below), in line with the hedging feature of the exchange.
    For $\mu =2$, Proposition \ref{prop:lcnormal} and its proof yield $\bo{p}^\theD= 1.97, {\bo{p}'} ^\theD= 2.02, $ and $\MC_\theD=0.43$  
    (with $\LC_\theD=0$ as per the first line in Table \ref{tab:SixCasesSummary}).\ \finenv\end{example}

As it can be seen from the analytical expressions   \eqref{eq:prepositionentropy}-\eqref{eq:postpositionentropy} and Table \ref{tab:practexample}, the covariance matrices $\Gamma_i, i \in \theD \cup \theD'$,  are the major driving factors for portfolio and price changes.
But the number and the risk preferences of the trading participants can also significantly affect these optimal quantities and the market cost:
\begin{example}\label{eg:lcnormal} 
  Let $\theD=\{1,\dots, n+1\}$, $d=\{n+1\}$,  $\varrho_i=\varrho_1$, and  $(R_i, P) \sim \mathcal{N}_2(\mu_i, \Gamma_i)$, $i \in \theD$.
  Suppose $\mathrm{cov}_i=0$, $i \in \{1,\dots, n-1\}$, and  $\mathrm{cov}_{n}=-\mathrm{cov}_{n+1}=\delta$.
  We consider two cases.
   {\rm\hfill\break \textbf{(i)}}  $ \theD'\setminus \theD =\varnothing$.
    By \eqref{eq:prepositionentropy} and  \eqref{eq:postpositionentropy}, the pre- and post-default portfolios are then given by
    \begin{equation*}
      \bo{q}_i=
      \begin{cases}
        0, & i=1,\dots,n-1, \\
       -\frac{\delta}{\sigma^2}, & i=n,\\
       \frac{\delta}{\sigma^2}, & i = n+1,
    \end{cases} \qquad \text{and} \qquad 
     \bo{q}'_i =
   \begin{cases}
      \frac{\delta}{n \sigma^2}, & i=1,\dots,n-1 \\
      -\frac{(n-1)\delta}{n \sigma^2},  & i = n.\\
    \end{cases}
   \end{equation*}
   The pre- and post-default prices are given by 
    \begin{equation*}
        \bo{p}^\theD=\mu \qquad \text{and} \qquad \bo{p}'^\theD=\mu-\frac{\varrho_1 \delta}{n}.
    \end{equation*}
   A further computation based on  \eqref{eq:mcnormalentro} yields
  \begin{equation*}
    \MC = \frac{\varrho_1}{2n}\left( \frac{\delta}{\sigma}\right)^2,
  \end{equation*}
which decreases to $0$ as the number $n$  of surviving members  to $\infty$.
 {\rm\hfill\break \textbf{(ii)}} $ \theD'\setminus \theD =\{n+2\}$ with  $\mathrm{cov}_{n+2}=\delta'$.
 In this case, the pre-default equilibrium is the same as in case {\rm (i)}, while the post-default equilibrium is given by
  \begin{equation*}
  \bo{q}'_i =
        \begin{cases}
        \frac{\delta+\delta'}{(n+1) \sigma^2}, & i=1,\dots,n-1, \\
        \frac{\delta'-n \delta}{(n+1) \sigma^2}, & i = n,   \\
       \frac{\delta-\delta'n}{(n+1) \sigma^2}, & i = n+2. 
       \end{cases} \qquad \text{and} \qquad \bo{p}'^\theD=\mu-\frac{\varrho_1 (\delta+\delta')}{n+1}.
  \end{equation*}
   A further computation based on 
   \eqref{eq:mcnormalentro} yields
  \begin{equation*}
    \MC= \frac{2\varrho_1 \delta^2 +2\varrho_1\delta'\delta-\varrho_1(\delta')^2n}{2\sigma^2 (n+1)},
  \end{equation*}
the sign of which depends on the value of the parameters. \ \finenv\end{example}

\subsection{Hedging on $\theD$}\label{ss:entropichedD}

We now turn to the ``hedging on own exchange $\theD$" case of Section  \ref{ss:he}, but with possibly new participants beyond the CCP $c$ in $\theD'$.

\begin{proposition}\label{prop:hcnormal}
  Let $(R_i,  P )\sim \mathcal{N}_{m+1}(\mu_i, \Gamma_i)$,  $i\in \theD \cup  \theD'$, with invertible covariance matrix $\Gamma$ of $ P $.
    When the CCP hedges the defaulter position $\bo{q}_d$ on its own exchange $\theD$, then  
   \beql{eq:hcnormalentro}
    &  \LC_\theD=-\bo{q}_d^\top (\bo{p}^\theD-\bo{p}'^\theD) ,\\
     & \MC_\theD  = \sum_{i \in \theD' } \varrho_i (\Delta \bo{q}_i)^\top  \Big[\mathrm{cov}_i + \Gamma \big(\bo{q}_i+\frac{1}{2}\Delta\bo{q}_i\big) \Big] + \Big( \frac{\varrho_c \mathbb{V}\mathrm{ar}(R_c)}{2} -\E  [R_c] \Big),
  \eeql
  where  $\varrho' =\left(\sum_{i\in   \theD'} \frac{1}{\varrho_i}\right)^{-1}$.
  In the absence of new participants (other than the CCP itself $c$) to $\theD'$,  \begin{equation}\label{eq:hcnormalentro2}
      \MC =\frac{\varrho^2(\varrho'-\varrho_\thec)}{2{\varrho_\thec}^2}\mathrm{cov}^\top   \Gamma^{-1} \mathrm{cov} +\frac{\varrho'}{2}\bo{q}_d^\top \Gamma \bo{q}_d-\frac{\varrho'\varrho}{\varrho_\thec}\bo{q}_d^\top \mathrm{cov}.
  \end{equation}
\end{proposition}

\proof 
By the clearing condition $\sum_{i\in \theD'}\bo{q}'_i =-\bo{q}_d$ (established like \eqref{eq:cchedgingonE}), \eqref{e:lmcD} yields $\LC_\theD = \sum_{i\in \theD'} (\bo{q}'_i)^\top (\bo{p}^\theD-\bo{p}'^\theD)=-\bo{q}_d^\top (\bo{p}^\theD-\bo{p}'^\theD)$.
Applying Proposition \ref{prop:entropy}
to $\anE=\theD$,
the pre-default equilibrium is uniquely given by \eqref{eq:prepositionentropy}.
As for the post-default equilibrium,  if the CCP hedges on $\theD$, then, in view of Remark \ref{rem:cchedging}, introducing the changed variable $\bo{z}'_i=\bo{q}'_i-\bo{q}_i$ and  $R'_i  = R_i+\bo{q}_i^\top P$ yields
  \begin{equation*}
   r'_i(z_i) \eqdef \rho_i(-R'_i  -z_i^\top  P )= \rho_i(-R_i - (z_i +\bo{q}_i)^\top  P ),
   \quad i \in \theD'.
  \end{equation*}
Following the proof of Proposition \ref{prop:entropy} with $r'_i$ here in the role of $r_i$ there, we obtain a unique post-default equilibrium 
   \begin{equation}\label{eq:postpositionentropyb}
\bo{q}'_i =\Gamma^{-1} \left(\frac{\varrho' }{\varrho_i} \mathrm{cov}' -\mathrm{cov}_i\right)-\frac{\varrho' }{\varrho_i}\bo{q}_d, \quad i \in \theD', \quad \text{and} \quad 
   \bo{p}'^\theD=\mu-\varrho'  \mathrm{cov}'  +\varrho'  \Gamma \bo{q}_d.
  \end{equation}
Hence, by \eqref{eq:entropicr},
  \begin{equation*}
      r_i(\bo{q}'_i\big) - r_i\big(\bo{q}_i) = -\Delta \bo{q}_i ^\top \mu + \varrho_i (\Delta \bo{q}_i)^\top  \big[\mathrm{cov}_i + \Gamma \big(\bo{q}_i + \frac{1}{2}\Delta\bo{q}_i\big) \big], \quad i\in\theD'.
  \end{equation*}
  This and $\sum_{i\in \theD'}\Delta\bo{q}_i =0$ reduce \eqref{eq:marketcost} to \eqref{eq:hcnormalentro}.

In the absence of new participants other than the CCP itself to $\theD'$,  \eqref{eq:prepositionentropy} and \eqref{eq:postpositionentropyb} yield
\begin{equation*}
      \Delta \bo{q}_i = \begin{cases}
          \frac{1}{\varrho_i} \Gamma^{-1} (\varrho'\mathrm{cov}' -\varrho\mathrm{cov})- \frac{\varrho'}{\varrho_i}\bo{q}_d & i \in \theD\setminus\{d\},\\
          \frac{1}{\varrho_i}\Gamma^{-1}\varrho'\mathrm{cov}'  -\bo{q}_d - \frac{\varrho'}{\varrho_i}\bo{q}_d & i=\thec.
      \end{cases}
  \end{equation*}
One can check that the value of $\varrho'\mathrm{cov}'$  given by  \eqref{eq:gammacov} also holds true for the hedging case.
Hence, using $\varrho'\mathrm{cov}'$  given by  \eqref{eq:gammacov}  and $\mathrm{cov}_\thec = \Gamma \bo{q}_d$, we obtain 
  \begin{equation}\label{eq:positionchange}
      \Delta \bo{q}_i = \begin{cases}
          \frac{\varrho'}{\varrho_i} \Gamma^{-1} \big(\Gamma \bo{q}_d -\frac{\varrho}{\varrho_\thec}\mathrm{cov}  \big), & i \in \theD\setminus\{d\},\\
           \frac{\varrho'}{\varrho_i} \Gamma^{-1} \big(\Gamma \bo{q}_d -\frac{\varrho}{\varrho_\thec}\mathrm{cov}  \big) +\frac{\varrho}{\varrho_i}\Gamma^{-1}\mathrm{cov} -\bo{q}_d, & i=\thec.
      \end{cases}
  \end{equation}
  Since $R_\thec =\bo{q}_d^\top (P-\bo{p}^\theD)$, we have  $\mathbb{V}\mathrm{ar}(R_\thec) =\bo{q}_d^\top \Gamma \bo{q}_d$ and $\E  [R_\thec] =\varrho\bo{q}_d^\top \mathrm{cov}$.
  Further computations using  $\Delta \bo{q}_i$ given by \eqref{eq:positionchange} reduce \eqref{eq:hcnormalentro} to \eqref{eq:hcnormalentro2}.
  \ \finproof

\begin{example}\label{ex:CCPHedging}
 In the ``CCP hedging on its own exchange'' case,
let $\theD=\{1,\dots,15\}$, $d=\{15\}$, $\theD'\setminus \theD =\{\mathrm{c}\}$, $\varrho_i=1=\varrho_\mathrm{c}$, and  $(R_i, P) \sim \mathcal{N}_2(\mu_i, \Gamma_i)$ with $\mathrm{cov}_i =\sigma c_i  \sqrt{\mathbb{V}\mathrm{ar} (R_i)}$, $\sigma= 0.2$, $c_i=(-1)^{i+1} 0.8$, and $\mathbb{V}\mathrm{ar} (R_i)=0.09i^2$, $i \in \theD$.
The corresponding pre- and post-default optimal positions computed from \eqref{eq:prepositionentropy} and \eqref{eq:postpositionentropyb} are given in Table \ref{tab:practexample2}.
    \begin{table}[htp]  
        \begin{centering}
            \begin{tabular}{@{}lcccccccc@{}}
                \toprule
                $\CM_i$ &$1$  &$2$  &$3$     &$4$     &$5$   &$6$  &$7$  &$8$ \\
                \midrule
               $\mathrm{cov}_i$ &$0.05$   &$-0.10$ &$0.14$   &$-0.19$ &$0.24$   &$-0.29$   &$0.34$   &$-0.38$ \\ 
                $\bo{q}_i$ &$-0.56$   &$3.04$  &$-2.96$  &$5.44$  &$-5.36$   &$7.84$   &$-7.76$    &$10.24$\\ 
                $\bo{q}'_i$ &$-1.76$  &$1.84$  &$-4.16$  &$4.24$ &$-6.56$  &$6.64$   &$-8.96$  &$9.04$ \\
                \bottomrule
                \\
                \toprule
                $\CM_i$ &$9$ &$10$   &$11$  &$12$   &$13$ &$14$  &$15$  &$\mathrm{c}$ \\
                \midrule
                $\mathrm{cov}_i$ &$ 0.43$ &$-0.48$   &$0.53$  &$-0.58$   &$ 0.62$   &$-0.67$  &$0.72$   &$-0.69$\\ 
                $\bo{q}_i$ &$-10.16$   &$12.64$   &$-12.56$  &$15.04$    &$-14.96$  &$17.44$    &$-17.36$ & \\ 
                $\bo{q}'_i$ &$-11.36$    &$11.44$    &$-13.76$   &$13.84$    &$-16.16$  &$16.24$  &{}  & {}$16.80$\\
                \bottomrule
            \end{tabular}
            \caption{Pre- and post-default optimal positions when the CCP hedges on its own exchange  $\theD$ with $\theD'\setminus\theD=\emptyset$ in the entropic case.}
        \label{tab:practexample2}
    \end{centering}
\end{table}
\noindent   
    For $\mu=2$, Proposition \ref{prop:hcnormal} and its proof yield $\bo{p}^\theD= 1.97, \bo{p}'^\theD= 2.02$, and
$\MC_\theD=0.42$ (with, by the third line in Table \ref{tab:SixCasesSummary}, $\LC_\theD=-0.83$).\ \finenv\end{example}

Table \ref{tab:comparison} displays the impacts of the default resolution of $d$ in terms of $\LC_i$ and $\Delta\brho_i, i\in\theD'$.
As can be seen from the table, the impact of the default resolution on the $\Delta\brho_i$ is almost the same in the liquidation and hedging cases, whereas its impact on the $\LC_i$ is significantly different in the two cases.
\begin{table}[htp] 
        \begin{centering}
        {
            \begin{tabular}{@{}llcccccccc@{}}
                \toprule
              &  $\CM_i$ &$1$  &$2$  &$3$     &$4$     &$5$   &$6$  &$7$  &$8$ \\
                \cmidrule{2-10}
         \multirow{2}{*}{Liquidation}    & $\LC_i$ &$0.09$  &$-0.09$  &$0.21$  &$-0.21$ &$0.33$  &$-0.33$   &$0.45$  &$-0.45$ \\
      &  $\Delta\brho_i$ &$-0.06$  &$0.1$  &$-0.18$  &$0.24$ &$-0.30$  &$0.36$   &$-0.42$  &$0.48$ \\ 
                \midrule
        \multirow{2}{*}{Hedging}      & $\LC_i$ &$0.03$  &$-0.15$  &$0.14$  &$-0.26$ &$0.26$  &$-0.38$   &$0.37$  &$-0.49$ \\
           &  $\Delta\brho_i$ & $-0.06$  & $0.12$   &$-0.17$   &$ 0.23$  &$-0.29$   &$0.35$    &$-0.40$   &$0.46$  \\ 
                \bottomrule 
                \\
                \toprule
            &    $\CM_i$ &$9$ &$10$   &$11$  &$12$   &$13$ &$14$  &$\thec$  &{} \\
                \cmidrule{2-9}
     \multirow{2}{*}{Liquidation}     & $\LC_i$ &$0.56$  &$-0.56$  &$0.68$  &$-0.68$ &$ 0.80$  &$-0.80$   &  &\\
      &  $\Delta\brho_i$ &$-0.53$  &$0.60$  &$-0.65$  &$0.72$ &$-0.77$  &$0.83$   &  & \\ 
                \cmidrule{1-9}
       \multirow{2}{*}{Hedging}        & $\LC_i$ &$0.49$  &$-0.61$  &$0.60$  &$-0.72$ &$0.72$  &$-0.84$   & $0$ & \\
      &  $\Delta\brho_i$ &$-0.52$   &$0.58$   &$-0.63$   &$0.69$  &$-0.75$  &$0.81$    &$0.83$ &   \\ 
                \bottomrule
            \end{tabular}
        }
        \caption{Impacts  of the default resolution on the $\LC_i $ and $\Delta\brho_i$ in the liquidation and hedging cases in the entropic risk measure examples \ref{ex:ccpliquidation}-\ref{ex:CCPHedging}.}
        \label{tab:comparison}
    \end{centering}
\end{table}
 
\section{Market Cost: the Case of Expected Shortfall}\label{sec:es}
Throughout this section, we assume that the risk preferences of each market participant is an expected shortfall $\mathbb{ES}_{\alpha_i}$ as per \eqref{e:ES}; each vector $(R_i, P )\sim \mathcal{E}_{m+1}(\mu_i, \Gamma_i, \psi)$ (or, sometimes, a more specific  $\mathcal{T}_{m+1}(\mu_i, \Gamma_i, \nu_i)$); the CCP only operates on its own exchange $\theD$.
Hence $\MC=\MC_\theD$ as in \eqref{eq:marketcost}, with, by \eqref{eq:eselliptical},
  \beql{e:rit}
  & r_i(q_i)=-\E  [R_i]-q_i^\top \mu+ \mathbb{ES}_{\alpha_i}(Z_i)\sqrt{\mathbb{V}\mathrm{ar}(R_i)+2q_i ^\top \mathrm{cov}_i+q_i^\top \Gamma q_i}\\&\qqq  \mbox{where $Z_i\sim \mathcal{E}_1(0,1,\psi)$ (or~$Z_i\sim \mathcal{T}_{1}(0, 1, \nu_i)$)}\sp i\in \theD\cup \theD'.
  \eeql

In what follows we provide numerical applications of Proposition \ref{prop:es} and \eqref{e:rit} in different cases.

\subsection{Liquidation on $\theD$}\label{ss:esliqD}
   
Let us first consider the liquidation case of Section \ref{sec:liqonE}, but with possibly new participants in $\theD'$.
 If $(R_i,  P )\sim \mathcal{E}_{m+1}(\mu_i, \Gamma_i, \psi)$ with $\Gamma_i$ positive definite, $i\in \theD \cup \theD'$, then, by Proposition \ref{rem:elliptical}, there exists a unique pre-default Radner equilibrium  $((\bo{q}_i)_{i \in \theD}, \bo{p}^\theD)$.
Following \eqref{eq:membopt} and \eqref{eq:smalrgeneral}, the pre-default member $i \in \theD$ optimality condition yields 
\begin{equation}\label{eq:loptimalprice}
    \bo{p}^\theD = \mu-\frac{\mathbb{ES}_{\alpha_i}(Z)}{\sqrt{\mathbb{V}\mathrm{ar}(R_i)+2\bo{q}_i ^\top\mathrm{cov}_i+\bo{q}_i ^\top\Gamma\bo{q}_i}} \left(\mathrm{cov}_i +\Gamma\bo{q}_i\right), \quad i\in \theD.
\end{equation}
The zero clearing  condition  and \eqref{eq:loptimalprice} imply
\begin{equation}\label{eq:loptimalposition1}
    \begin{cases}
        \sum_{i \in \theD} \bo{q}_i =0\\
        \frac{\mathbb{ES}_{\alpha_1}(Z)} {\mathbb{ES}_{\alpha_i}(Z) } \sqrt{\frac{\mathbb{V}\mathrm{ar}(R_i)+2\bo{q}_i ^\top\mathrm{cov}_i+\bo{q}_i ^\top\Gamma\bo{q}_i}{\mathbb{V}\mathrm{ar}(R_1)+2\bo{q}_1^\top \mathrm{cov}_1+\bo{q}_1^\top \Gamma\bo{q}_1}}\left(\mathrm{cov}_1 +\Gamma\bo{q}_1\right) -\mathrm{cov}_i -\Gamma\bo{q}_i=0, & i \in \theD.
    \end{cases}
\end{equation}
Therefore, computing the optimal position is equivalent to finding the root of a vector function from $\mathbb{R}^{m|\theD|}$ into $\mathbb{R}^{m|\theD|}$.
Likewise, the post-default optimal positions solve
\begin{equation}\label{eq:loptimalposition2}
    \begin{cases}
        \sum_{i\in \theD' }  \bo{q}'_i =0\\
        \frac{\mathbb{ES}_{\alpha_1}(Z)} {\mathbb{ES}_{\alpha_i}(Z) } \sqrt{\frac{\mathbb{V}\mathrm{ar}(R_i)+2(\bo{q}'_i)^\top\mathrm{cov}_i+(\bo{q}'_i)^\top\Gamma\bo{q}'_i}{\mathbb{V}\mathrm{ar}(R_1)+2{\bo{q}'_1}^\top \mathrm{cov}_1+{\bo{q}'_1}^\top \Gamma\bo{q}'_1}}\left(\mathrm{cov}_1 +\Gamma\bo{q}'_1\right) -\mathrm{cov}_i -\Gamma\bo{q}'_i=0, & i\in \theD'
    \end{cases}
\end{equation}
and the post-default price is equal to
\begin{equation*}
  \bo{p}'^\theD= \mu-\frac{\mathbb{ES}_{\alpha_i}(Z_i)}{\sqrt{\mathbb{V}\mathrm{ar}(R_i)+2(\bo{q}'_i)^\top\mathrm{cov}_i+(\bo{q}'_i)^\top \Gamma{\bo{q}'_i}}} \left(\mathrm{cov}_i +\Gamma\bo{q}'_i\right), \quad i\in\theD'.
\end{equation*}

In the remainder of this subsection, we assume that $m=1$, $\alpha_i=0.975$ for each market participant, $\mu=2$, and  $\sigma = 0.2$.

\begin{example}[\textbf{Multivariate normal}]\label{eg:lces1}
    Let $\theD=\{1,\dots,15\}$, $d=\{15\}$, $\theD'\setminus\theD=\emptyset$, $\varrho_i=1$,   $(R_i, P) \sim \mathcal{N}_2(\mu_i, \Gamma_i)$, $\mathrm{cov}_i = c_i \sigma \sqrt{\mathbb{V}\mathrm{ar} (R_i)}$ with $c_i = (-1)^{i+1} 0.8$, and $\mathbb{V}\mathrm{ar} (R_i)=0.09i^2$, $i\in \theD$.
    Solving \eqref{eq:loptimalposition1}-\eqref{eq:loptimalposition2} yields the pre- and post-default optimal positions displayed in Table \ref{tab:practexample3}.
     \begin{table}[htp]  
        \begin{centering}
            \begin{tabular}{@{}lcccccccc@{}}
                \toprule
                $\CM_i$ &$1$ &$2$  &$3$  &$4$  &$5$  &$6$   &$7$  &$8$ \\
                \midrule
                $\mathrm{cov}_i$ &$0.05$   &$-0.10$ &$0.14$   &$-0.19$ &$0.24$   &$-0.29$   &$0.34$   &$-0.38$ \\ 
                $\bo{q}_i$  &$-1.12$  &$2.56$  &$-3.36$  &$5.12$   &$-5.60$  &$7.68$  &$-7.84$   &$10.24$ \\ 
                $\bo{q}'_i$ &$-1.28$  &$2.24$  &$-3.84$  &$4.48$  &$-6.40$  &$6.72$  &$-8.96$  &$8.96$ \\
                \bottomrule
                \\
                \toprule
                $\CM_i$ &$9$  &$10$  &$11$   &$12$  &$13$  &$14$ &$15$  &{} \\
                \midrule
                $\mathrm{cov}_i$ &$ 0.43$ &$-0.48$   &$0.53$  &$-0.58$   &$ 0.62$   &$-0.67$  &$0.72$   & {}\\ 
                $\bo{q}_i$  &$-10.08$   &$12.80$   &$-12.32$  & $15.36$   &$-14.56$   &$17.92$  &$-16.80$   &\\ 
                $\bo{q}'_i$ &$-11.52$   &$11.20$   &$-14.08$  &$13.44$  &$-16.64$   &$15.68$   & \\
                \bottomrule
            \end{tabular}
            \caption{Pre- and post-default optimal positions when the CCP liquidates on its own exchange  $\theD$ with  $\theD'\setminus\theD=\emptyset$ in the expected shortfall case.}
        \label{tab:practexample3}
    \end{centering}
\end{table}
\noindent
The above results also yield  $\bo{p}^\theD = 1.96, \bo{p}'^\theD = 2.04$, and
 $\MC_\theD=0.69$ (with $\LC=0$ as per the first line in Table \ref{tab:SixCasesSummary}).\ \finenv\end{example}

\begin{example}[\textbf{Multivariate Student $t$}]
 Reconsider Example \ref{eg:lces1} but for $(R_i, P) \sim \mathcal{T}_2(\mu_i, \Gamma_i, \nu_i)$ with $\nu_i= \nu=2.5$, $i \in \theD$.
   \eqref{e:rit} yields
  \begin{align*}
    &  r_i(q_i) =  -\E  [R_i]-q_i^\top  \mu+ \mathbb{ES}_{\alpha_i}(Z)\sqrt{\mathbb{V}\mathrm{ar}(R_i)+2q_i ^\top \mathrm{cov}_i+ q_i^\top \Gamma q_i} \\& \qqq\mbox{with $Z \sim \mathcal{T}_1(0,1, \nu) $}\sp i\in\theD.
  \end{align*}
By \citet[Example $2.15$, page $71$]{McNeilFreyEmbrechts2015}, we obtain
$
   \mathbb{ES}_{\alpha_i}(Z)=\sqrt{\frac{\nu-2}{\nu}} \quad \frac{t_{\nu}\left( T^{-1}_\nu(\alpha_i)\right) \left[ \nu +\left(T^{-1}_\nu(\alpha)\right)^2\right]}{(1-\alpha)(\nu-1)}.
$
By inspection, $\bo{q}$ and $\bo{q}'$ are the same as in Example \ref{eg:lces1}, given by Table \ref{tab:practexample3}.
The above result also yield  $\bo{p}^\theD = 1.94 , \bo{p}'^\theD= 2.06$, and  $\MC_\theD =1.06$ (with $\LC_\theD=0$ as per the first line in Table \ref{tab:SixCasesSummary}).\ \finenv\end{example}

\subsection{Hedging on $\theD$}\label{ss:eshedD}

We now turn to the ``hedging on own exchange $\theD$" case of Section  \ref{ss:he}, but with possibly new participants beyond the CCP $c$ in $\theD'$.  
Let $(R_i,  P )\sim \mathcal{E}_{m+1}(\mu_i, \Gamma_i, \psi)$ with $\Gamma_i$ positive definite, for each $i \in  \theD \cup( \theD'\setminus\{\mathrm{c}\})$.
By Proposition \ref{prop:es}, there exists a unique pre-default Radner equilibrium $((\bo{q}_i)_{i \in \theD}, \bo{p}^\theD)$, which can be computed by  \eqref{eq:loptimalprice} and \eqref{eq:loptimalposition1}.
Since  $R_\thec = \bo{q}_d^\top (P-\bo{p}^\theD)$, letting $z = (z_1,z_2,\dots,z_{m+1}) = (z_1,\hat{z})$, we obtain 
\begin{equation*}
     \E \Big[e^{{\rm i} z^\top (R_\thec, P) }\Big] =  \E \Big[e^{{\rm i} \big((z_1 \bo{q}_d+ \hat{z})^\top P-z_1\bo{q}_d^\top \bo{p}^\theD\big)}\Big] = \exp({\rm i}\ z^\top \mu_\thec )\psi\left(\frac{1}{2} z^\top   \Gamma_\thec z\right), \quad z \in \mathbb{R}^{m+1},
  \end{equation*}
as $
    (z_1\bo{q}_d+ \hat{z})^\top P-z_1\bo{q}_d^\top \bo{p}^\theD \sim \mathcal{E}_1\Big(z^\top\mu_c, z^\top\Gamma_\thec z, \psi\Big).
$
Hence $(R_\thec, P) \sim \mathcal{E}_{m+1}(\mu_\thec, \Gamma_\thec, \psi)$.
However, as $R_\thec$ is in the span of $P$, $\Gamma_\thec$ is not positive definite (see Remark \ref{rem:spancov}).
To nevertheless ensure a unique post-default equilibrium (beyond the setup of Proposition \ref{prop:es}), we assume that $ \theD'\setminus \{\thec\} \neq \varnothing$.
By Theorem \ref{thm:uniqueness}, we have a unique post-default price $\bo{p}'^\theD$  and a unique post-default portfolio $\bo{q}'_i, i\in \theD'\setminus\{\thec\}$.
Hence by the post-default clearing condition, we also have a unique post-default position $\bo{q}'_\thec$.

Letting $R'_i = R_i + \bo{q}_i^\top  P$ play the role of $R_i$ in \eqref{eq:smallr}, we obtain
\begin{multline*}
    r'_i(z_i) \eqdef \mathbb{ES}_{\alpha_i}(-R'_i-z_i^\top P)\\= -\E  [R_i]-(z_i+\bo{q}_i)^\top   \mu+ \mathbb{ES}_{\alpha_i}( Z)\sqrt{(1,z_i+\bo{q}_i)^\top \Gamma_i (1,z_i+\bo{q}_i)}, \quad i\in \theD'
\end{multline*} 
with $Z \sim \mathcal{E}_1(0, 1, \psi)$. 
In view of Remark \ref{rem:cchedging}, by change of variable $\bo{z}'_i = \bo{q}'_i-\bo{q}_i$, 
the proof of Proposition \ref{prop:es} shows that, for $i \in  \theD'  \setminus\{\mathrm{c}\},$  $r'_i$  is differentiable and strictly convex with
\begin{multline*}
  -\bo{p}'^\theD = \nabla r'_i (\bo{z}'_i) = -\mu + \frac{\mathbb{ES}_{\alpha_i}(Z)}{\sqrt{(1,\bo{z}'_i+\bo{q}_i)^\top  \Gamma_i (1,\bo{z}'_i+\bo{q}_i)}} (\mathrm{cov}_i +\Gamma \bo{z}'_i +\Gamma \bo{q}_i)\\
  = -\mu + \frac{\mathbb{ES}_{\alpha_i}(Z)}{\sqrt{\mathbb{V}\mathrm{ar}(R_i)+2(\bo{q}'_i)^\top\mathrm{cov}_i+(\bo{q}'_i)^\top \Gamma\bo{q}'_i}} \left(\mathrm{cov}_i +\Gamma\bo{q}'_i\right). \hspace{3cm}
\end{multline*}
For the CCP, the member $c$ optimality condition  gives $-\bo{p}'^\theD \in \partial r'_\thec(\bo{z}'_\thec) = \partial r'_\thec(\bo{q}'_\thec)$, i.e. $ (r'_\mathrm{c})^\ast (-\bo{p}'^\theD )=-{\bo{q}'_{\mathrm{c}}}^\top   \bo{p}'^\theD - r'_c(\bo{q}'_\mathrm{c}) $.
Hence computing the optimal post-default position reduces to the root-finding problem
\begin{equation}\label{eq:hoptimalposition2}
    \begin{cases}
        \sum_{i\in\theD' } \bo{q}'_i +\bo{q}_d =0\\
        \frac{\mathbb{ES}_{\alpha_1}(Z)} {\mathbb{ES}_{\alpha_i}(Z) } \sqrt{\frac{\mathbb{V}\mathrm{ar}(R_i)+2(\bo{q}'_i)^\top\mathrm{cov}_i+(\bo{q}'_i)^\top \Gamma\bo{q}'_i}{\mathbb{V}\mathrm{ar}(R_1)+2{\bo{q}'_1}^\top \mathrm{cov}_1+{\bo{q}'_1}^\top \Gamma\bo{q}'_1}}\left(\mathrm{cov}_1 +\Gamma\bo{q}'_1\right) -\mathrm{cov}_i -\Gamma\bo{q}'_i=0, & i\in \theD'\setminus\{\thec\}\\
         r'_c(\bo{q}'_\mathrm{c}) + {\bo{q}'_{\mathrm{c}}}^\top   \bo{p}'^\theD + (r'_\mathrm{c})^\ast (-\bo{p}'^\theD )=0. & 
    \end{cases}
\end{equation}

\begin{example}\label{ex:CCPHedgingees}
In the ``hedging on own exchange $\theD$" case with $\theD'\setminus \theD=\{\mathrm{c}\}$ (as per Section \ref{ss:he}) and expected shortfall risk measures,
let $\theD=\{1,\dots,15\}$, $d=\{15\}$, $\alpha_i=0.975 $ and $(R_i, P) \sim \mathcal{N}_2(\mu_i, \Gamma_i)$ (so $m=1$) with $\mathrm{cov}_i =\sigma  c_i   \sqrt{\mathbb{V}\mathrm{ar} (R_i)} $, $\sigma= 0.2$, $c_i=(-1)^{i+1} 0.8$, and $\mathbb{V}\mathrm{ar} (R_i)=0.09i^2$, $i \in \theD $.
The corresponding pre- and post-default optimal positions computed by \eqref{eq:loptimalposition1} and \eqref{eq:hoptimalposition2} are given in Table \ref{tab:practexample2es}.
    \begin{table}[htp] 
        \begin{centering}
            \begin{tabular}{@{}lcccccccc@{}}
                \toprule
                $\CM_i$ &$1$  &$2$  &$3$     &$4$     &$5$   &$6$  &$7$  &$8$ \\
                \midrule
                $\mathrm{cov}_i$ &$0.05$   &$-0.10$ &$0.14$   &$-0.19$ &$0.24$   &$-0.29$   &$0.34$   &$-0.38$ \\ 
                $\bo{q}_i$  &$-1.12$  &$2.56$  &$-3.36$  &$5.12$   &$-5.60$  &$7.68$  &$-7.84$   &$10.24$ \\ 
                $\bo{q}'_i$ &$-1.28$  &$2.24$  &$-3.84$  &$ 4.48$ &$-6.40$  &$6.72$   &$-8.96$  &$8.96$ \\
                \bottomrule
                \\
                \toprule
                $\CM_i$ &$9$ &$10$   &$11$  &$12$   &$13$ &$14$  &$15$  &$\mathrm{c}$ \\
                \midrule
                $\mathrm{cov}_i$ &$ 0.43$ &$-0.48$   &$0.53$  &$-0.58$   &$ 0.62$   &$-0.67$  &$0.72$   &$-0.69$\\ 
                $\bo{q}_i$  &$-10.08$   &$12.80$   &$-12.32$  & $15.36$   &$-14.56$   &$17.92$  &$-16.80$   &\\ 
                $\bo{q}'_i$ &$-11.52$    &$11.20$    &$-14.08$   &$13.44$    &$-16.64$  &$15.68$  &{}  &$16.80$\\
                \bottomrule
            \end{tabular}
            \caption{Pre- and post-default optimal positions when the CCP hedges on its own exchange  $\theD$ with  $\theD'\setminus\theD=\{\thec\}$ in the expected shortfall case.}
        \label{tab:practexample2es}
    \end{centering}
\end{table}
\noindent
    For $\mu=2$, we obtain $\bo{p}^\theD=  1.96, \bo{p}'^\theD= 2.04$, and $\MC_\theD=0.69$ (with, by the third line in Table \ref{tab:SixCasesSummary}, $  \LC_\theD= -1.39$).\ \finenv\end{example}

Table \ref{tab:comparison2} is the expected shortfall analog of Table \ref{tab:comparison}, with qualitatively similar conclusions.
\begin{table}[htp]  
        \begin{centering}
        \resizebox{\columnwidth}{!}{
            \begin{tabular}{@{}llcccccccc@{}}
                \toprule
              &  $\CM_i$ &$1$  &$2$  &$3$     &$4$     &$5$   &$6$  &$7$  &$8$ \\
                \cmidrule{2-10}
       \multirow{2}{*}{Liquidation}     & $\LC_i$ &$0.11$  &$-0.18$  &$0.32$  &$-0.37$ &$0.53$  &$-0.56$   &$0.74$  &$-0.74$ \\ 
         &  $\Delta\brho_i$ &$-0.10$  &$0.20$  &$-0.30$  &$0.40$ &$-0.50$  &$0.60$   &$-0.70$  &$0.80$ \\
                \midrule
       \multirow{2}{*}{Hedging}       & $\LC_i$ &$0.09$  &$-0.21$  &$0.28$  &$-0.42$ &$0.46$  &$-0.64$   &$0.65$  &$-0.85$ \\  
       &  $\Delta\brho_i$ &$-0.10$  &$0.20$  &$-0.30$  &$0.40$ &$-0.50$  &$0.60$   &$-0.70$  &$0.80$ \\ 
                \bottomrule
                \\
                \toprule
            &    $\CM_i$ &$9$ &$10$   &$11$  &$12$   &$13$ &$14$  &$\thec$  &{} \\
                 \cmidrule{2-9}
     \multirow{2}{*}{Liquidation}     & $\LC_i$ &$0.95$  &$-0.93$  &$1.167$  &$-1.11$ &$ 1.38$  &$-1.30$   &  &\\
          &  $\Delta\brho_i$ &$-0.89$  &$0.99$  &$-1.09$  &$1.19$ &$-1.29$  &$1.39$   &  & \\ \cmidrule{1-9}
           \multirow{2}{*}{Hedging}    & $\LC_i$ &$0.84$  &$-1.06$  &$1.02$  &$-1.27$ &$ 1.21$  &$-1.48$   & $0$ &\\
         &  $\Delta\brho_i$ &$-0.89$  &$0.99$  &$-1.09$  &$1.19$ &$-1.29$  &$1.39$   &  $1.39$ &\\ 
                \bottomrule
            \end{tabular}
        }
        \caption{Impacts  of the default resolution on the $\LC_i $ and $\Delta\brho_i$ in the liquidation and hedging cases in the expected shortfall examples \ref{eg:lces1}-\ref{ex:CCPHedgingees}.}
        \label{tab:comparison2}
    \end{centering}
\end{table}

\section{Credit Cost}\label{s:landc}
   
The MC term
\eqref{e:mastergen}
only addresses the impact of the considered default resolution strategy in terms of mis-hedge of market risk. 
It remains to address its credit cost ($\CC$).
By this we need to account for counterparty credit risk in a broad sense including the implications of this risk in terms of capital and funding costs which we quantify by XVA metrics as per \cite{BastideCrepeyDrapeauTadese21a}.
The overall impact of a clearing member default's resolution strategy should then be assessed in terms of the following all-inclusive funds transfer price:  
\begin{equation}\label{e:theftp}
    \FTP= \MC + \CC,
\end{equation}
which should thus supersede $\MC$ in the default resolution approach depicted in Section \ref{ss:methodo}.

\subsection{Structure of the Exchanges}\label{ss:StructExch}

As depicted in Figure \ref{fig:splitingclvsotcbilat}, clearing member bank $a$'s trades with a CCP are divided into proprietary or house trades $\bf{q}_a$, which are in effect hedges of the bank's OTC bilateral trading exposures $R_a=\sum_{o\in O}R_a^o$ (where the non-cleared, end-clients $ o$ are ``outside'' of the exchange), and back-to-back hedges $\bf{q}_b^a$ of intermediated cleared client trades, through which non-member clients $b$ (simple participants to the exchange) can access the CCP clearing services.
\begin{figure}[htp]
\begin{center}
\begin{tikzpicture}
[
roundnodeC/.style={circle, draw=white, fill=white, thick, minimum size=1mm},
roundnodeBC/.style={circle, draw=orange!60, fill=green!5, very thick, minimum size=10mm},
roundnodeCC/.style={circle, draw=blue!60, fill=white!5, very thick, minimum size=10mm},
roundnodeCM/.style={circle, draw=green!60, fill=white!5, very thick, minimum size=12mm},
roundnodeCH/.style={circle, draw=black, fill=red!5, very thick, minimum size=25mm},
scale=0.8, transform shape]
\node at (-8,2) [roundnodeC]   (c1)     {};
\node at (-3.5,-0.6) [roundnodeC]   (c2)     {};
\node at (-4,2) [roundnodeCC]   (cc0)     {$b$};
\node at (0.2, 1.75) [roundnodeCM]   (cm0)     {$a$}; 
\node at (5.5,1.75) [roundnodeCH]  (ccp)    {CCP};
\draw[thick, blue, ->] (c1.west) -- (cc0.180);
\node at (-6.5,2.3) [blue] {$R_b$};
\draw[thick, blue, ->] (cc0.east) -- (cm0.150);
\node at (-1.8,2.3) [blue] {$\bo{q}_b^a$};
\draw[thick, blue, ->] (cm0.30) -- (ccp.167);
\node at (2.5,2.3) [blue] {$\bo{q}_b^a$};
\draw[thick, blue, ->] (cm0.-30) -- (ccp.-167);
\node at (2.5,1.2) [blue] {$\bo{q}_a$};
\draw[thick, blue, ->] (c2.90) -- (cm0.-150);
\node at (-1.5,0.1) [rotate=30, blue] {$R_a=\displaystyle\sum_{o\in O}R_a^o$};
\end{tikzpicture}
\end{center}
\caption{Client clearing ($\bo{q}_b^a$ transits from $b$ to the CCP  via $a$) versus bilateral hedging ($a$ hedges $R_a=\sum_{o\in O}R_a^o$ by a proprietary trading position $\bo{q}_a$ with the CCP).}
\label{fig:splitingclvsotcbilat}
\end{figure}

Our next task is to clarify the nature of the Radner equilibrium on an exchange $E$ accounting for this distinction between proprietary and cleared deals.
Let $\anE=\CMset\cup\CLset$, with $\CMset\cap\CLset=\varnothing$, be the split between the set $\CMset$ of those participants $a$ to the exchange that are also clearing members of the CCP and the set $\CLset$ of simple participants (non clearing members) $b$, having recourse to the clearing members for intermediating trades with the CCP.  
Let ${q}_a$ be the proprietary position of member $a$ and ${q}_b^a$ be the position of client (simple participant to the exchange) $b$ cleared by member $a$. 
As depicted in \hyperref[fig:splitingclvsotcbilat]{Figure \ref{fig:splitingclvsotcbilat}}, the position ${q}_b^a$ only transits from $b$ to $a$ and then passes from $a$ to the CCP.
Hence, even though the total position of $a$ vis-à-vis the CCP is ${q}_a+\sum_{b\in\CLset}{q}_b^a$, as $a$ holds $-\sum_{b\in\CLset}{q}_b^a$ vis-à-vis  the clients it clears for, the respective positions of $a$ and $b$ involved in the Radner equilibrium on $E$ are ${q}_a$ and $q_b=\sum_{a\in\CMset}{q}_b^a.$
Once $((\bo{q}_{i})_{ i \in E=A\cup B}, \bo{p}^E)$ has been obtained as the solution of the corresponding Radner equilibrium   \eqref{eq:mogeneral}-\eqref{eq:zclearing},  
the splits $\bo{q}_b=\sum_{a\in A} \bo{q}_b^a$ (for each $b\in B$) should follow in a second stage from pure credit risk considerations, quantifiable by suitable XVA metrics. Similar comments apply ``with $\cdot'$ everywhere" to any post-default Radner equilibrium on $E'=A'\cup B'$.

\subsection{Credit Costs XVA Framework}

The settlement of a CCP portfolio after a clearing member's default involves both market and counterparty credit risks, depending on the portfolio. A distinction is made between house and intermediated deals, with intermediated client transactions fully hedged by corresponding intermediary trades as depicted in Figure \ref{fig:rewiring-vs-liquidation} and Table \ref{t:costs}. In case of a default, these transactions, along with their hedge, are transferred to a surviving clearing member, without market impact but involving a shift in counterparty credit risk. This transfer of risk is quantified by XVA costs, as explained by \citet[Section 7]{BastideCrepeyDrapeauTadese21a}.

Table \ref{t:costs} distinguishes between delta-one financial instruments, such as repo transactions and equity swaps, which are rolled over without upfront payments, and upfront derivatives like multi-leg swaps and options.
While delta-one transactions have minimal counterparty credit risk due to their short reset periods, they can still involve significant liquidation or hedging costs.
For these assets, market cost analysis is often sufficient.
For swaps or derivatives portfolios, instead, XVA implications play a crucial role in the default resolution strategy, which is the primary focus of this section.

\begin{figure}[htp]
\begin{center}
\begin{subfloat}[case of a house portfolio]{\label{fig:subfigA}
\centering
\begin{tikzpicture}
[
roundnodeC/.style={circle, draw=white, fill=white, thick, minimum size=1mm},
roundnodeBC/.style={circle, draw=orange!60, fill=green!5, very thick, minimum size=10mm},
roundnodeCC/.style={circle, draw=blue!60, fill=white!5, very thick, minimum size=10mm},
roundnodeCM/.style={circle, draw=green!60, fill=white!5, very thick, minimum size=10mm},
roundnodeCH/.style={circle, draw=black, fill=red!5, very thick, minimum size=25mm},
scale=0.52, transform shape]
\Large
\draw[thick, dashed, violet] (4.5,-3.3) arc (80:-80:-5.5);

\node at (2,6.8) [rotate=40,violet] {CCP};

\node at (-2,-1.5) [roundnodeC]   (c1)     {};
\node at (-1.75,-2.5) [roundnodeC]   (c2)     {};
\node at (-1.5,-3.5) [roundnodeC]   (c3)     {};
\node at (0.2, 3.5) [roundnodeCM]   (cm0)     {$a$};
\node at (1.5,-1.5) [roundnodeCM]   (cm1)     {$d$};
\node at (4.5,1.75) [roundnodeCH]  (ccp)    {CCP};
 
\draw[dashed, red, ->] (c1.east) -- (cm1.west);
\draw[dashed, red, ->] (c2.east) -- (cm1.-140);
\draw[dashed, red, ->] (c3.east) -- (cm1.-110);

\draw [decorate,
    decoration = {calligraphic brace}] (-1.5,-4) --  (-2,-1.4);
\node at (-3.3,-3) [blue] {$\displaystyle\sum_{o\in O} R_d^o=R_d$};

\draw[thick, red, ->] (cm0) -- (ccp.170);
\draw[dashed, red, ->] (cm1.30) -- (ccp.-110);

\node at (2.1,3.1) [rotate=-27, blue] {$\bo{q}_d^\top  ({\bo{p}'^\theD }%
-\bo{P})$};
\node at (3.3,-0.6) [rotate=40, blue] {$\bo{q}_d^\top  (\bo{p}^\theD 
-\bo{P})$};
\end{tikzpicture}
}
\end{subfloat}\qquad\qquad
\begin{subfloat}[case of a client portfolio]{\label{fig:subfigB}
\centering
\begin{tikzpicture}[ 
roundnodeCC/.style={circle, draw=blue!60, fill=white!5, very thick, minimum size=10mm}, 
roundnodeCM/.style={circle, draw=green!60, fill=white!5, very thick, minimum size=10mm},
roundnodeCH/.style={circle, draw=black, fill=red!5, very thick, minimum size=25mm}, 
scale=0.52, transform shape]
\Large
\draw[thick, dashed, violet] (2.3,-3.3) arc (60:-60:-6.5);

\node at (0.4,6.8) [rotate=50,violet] {CCP};

\node at (-3.7,0.1) [roundnodeCC]   (cc1)     {$b$}; 
\node at (-0.8, 3.5) [roundnodeCM]   (cm0)     {$a$};
\node at (0.5,-1.5) [roundnodeCM]   (cm1)     {$d$};
\node at (5.5,1.75) [roundnodeCH]  (ccp)    {CCP};

\draw[thick, red, ->] (cc1.20) -- (cm0.-125);
\draw[dashed, red, ->] (cc1.-10) -- (cm1.160);

\draw[thick, red, ->] (cm0.east) -- (ccp.150); 
\draw[dashed, red, ->] (cm1.30) -- (ccp.-130);

\node at (-2.6,1.9) [blue] {$R_d^b$};
\node at (-1.7,-1.2) [blue] {$R_d^b$};

\node at (2,3.4) [rotate=-13, blue]
{$R_d^b$};
\node at (3.1,-0.7) [rotate=30, blue] {$(\bo{q}_d^b)^\top (\bo{p}^\theD
-\bo{P})=R_d^b$};

\end{tikzpicture}
}
\end{subfloat}
\end{center}

\caption{
Default management of a CCP portfolio after clearing member $d$ defaults with pre-default positions shown with dashed red lines and post-default positions with solid red lines: ({\em Left}) The default member $d$'s receivable $R_d$ from OTC bilateral counterparties is excluded from the default resolution, its house hedging portfolio $\bo{q}_d$ is transferred to surviving member $a$; ({\em Right}) Both $d$'s client account $\bo{q}_d^b$ and the corresponding receivable $R_d^b$ from client $b$ are transferred as a package to surviving clearing member $a$.
\\
{\footnotesize{Notations as detailed below and in Sections \ref{sec:re}-\ref{sec:model}-\ref{app:xva}.}
}} 
\label{fig:rewiring-vs-liquidation}
\end{figure}

\begin{table}[htbp]
    \centering
    \begin{tabular}{m{.30\textwidth}|m{.33\textwidth}|m{.28\textwidth}}
    \toprule
        \diagbox{\theadfont{Account}}{\theadfont{Assets}}& swaps and options & \r{delta-one} ($\subseteq$ house account) \\
        \midrule 
    house   & liquidity \b{(MC)} and credit \b{($\CC= \sum_{ i\neq d} \Delta \XVA_i+\AC$)}  & liquidity \b{(MC)}, \r{no credit} \\  
    \hline
    \g{client} ($\subseteq$ pure auctioning)  &  \g{no liquidity}, credit \b{(AC)} &  $\varnothing$  \\
         \bottomrule
    \end{tabular}
     \caption{Costs of the CCP for settling a netting set of deals of a defaulted clearing member, depending on the nature of the defaulted portfolio.\\
\footnotesize{
The transactions represented in the lower right cell would involve neither liquidity nor credit risk; however, this cell is actually empty because delta-one transactions do not require intermediation.
}}
\label{t:costs}
\end{table}

Our XVA metrics are computed under the premise that the (random) loss triggered by the default of a market participant in the future is allocated between the surviving members of its CCP, pro-rata of their default fund contribution to the CCP (see e.g. after \eqref{e:SimpleTdgLossWithCr}). 
At time ``$0-$'', i.e. ``right before'' the instant default of the clearing member $d$ at time 0, the participants $i$ to the exchanges charge to their clients the expectation of their ensuing losses, in the form of their CVA$_i$, as well as their rehypothecated and segregated collateral funding costs FVA$_i$ and MVA$_i$, and costs of capital KVA$_i$. These costs sum up to $\XVA_i=\CVA_i+\FVA_i+\MVA_i+\KVA_i$, computed for each participant  $i$ the way detailed for $i=0$ in Section \ref{app:xva}, based on the pre-default Radner equilibrium quantities and prices on each exchange. 
To quantify the XVA impact of a given default resolution procedure for a CCP portfolio of $d$, we also compute the time 0, post-default  $\XVA'_i$, for any participant
$i\neq d$ to the exchanges.
The credit cost of the settlement of the defaulted portfolio, coming on top of the already computed $\MC$, is 
\beql{e:CCexpr}
\CC=\sum_{i\neq d}\underbrace{ (\XVA'_i - \XVA_i )}_{\Delta \XVA_i}  +\AC ,
\eeql
where the auctioning cost  AC
 is another XVA incremental impact corresponding to the FTP in \citet[Section 7]{BastideCrepeyDrapeauTadese21a} (which only involved credit costs), i.e. the XVA impact of auctioning any hedged (as opposed to liquidated) positions. Indeed, a CCP cannot keep the defaulted portfolio and the corresponding hedge on its book, it will look at auctioning them \citep*{ferera2020,oleschak2019}.

In the end, the all-inclusive FTP \eqref{e:theftp} that emerges from the present paper (where both market and credit costs are involved) can be detailed as
\beql{e:FTPexpr}
\FTP=\underbrace{\LC +\sum_{E'}\sum_{i\in E'}\Delta\brho_i}_{\MC}+\underbrace{\sum_{i\neq d} \Delta \XVA_i+\AC}_{\CC}.
\eeql 
\begin{remark}\label{r:auc}
Under a pure auctioning default resolution strategy, only the last term remains.
But this AC term could be very expensive in the case of an unhedged portfolio 
(see the first quotation in Section \ref{sec:model}).\ \finenv\end{remark}

\subsection{\hyperref[eg:lces1]{Example \ref{eg:lces1}} Continued}\label{ss:econt}
We complete from the credit costs perspective the ``liquidation on own exchange'' \hyperref[eg:lces1]{Example \ref{eg:lces1}}. In the case of default resolution procedures implemented on the own exchange $\theD$ of the CCP, \eqref{e:FTPexpr} boils down to 
\beql{e:FTPexprD}
\FTP=\underbrace{\LC_{\theD} + \sum_{i\in \theD\setminus\{d\}}\Delta\brho_i}_{\MC}+\underbrace{\sum_{i\in\theD\backslash\{d\}} \Delta \XVA_i+\AC}_{\CC}.
\eeql

\noindent We endorse the Gaussian latent factors XVA setup of Section \ref{ss:xvad}.
\subsubsection{Liquidation}
Table \ref{tab:practexampleFTP} provides the resulting ``time  $0-$'' $\XVA_i$ (with all members, $d$ included)  and time 0  $\XVA'_i$ (without $d$),
 using the allocated positions of \hyperref[tab:practexample3]{Table \ref{tab:practexample3}} and the XVA 
 specifications of Table \ref{tab:XVAconfig1CCP20Mbs} along with $\IM_i =\DF_i=0$, whilst Table \ref{tab:practexampleFTP2} presents the same results for $\IM_i$ and $\DF_i$ set at the 75\% and 80\% confidence level.
 Note that the chosen period length of $T=5$ years covers the bulk (if not the final maturity) of most realistic CCP portfolios\footnote{most OTC derivatives have a maturity of less than 5 years \citep[Graphs A.2--4]{BIS2022}.}.
The aggregated XVA cost \eqref{e:CCexpr} of liquidating the defaulted portfolio\footnote{note that there is no auction in this (liquidation) case.} is $\CC=\sum_{i\in\theD\setminus\{d\}} \Delta\XVA_i$, namely ${-0.77}$ for the case $\IM_i =\DF_i=0$ and $-0.25$ with $\IM_i$ and $\DF_i$ set at the 75\% and 80\% confidence level, coming on top of the market cost of $\MC = 0.70$ already obtained in \hyperref[eg:lces1]{Example \ref{eg:lces1}}.   
 \begin{table}[htp] 
        \begin{centering}
        \resizebox{\columnwidth}{!}{
            \begin{tabular}{@{}lccccccccccccccc@{}}
                \toprule
                $\CM_i$ &$1$ &{} &$2$  &{}   &$3$    &{} &$4$ &{}    &$5$    &{} &$6$ &{}   &$7$     &{} &$8$ \\
                \midrule
                $\XVA_i$ & $0.81$ &{} & $1.21$ &{}& $1.36$ &{}& $1.70$ &{}& $1.74$ &{}& $2.05$ &{}& $2.02$ &{}& $2.37$ \\
                $\XVA'_i$ & $0.92$ &{} & $1.17$ &{}& $1.47$ &{}& $1.61$ &{}& $1.82$ &{}& $1.93$ &{}& $2.07$ &{}& $2.19$ \\
                $\Delta\XVA_i$ & $0.11$ &{} & $-0.04$ &{}& $0.12$ &{}& $-0.09$ &{}& $0.08$ &{}& $-0.12$ &{}& $0.05$ &{}& $-0.18$ \\
                \bottomrule
                \\
                \toprule
                $\CM_i$ &$9$ &{}  &$10$  &{}  &$11$  &{}  &$12$  &{}  &$13$  &{}  &$14$  &{} &$15$   &{}  &  \\
                \midrule
                $\XVA_i$ & $2.25$ &{} & $2.62$ &{}& $2.42$ &{}& $2.84$ &{}& $2.57$ &{}& $3.04$ &{}& $2.69$ &{}& {} \\
                $\XVA'_i$ & $2.28$ &{} & $2.42$ &{}& $2.42$ &{}& $2.60$ &{}& $2.54$ &{}& $2.77$ &{}& {} &{}& {} \\
                $\Delta\XVA_i$ & $0.03$ &{} & $-0.20$ &{}& $-0.01$ &{}& $-0.24$ &{}& $-0.03$ &{}& $-0.27$ &{}& {} &{}& {} \\
                \bottomrule
            \end{tabular}
        }
        \caption{The pre- and post-default $\XVA$s  
        computed from \eqref{e:CVApostdef}  when the CCP liquidates $d$ on its own exchange, $\theD$, in the expected shortfall case with  $\IM_i =\DF_i=0.$}.
        \label{tab:practexampleFTP}
    \end{centering}
\end{table}

\begin{table}[htp] 
        \begin{centering}
        \resizebox{\columnwidth}{!}{
            \begin{tabular}{@{}lccccccccccccccc@{}}
                \toprule
                $\CM_i$ &$1$ &{} &$2$  &{}   &$3$    &{} &$4$ &{}    &$5$    &{} &$6$ &{}   &$7$     &{} &$8$ \\
                \midrule
                $\XVA_i$ & $0.37$ &{} & $0.59$ &{}& $0.72$ &{}& $0.89$ &{}& $1.00$ &{}& $1.12$ &{}& $1.21$ &{}& $1.34$ \\
                $\XVA'_i$ & $0.42$ &{} & $0.56$ &{}& $0.80$ &{}& $0.84$ &{}& $1.07$ &{}& $1.05$ &{}& $1.28$ &{}& $1.22$ \\
                $\Delta\XVA_i$ & $0.05$ &{} & $-0.03$ &{}& $0.08$ &{}& $-0.06$ &{}& $0.07$ &{}& $-0.07$ &{}& $0.07$ &{}& $-0.11$ \\
                \bottomrule
                \\
                \toprule
                $\CM_i$ &$9$ &{}  &$10$  &{}  &$11$  &{}  &$12$  &{}  &$13$  &{}  &$14$  &{} &$15$   &{}  &  \\
                \midrule
                $\XVA_i$ & $1.42$ &{} & $1.50$ &{}& $1.57$ &{}& $1.64$ &{}& $1.72$ &{}& $1.77$ &{}& $1.85$ &{}& {} \\
                $\XVA'_i$ & $1.49$ &{} & $1.37$ &{}& $1.63$ &{}& $1.49$ &{}& $1.78$ &{}& $1.60$ &{}& {} &{}& {} \\
                $\Delta\XVA_i$ & $0.07$ &{} & $-0.12$ &{}& $0.06$ &{}& $-0.15$ &{}& $0.06$ &{}& $-0.17$ &{}& {} &{}& {} \\
                \bottomrule
            \end{tabular}
        }
        \caption{The pre- and post-default $\XVA$s  
        computed from \eqref{e:CVApostdef} when the CCP liquidates $d$ on its own exchange, $\theD$, in the expected shortfall case with $\IM_i$ and $\DF_i$ set at the 75\% and 80\% confidence level.}
        \label{tab:practexampleFTP2}
    \end{centering}
\end{table}

\subsubsection{Pure Auctioning}
Instead of liquidation on the CCP's own exchange $\theD$, we now consider another default resolution strategy, in the form of an (idealized) auction inducing the taker giving rise to the least auction cost $\AC$ among all possible takers $i\in  \theD\setminus\{d\}$\footnote{This approach developed in \citep[Section~7]{BastideCrepeyDrapeauTadese21a} can indeed be seen as rendering the output of an idealized, efficient auction used for closing out the account of a defaulted clearing member (cf.~\citet*[Section~3.3]{oleschak2019}).}. The results are displayed in \hyperref[tab:FTPresults]{Table \ref{tab:FTPresults}} for the case $\IM_i =\DF_i=0$ and in \hyperref[tab:FTPresults2]{Table \ref{tab:FTPresults2}} for $\IM_i$ and $\DF_i$ set at the 75\% and 80\% confidence level.
\begin{table}[htp]
\begin{centering}
{
\begin{tabular}{cccc}
\toprule
$\CM_i$ &  {$\displaystyle\sum_{i\in \theD\setminus\{d\}}(\CVA'_i- \CVA_i)$} &  {$\displaystyle\sum_{i\in \theD\setminus\{d\}}(\KVA'_i- \KVA_i)$} & $\AC$\\ 
\midrule
14  & -1.00 (-0.38) & -3.52 (-0.63) & -4.52 (-1.01)\\
12  & -0.85 (-0.34) & -2.93 (-0.57) & -3.78 (-0.91)\\ 
10  & -0.58 (-0.23) & -2.02 (-0.42) & -2.60 (-0.64)\\ 
8  & -0.31 (-0.11) & -1.13 (-0.24) & -1.44 (-0.35)\\ 
6  & -0.04 (0.02) & -0.30 (-0.04) & -0.34 (-0.02)\\
4  & 0.23 (0.15) & 0.42 (0.20) & 0.65 (0.35)\\ 
2  & 0.50 (0.29) & 0.95 (0.52) & 1.45 (0.81)\\ 
1  & 0.77 (0.43) & 1.59 (0.86) & 2.35 (1.29)\\ 
3  & 0.77 (0.39) & 2.09 (0.62) & 2.85 (1.01)\\ 
5  & 0.77 (0.36) & 2.47 (0.47) & 3.23 (0.82)\\ 
7  & 0.77 (0.32) & 2.70 (0.37) & 3.46 (0.70)\\ 
9  & 0.77 (0.29) & 2.84 (0.30) & 3.60 (0.60)\\ 
11  & 0.77 (0.26) & 2.92 (0.27) & 3.69 (0.53)\\ 
13  & 0.77 (0.23) & 2.96 (0.25) & 3.73 (0.48)\\
\bottomrule
\end{tabular}
        }
        \caption{Auctioning costs $\AC$ corresponding to the different possible takers of the portfolio of the defaulted member $d={15}$, ranked by increasing value, for  $\IM_i =\DF_i=0$. In parenthesis, the contributions to $\AC$ of the considered possible taker itself.}
\label{tab:FTPresults}
    \end{centering}
\end{table}
\begin{table}[htp]
\begin{centering}
\resizebox{\columnwidth}{!}{
\begin{tabular}{ccccc}
\toprule
$\CM_i$ &  {$\displaystyle\sum_{i\in \theD\setminus\{d\}}(\MVA'_i- \MVA_i)$}&  {$\displaystyle\sum_{i\in \theD\setminus\{d\}}(\CVA'_i- \CVA_i)$} &  {$\displaystyle\sum_{i\in \theD\setminus\{d\}}(\KVA'_i- \KVA_i)$} & $\AC$\\ 
\midrule
14 & -0.06 (-0.15) & -0.49 (-0.13) & -2.33 (-0.42) & -2.88 (-0.71)\\
12 & -0.01 (-0.11) & -0.45 (-0.12) & -2.00 (-0.39) & -2.46 (-0.63)\\ 
10 & 0.06 (-0.04) & -0.36 (-0.08) & -1.47 (-0.27) & -1.77 (-0.40)\\ 
8 & 0.13 (0.03) & -0.27 (-0.04) & -0.93 (-0.14) & -1.07 (-0.16)\\ 
6 & 0.20 (0.10) & -0.17 (0.00) & -0.44 (0.01) & -0.41 (0.11)\\
4 & 0.26 (0.17) & -0.07 (0.04) & 0.00 (0.19) & 0.20 (0.40)\\ 
2 & 0.33 (0.23) & 0.03 (0.09) & 0.38 (0.41) & 0.74 (0.74)\\ 
1 & 0.41 (0.31) & 0.11 (0.14) & 0.76 (0.64) & 1.28 (1.09)\\ 
3 & 0.42 (0.31) & 0.10 (0.13) & 0.92 (0.51) & 1.44 (0.95)\\ 
5 & 0.42 (0.31) & 0.09 (0.11) & 1.06 (0.42) & 1.58 (0.85)\\ 
7 & 0.43 (0.32) & 0.08 (0.10) & 1.16 (0.37) & 1.67 (0.78)\\
9 & 0.44 (0.32) & 0.07 (0.09) & 1.23 (0.31) & 1.74 (0.72)\\ 
11 & 0.45 (0.32) & 0.06 (0.08) & 1.28 (0.28) & 1.78 (0.68)\\
13 & 0.45 (0.33) & 0.05 (0.06) & 1.29 (0.26) & 1.80 (0.65)\\
\bottomrule
\end{tabular}
        }
        \caption{Auctioning costs $\AC$ corresponding to the different possible takers of the portfolio of the defaulted member $d={15}$, ranked by increasing value, for $\IM_i$ and $\DF_i$ set at the 75\% and 80\% confidence level. In parenthesis, the contributions to $\AC$ of the considered possible taker itself.}
\label{tab:FTPresults2}
\end{centering}
\end{table}
From Tables \ref{tab:FTPresults}-\ref{tab:FTPresults2}, participant $14$ is the survivor taker leading to the smallest auctioning cost $\AC$, namely$ {-4.52}$ and $-2.88$, when taking over the defaulted portfolio of $\mathrm{CM}_{15}$ (and there are in this case no additional costs to consider, cf.~\hyperref[r:auc]{Remark \ref{r:auc}}). 
Such take-over makes intuitive sense given the pre-default position $\bo{q}_{15}=-16.8$ of the defaulting member $d=15$, compared with the position of member $14$ at $\bo{q}_{14}=17.92$, an almost offset of $\bo{q}_{15}$. 
Member $12$ shows the second best solution with an auctioning cost $\AC$ close to zero, also justifiable by his offsetting position at $\bo{q}_{12}=15.36$.

\subsubsection{Hedging then auctioning}
Finally, we consider one more default resolution strategy, where the CCP hedges (as per \hyperref[ss:he]{Section \ref{ss:he}}) the defaulted portfolio before auctioning all its positions. 
In the hedging case, with the CCP $c$ contributing to the post-default quantities and price discovery $((\bo{q}'_i)_{i\in \theD ' = (  \theD\setminus \{d\} )\cup \{c\}},\bo{p}'^\theD)$, resolving \eqref{eq:mo} (for $E=\theD$) and 
\eqref{eq:ccpendowment}--\eqref{eq:cchedgingonE} 
in the configuration of \hyperref[eg:lces1]{Example \ref{eg:lces1}} (except for the new member $c$) leads to $\bo{q}_c'=-\bo{q}_d$. 
Coincidentally, this hedging resolution thus leads to a perfect replication as per Section \ref{rem:fullreplication}. 
The corresponding $\sum_{i\neq d}\Delta\XVA_i$, detailed in Tables  \ref{tab:practexampleFTPhedg} and \ref{tab:practexampleFTPhedg2}, is $-14.45$ for $\IM_i =\DF_i=0$ and $-6.80$ for $\IM_i$ and $\DF_i$ set at the 75\% and 80\% confidence levels. 
As the residual market risk is null in such a replication case, taking over the defaulted portfolio and its hedge  does not generate any market risk.
Hence in these cases no additional cost is generated by the auctioning process, i.e.~$\AC=0$.
\begin{table}[htp] 
        \begin{centering}
        \resizebox{\columnwidth}{!}{
            \begin{tabular}{@{}lccccccccccccccc@{}}
                \toprule
                $\CM_i$ &$1$ &{} &$2$  &{}   &$3$    &{} &$4$ &{}    &$5$    &{} &$6$ &{}   &$7$     &{} &$8$ \\
                \midrule
                $\XVA_i$ & $0.81$ &{} & $1.21$ &{}& $1.36$ &{}& $1.7$ &{}& $1.74$ &{}& $2.05$ &{}& $2.02$ &{}& $2.37$ \\
                $\XVA'_i$ & $0.19$ &{} & $0.34$ &{}& $0.54$ &{}& $0.63$ &{}& $0.83$ &{}& $0.89$ &{}& $1.07$ &{}& $1.13$ \\
                $\Delta\XVA_i$ & $-0.62$ &{} & $-0.87$ &{}& $-0.81$ &{}& $-1.06$ &{}& $-0.91$ &{}& $-1.15$ &{}& $-0.94$ &{}& $-1.24$ \\
                \bottomrule
                \\
                \toprule
                $\CM_i$ &$9$ &{}  &$10$  &{}  &$11$  &{}  &$12$  &{}  &$13$  &{}  &$14$  &{} &$15$   &{}  & $c$ \\
                \midrule
                $\XVA_i$ & $2.25$ &{} & $2.62$ &{}& $2.42$ &{}& $2.84$ &{}& $2.57$ &{}& $3.04$ &{}& $2.69$ &{}& \\
                $\XVA'_i$ & $1.28$ &{} & $1.34$ &{}& $1.44$ &{}& $1.54$ &{}& $1.59$ &{}& $1.72$ &{}& {} &{}& $2.39$ \\
                $\Delta\XVA_i$ & $-0.98$ &{} & $-1.27$ &{}& $-0.98$ &{}& $-1.3$ &{}& $-0.98$ &{}& $-1.32$ &{}& {} &{}& {} \\
                \bottomrule
            \end{tabular}
        }
         \caption{The pre- and post-default $\XVA$s computed from \eqref{e:CVApostdef} when the CCP hedges the portfolio of the defaulted member $d=15$ on its own exchange $\theD$, in the expected shortfall case with $\IM_i =\DF_i=0$.}
        \label{tab:practexampleFTPhedg}
    \end{centering}
\end{table}
\begin{table}[htp] 
        \begin{centering}
        \resizebox{\columnwidth}{!}{
            \begin{tabular}{@{}lccccccccccccccc@{}}
                \toprule
                $\CM_i$ &$1$ &{} &$2$  &{}   &$3$    &{} &$4$ &{}    &$5$    &{} &$6$ &{}   &$7$     &{} &$8$ \\
                \midrule
                $\XVA_i$ & $0.37$ &{} & $0.59$ &{}& $0.72$ &{}& $0.89$ &{}& $1$ &{}& $1.12$ &{}& $1.21$ &{}& $1.34$ \\ 
                $\XVA'_i$ & $0.13$ &{} & $0.21$ &{}& $0.38$ &{}& $0.4$ &{}& $0.6$ &{}& $0.56$ &{}& $0.79$ &{}& $0.72$ \\ 
                $\Delta\XVA_i$ & $-0.24$ &{} & $-0.38$ &{}& $-0.34$ &{}& $-0.5$ &{}& $-0.4$ &{}& $-0.56$ &{}& $-0.42$ &{}& $-0.62$ \\
                \bottomrule
                \\
                \toprule
                $\CM_i$ &$9$ &{}  &$10$  &{}  &$11$  &{}  &$12$  &{}  &$13$  &{}  &$14$  &{} &$15$   &{}  & $c$ \\
                \midrule
                $\XVA_i$ & $1.42$ &{} & $1.50$ &{}& $1.57$ &{}& $1.64$ &{}& $1.72$ &{}& $1.77$ &{}& $1.85$ &{}& \\ 
                $\XVA'_i$ & $0.97$ &{} & $0.85$ &{}& $1.12$ &{}& $0.98$ &{}& $1.26$ &{}& $1.09$ &{}& {} &{}& $1.30$ \\
                $\Delta\XVA_i$ & $-0.45$ &{} & $-0.64$ &{}& $-0.45$ &{}& $-0.67$ &{}& $-0.46$ &{}& $-0.68$ &{}& {} &{}& {} \\
                \bottomrule
            \end{tabular}
        }
        \caption{The pre- and post-default $\XVA$s  computed from \eqref{e:CVApostdef} when the CCP hedges the portfolio of the defaulted member $d=15$ on its own exchange $\theD$, in the expected shortfall case with $\IM_i$ and $\DF_i$ at the 75\% and 80\% confidence level.}
        \label{tab:practexampleFTPhedg2}
    \end{centering}
\end{table}

\subsubsection{Synthesis}
We sum up in Tables \ref{tab:res_num_sum_up_old} and \ref{tab:res_num_sum_up} the FTP of each considered default management scheme without and with collateral (in the sense here of initial margins and default fund contributions),  from the cheapest to the dearest one (again, in this example, hedging then auctioning with $\theD'\setminus\theD=\{c\}$ happens to coincide with replicating then auctioning  with $\theD'\setminus\theD=\varnothing$). 
The FTPs of the hedging then auctioning scheme provides much larger gains than the pure auctioning strategy, which itself provides more gains than the full liquidation strategy. 
However, our approach only endorses the point of view of the participants to the exchange. Indeed, our costs of settling the house portfolio of a defaulted clearing member
ignore the damage of the default to the ``outer'' actors $o$ (end-users external to the exchanges).
From this viewpoint, whenever available, centrally cleared trading is preferable to bilateral trading  (but, as per today, centrally cleared trading can only concern the standardized half of the market).

\begin{table}[htp]
    \centering
    {
    \begin{tabular}{lccccc}
    \toprule
        & $\LC_\theD$ &  
        $\displaystyle\sum_{i\in \theD'} \Delta\brho_i$ & $\displaystyle\sum_{i\in \theD\setminus\{d\}}\Delta\XVA_i$ & $\AC$ & $\FTP$ \\
        \midrule 
          liquidating   &  0	  &  0.70		 & -0.77	 &  0		 &  -0.08 \\
           auctioning  &   0  &  0  &  0  & -4.52	 &	-4.52\\
                 hedging then auctioning   &  -1.39	 & 	2.09	 & 	-14.45	 	&  0  &  	 	-13.75  \\
         \bottomrule
    \end{tabular}
    \caption{$\FTP$s of different default management schemes on $\theD$ split as per \eqref{e:FTPexprD} for $\IM_i =\DF_i=0$.} 
    \label{tab:res_num_sum_up_old}
    }
\end{table}

\begin{table}[htp]
    \centering
    {
    \begin{tabular}{lccccc}
    \toprule
        & $\LC_\theD$ &  
        $\displaystyle\sum_{i\in \theD'} \Delta\brho_i$ & $\displaystyle\sum_{i\in \theD\setminus\{d\}}\Delta\XVA_i$ & $\AC$ & $\FTP$ \\
        \midrule 
          liquidating   &  0  &  0.70 & -0.25 &  0  &  0.45\\
           auctioning  &   0  &  0  &  0  & -2.88 & -2.88 \\
                 hedging then auctioning   &  -1.39	 & 2.09 & 	-6.80&  0  &  	 	-6.11   \\
         \bottomrule
    \end{tabular}
    \caption{$\FTP$s of different default management schemes on $\theD$ split as per \eqref{e:FTPexprD} for $\IM_i$ and $\DF_i$ set at the 75\% and 80\% confidence level.\\ \footnotesize{In this example, hedging then auctioning with $\theD'\setminus\theD=\{c\}$ happens to coincide with replicating then auctioning  with $\theD'\setminus\theD=\varnothing$.}} 
    \label{tab:res_num_sum_up}
    }
\end{table}

\bibliographystyle{chicago}
\bibliography{biblio}

\newpage
\appendix

\section{Some Proofs in Section \ref{sec:re}}\label{app:proof}

\subsection{Proof of Lemma \ref{lem:radnerequivalence} \label{app:proof1}.}

Let $((\bo{q}_i)_{i\in \anE}, \bo{p} )$ be a Radner equilibrium as per Definition \ref{defi:radner}.
The zero clearing condition \eqref{eq:zclearing} yields (iii).
  By \citet[Theorem $16.4$, page $145$]{rockafellar1970}, the convex conjugate of the inf-convolution of proper convex functions is the sum of the corresponding conjugates, i.e.
  \begin{equation*}
      r^\ast(-\bo{p}) = \sum_{i\in \anE} r^\ast_i(-\bo{p}).
  \end{equation*}
Summing the expression \eqref{eq:membopt2} across all $r_i$ and using (iii) gives
  \begin{equation*}
      \sum_{i\in \anE}r_i(\bo{q}_i) =  0 - r^\ast(-\bo{p}) \leq r(0),
  \end{equation*}
  where the inequality holds by definition \eqref{defi:convconj}  of the convex conjugate of $r$.
  By definition  of $r(0)$, the above inequality becomes equality, i.e.
  \begin{equation*}
      r(0) =   - r^\ast(-\bo{p})=\sum_{i\in \anE}r_i(\bo{q}_i).
  \end{equation*}
  Hence (ii) holds and so does also (i), in view of the equivalence between \eqref{eq:membopt} and \eqref{eq:membopt2}, here applied to $r$ (instead of $r_i$ there).

  Conversely, suppose that $((\bo{q}_i)_{i\in \anE}, \bo{p})$ satisfies (i)--(iii).
  (iii) is the  zero clearing condition \eqref{eq:zclearing},
whereas (i) implies via \eqref{eq:membopt2} applied to $r$ that
  \begin{equation}\label{e:FenchLegEq2}
      r(0) =   - r^\ast(-\bo{p}) = \sum_{i\in \anE}\left(-\bo{q}_i^\top  \bo{p}  -\ r^\ast_i (-\bo{p})\right).
  \end{equation}
  By \eqref{eq:membopt3} and \eqref{e:FenchLegEq2},
  $-\bo{p}  \notin \partial r_i(\bo{q}_i)$ for some $i\in \anE$ would imply that $r(0) <\sum_{i\in \anE}r_i(\bo{q}_i)$, contradicting (ii).

 Hence \eqref{eq:membopt}, which is equivalent to the member $i$ optimality condition \eqref{eq:zclearing}, holds for each $i\in E$. 

  \subsection{Proof of Proposition \ref{prop:entropy}  \label{app:proof2}.}
  
  By  \hyperref[thm:largeloss]{Theorem \ref{thm:largeloss}} and Remark \ref{rem:largeloss}, there exists a Radner equilibrium. In view of 
 \eqref{e:ER},
  \begin{equation}\label{eq:linearcomb}
    -R_i-\bo{q}_i^\top   P \sim \mathcal{N}_1\big(-\E  [R_i]-\bo{q}_i^\top\mu,\mathbb{V}\mathrm{ar}(R_i) +2\bo{q}_i^\top \mathrm{cov}_i+\bo{q}_i^\top \Gamma\bo{q}_i\big).
  \end{equation}
The moment generating function of  a standard normal variate $L$ is
$ \mathbb{R}\ni z \mapsto \E  [\exp(zL)]= \exp\left( z\E  [L] + \mathbb{V}\mathrm{ar}(L)z^2/2\right)$, hence $\rho_i(L)=\E  [L] + \varrho_i \mathbb{V}\mathrm{ar}(L)/2$.
  This and \eqref{eq:linearcomb} yield
  \begin{equation}\label{eq:entropicr}
   r_i(q_i)=-\E  [R_i]-q_i^\top \mu +\frac{\varrho_i \mathbb{V}\mathrm{ar}(R_i)}{2}+\varrho_i q_i ^\top \mathrm{cov}_i+\frac{1}{2}\varrho_i q_i ^\top\Gamma q_i
  \end{equation}
 and
  \begin{equation}\label{e:Nabla_r_i}
    \bm{\nabla} r_i (q_i)=-\mu+\varrho_i\mathrm{cov}_i+\varrho_i\Gamma q_i,  \quad  i\in E.
  \end{equation}
The optimality condition relative to the participant $i\in \anE$ yields
  \begin{equation*}
    -\bo{p}  = \bm{\nabla}r_i(\bo{q}_i)=-\mu+ \varrho_i\mathrm{cov}_i+\varrho_i\Gamma \bo{q}_i,
  \end{equation*}
  hence
  \beql{e:qG}
    \bo{q}_i=\Gamma^{-1}\left(\frac{1}{\varrho_i}(\mu-\bo{p} )-\mathrm{cov}_i\right).
  \eeql
  On the other hand, the clearing condition yields
  \begin{equation*}
    \sum_{i\in \anE}\bo{q}_i=0=\Gamma^{-1}\left(\frac{1}{\varrho}(\mu-\bo{p} )-\mathrm{cov}\right),
  \end{equation*}
  which is equivalent to
  \begin{equation*}
    \bo{p} =\mu - \varrho \mathrm{cov}.
  \end{equation*}
\eqref{e:qG} in turn gives \eqref{e:ps}.

  \subsection{Proof of Proposition \ref{prop:es}  \label{app:proof3}.}
  
  We divide the proof in three steps.

  \subsubsection*{Existence}
          For a univariate normally distributed $L$,  by \citet[Example $2.14$, page $70$]{McNeilFreyEmbrechts2015}, $\mathbb{ES}_\alpha(L)=\E  [L] + \sqrt{\mathbb{V}\mathrm{ar}(L)}$ $\mathbb{ES}_\alpha(Z)$ with $ Z \sim \mathcal{N}_1(0, 1),$ $\mathbb{ES}_\alpha(Z)= \frac{\phi\left(\Phi^{-1}(\alpha)\right)}{1-\alpha}$.
          This and \eqref{eq:linearcomb}   yield
          \begin{equation}\label{eq:smallr}
           r_i(q_i)=-\E  [R_i]-q_i^\top \mu +\mathbb{ES}_{\alpha_i}(Z) \sqrt{\mathbb{V}\mathrm{ar}(R_i)+2q_i^\top \mathrm{cov}_i+q_i ^\top\Gamma q_i}   , \quad q_i\in\mathbb{R}^m.
          \end{equation}
          By \citet[Corollary $8.5.2$]{rockafellar1970}, the recession function of $r_i$ is given by
          \begin{equation*}
            (r_i0^+)(y)=\lim_{\lambda \searrow 0}\lambda r_i(y/\lambda)=-y^\top  \mu+ \sqrt{y^\top \Gamma y}\;\mathbb{ES}_{\alpha_i}(Z).
          \end{equation*}
          Let $q_1, \dots, q_{|E|}$ be vectors in $\mathbb{R}^m$ such that
          \begin{equation*}
            \sum_{i\in \anE}(r_i0^+)(q_i) \leq 0 \quad \text{and} \quad \sum_{i\in \anE}(r_i0^+)(-q_i)>0,
          \end{equation*}
          i.e.
          \begin{equation*}
            -\mu^\top  \left(\sum_{i\in \anE} q_i\right)< \sum_{i\in \anE} \sqrt{q_i^\top  \Gamma q_i}\;\mathbb{ES}_{\alpha_i}(Z) \leq \mu^\top  \left(\sum_{i\in \anE}q_i\right).
          \end{equation*}
          Thus  $\sum_{i\in \anE}q_i\neq 0$.
          By \citet[Corollary $9.2.1$, page $76$]{rockafellar1970}, 
         the inf-convolution of real valued convex functions is a real valued convex function.
          Hence
          the inf-convolution $r$ is attained on $\mathbb{R}^m$ and,
     by Lemma \ref{lem:radnerequivalence}, there exists a Radner equilibrium $((\bo{q}_i)_{i\in \anE}, \bo{p})$.

      \subsubsection*{Unique price} We know that $-R_i-\bo{q}_i^\top P $, $i\in \anE$, is a continuous random variable.
          By \citet*[Theorem $4.3$ and Section $5.2$]{kalkbrener2005}, an expected shortfall is differentiable at continuous random variables.
          Therefore, by Remark \ref{rem:differentiablity}, the optimal price $\bo{p} $ is unique.
 \subsubsection*{Unique portfolio}
          If $\Gamma_i$ is positive definite, then,
          \begin{equation*}
            (1,q_i)^\top \Gamma_i (1,q_i)= \mathbb{V}\mathrm{ar}(R_i)+2q_i^\top \mathrm{cov}_i+q_i^\top \Gamma q_i>0, \quad q_i\in \mathbb{R}^m.
          \end{equation*}
          This  and \eqref{eq:smallr} implies that $r_i$ is differentiable such that 
          \begin{equation*}
            \bm{\nabla}r_i(q_i)=-\mu + \frac{\mathbb{ES}_{\alpha_i}(Z)}{\sqrt{(1,q_i)^\top \Gamma_i (1,q_i)}} \left( \mathrm{cov}_i + \Gamma q_i\right), \quad  q_i\in \mathbb{R}^m.
          \end{equation*}
          Following \citet*[Theorem $2.14$, page $47$]{rockafellar1998}, the strict convexity of $r_i$ is equivalent to
          \begin{equation*}
           r_i(y)>r_i(q_i)+\bm{\nabla}r_i(q_i)^\top  (y-q_i), \quad  q_i\neq y.
          \end{equation*}
          A simple computation reduces this first order condition to
          \begin{equation}\label{eq:strictconv}
            \sqrt{\left[(1,q_i)^\top \Gamma_i (1,q_i)\right] \left[(1,y) ^\top\Gamma_i (1,y) \right] } > (1,q_i)^\top \Gamma_i (1,y),\quad  q_i\neq y.
          \end{equation}
          If $y\neq q_i$, then $(1,y)$ is not colinear to $(1,q_i)$.
          Hence,  by \citet*[Eqn.~(2.49), page 79]{richard2007} applied with $\bo{b} = (1,q_i),\bo{d}=\Gamma_i(1,y)$,  and $\bo{B}=\Gamma_i$ (hence  $(1,q_i)=\bo{b}\neq c\bo{B}^{-1} \bo{d} =c(1,y)$ for any constant $c$), 
          \begin{equation*}
            \left[(1,q_i)^\top \Gamma_i (1,q_i)\right] \left[(1,y) ^\top\Gamma_i (1,y) \right]> \left[(1,q_i)^\top \Gamma_i (1,y)\right]^2\quad \text{holds for any} \quad q_i\neq y.
          \end{equation*}
          This in turn yields \eqref{eq:strictconv}. 
          Hence by Theorem \ref{thm:uniqueness} there exists a unique equilibrium. 
 
\section{Regulatory guidelines for CCP default management process}\label{s:RegFmwk}
When a clearing member of a CCP defaults, its position is taken over by the CCP.
The CCP should then close the defaulter's positions in a way that does not harm the other members or the CCP itself. 
As outlined in \citep{bis2020}, the CCP can settle the defaulter's positions via an auction organized by the CCP between the surviving members (and sometimes invited participants).
According to \citep[page $7$]{bis2020}, the  chance of a successful auction is increased by hedging the defaulted portfolio's risks prior to the auction:
\begin{quote}
    {A CCP should establish a framework for its approach to hedging risks from a defaulted participant's portfolio prior to a default management auction to increase the chance of a successful auction. 
    [...] The goals of a CCP's hedging strategy are generally to minimise the CCP's exposure to the defaulted participant's portfolio and to decrease the overall risk that the portfolio may pose to the CCP and the auction participants. Portfolios with less risk exposure lessen the potential effects of market volatility on the portfolio
    [...] and time dependency of valuations by auction participants.}
\end{quote}
A close-out procedure can also involve some liquidation on open markets.
As different positions are liquidated separately, hedging prior to liquidation would entail additional  costs for liquidating the hedging side of the portfolio. 
The main default resolution strategies are thus liquidation versus hedging then auctioning.
As pointed out in \citet*{oleschak2019}, 
\begin{quote}{in cases where the position to be transferred is large in relation to market liquidity or where a central market does not exist, auctions with the surviving agents as bidders is the mechanism of choice.}
\end{quote}
In any case, the CCP deals with the losses incurred throughout the close-out period by using the collateral of the defaulter, its own resources (skin in the game), and financial resources pooled between the clearing members in the form of a default fund  \citep{gregor2014,bruno2016,oleschak2019}.
The CCP should assess the adequacy of these financial resources by a careful estimation of the close-out costs of the defaulters' positions, which is the focus of this paper.

\section{Examples of Default Resolution Strategies When the CCP Operates on External Exchanges}\label{app:examples}
In this section, we consider default resolution strategies where the CCP also operates on some external exchanges (as opposed to its own exchange only in Section \ref{ss:fourcases}).

\subsection{The CCP fully  liquidates on another exchange}\label{sec:liqonEbar}

If the CCP liquidates $\bo{q}_d$ on some exchange $\theE \notni d$ (hence $\theE \neq \theD$),
then $\MC_\anE=0, \anE\neq \theE,\theD,$ and 
$ \sum_{i\in \theE' =\theE }\Delta  {\bo{q}}_i = \bo{q}_d$.
As $\sum_{i\in {\theE}} {\bo{q}}_i=0$, the ensuing the post-default equilibrium clearing condition on $ \theE '$ is
\begin{equation}\label{eq:ccdirectc}
  \sum_{i\in \theE '=\theE  } {\bo{q}}_i' =\bo{q}_d.
\end{equation} 
\begin{remark}\label{rem:equilibriumtrans}
By change of variables $\bo{z}'_i= {\bo{q}}_i'-k_i\bo{q}_d$ and $ {R}'_i= {R}_i+k_i\bo{q}_d^\top  P$, for reals $k_i$ such that $\sum_{i\in  {\theE'}  } k_i =1,$  the clearing condition \eqref{eq:ccdirectc} and the optimality condition \eqref{eq:mo} relative to the post-default equilibrium  $((\bo{q}'_i)_{i\in    \theE'}, \bo{p}'^\theE)$ become  $ \sum_{i\in  \theE'}\bo{z}'_i=0$   and
  \begin{equation*}
     {\rho}_i(- {R}'_i + (\bo{z}_i')^\top  ( {\bo{p}}'^\theE-P)) \leq  {\rho}_i(- {R}'_i +z_i^\top  ( {\bo{p}}'^\theE-P))\sp  z_i\in\mathbb{R}^m.
  \end{equation*}
 On $\theE '$, we thus recover a zero clearing condition and 
 member optimally conditions  formally similar to  Definition \ref{defi:radner}.\ \finenv\end{remark}

On the exchange $\theD$ of the CCP, we have  $\sum_{i\in \theD' = \theD\setminus\{d\}}\Delta \bo{q}_i =0$, whence the post-default clearing condition
\begin{equation}
\label{eq:cchedgingonEown}
  \sum_{i\in \theD' = \theD\setminus\{d\}}\bo{q}'_i =-\bo{q}_d.
\end{equation}
Therefore  
\begin{equation*}
   \MC=\MC_\theE +\MC_\theD, 
\end{equation*}
where 
\begin{equation*}
    \MC_\theE = \underbrace{\bo{q}_d^\top (\bo{p}^\theE-\bo{p}'^\theE)}_{\LC_\theE}+\sum_{i\in\theE'=\theE}\Delta\brho_i \sp \quad  \MC_\theD = \underbrace{-\bo{q}_d^\top (\bo{p}^\theD-\bo{p}'^\theD)}_{\LC_\theD}+\sum_{i\in\theD'=\theD\setminus\{d\}}\Delta\brho_i.
\end{equation*}
\begin{remark}\label{rem:equilibriumtrans2}
  By change of variable $\bo{z}'_i =\Delta \bo{q}_i$ and $R'_i  = R_i+\bo{q}_i^\top P $, $i\in \theD'=\theD\setminus\{d\}$, the clearing condition \eqref{eq:cchedgingonEown} and optimality condition \eqref{eq:mo} relative to the post-default equilibrium  $((\bo{q}'_i)_{i\in \theD'}, \bo{p}'^\theD)$ become  $\sum_{i\in  \theD '} \bo{z}'_i =0$ and 
  \begin{equation*}
    \rho_i\big(-R'_i + (\bo{z}_i')^\top  (\bo{p}'^\theD-P)\big) \leq \rho_i\big(-R'_i +z_i^\top  (\bo{p}'^\theD-P)\big)\sp  z_i\in\mathbb{R}^m.\ \finenv
  \end{equation*}
  \end{remark}

\subsection{The CCP fully  hedges on another exchange}\label{ss:HedgingOtherExch}

The considered CCP of $d$ can also hedge the portfolio $\bo{q}_d$ that it inherit from member $d$ (if not liquidated) by trading on an exchange  $\theE\notni d$, in which case  $\theE'=\theE \cup \{\thec\}$ and $R_\mathrm{c}=\bo{q}_d^\top  \big( P -\bo{p}^{\theD}\big)$ 
(arising from the pre-default Radner equilibrium on the exchange $\theD$ of the CCP).
In this case, $\MC_E=0, E\neq \theE,\theD$, and $ \sum_{\theE'= \theE \cup\{\thec\} }\Delta  {\bo{q}}_i  =0$ (the amount demanded must be equal to the amount supplied on the post-default exchange $\anE'$ where the hedge is implemented).
As $\sum_{i\in   {\theE}} {\bo{q}}_i  =0$ and $\Delta \bo{q}'_c =\bo{q}'_c$, the ensuing post-default equilibrium clearing condition on $ \theE'$ is
\begin{equation}\label{e:chE}
     \sum_{i\in  \theE' = \theE \cup \{\thec\}} {\bo{q}}'_i   =0.
\end{equation}
The corresponding member optimality condition \eqref{eq:mo} for the CCP $c$ is 
\begin{equation*} 
\rho_\thec\left( \bo{q}_d ^\top  (\bo{p}^\theD-P)+( \bo{q}'_\thec )^\top (\bo{p}'^\theE-P) \right) \leq \rho_\thec\left(\bo{q}_d ^\top  (\bo{p}^\theD-P)+ q_c^\top (\bo{p}'^\theE-P)\right), \quad q_\thec \in \mathbb{R}^d.
\end{equation*}

\noindent On the own exchange $\theD$ of the CCP, we have $\sum_{i\in \theD'=  \theD\setminus\{d\} }\Delta\bo{q}_i =0$, whence  the post-default clearing condition  
\begin{equation}
\label{eq:cchedgingonEownbis}
  \sum_{i\in \theD'=  \theD\setminus\{d\} }\bo{q}'_i =-\bo{q}_d.
\end{equation}

\noindent Therefore $\MC=\MC_{\theE}+\MC_{\theD}$,
where 
\begin{equation*}
    \MC_\theE = \underbrace{0}_{\LC_\theE}+ \sum_{\theE'= \theE \cup \{\thec\}}\Delta\brho_i,  \quad  \MC_\theD =  \underbrace{-\bo{q}_d ^\top  (\bo{p}^\theD-\bo{p}'^\theD)}_{\LC_\theD}+\sum_{\theD' = \theD\setminus\{d\}}\Delta\brho_i.
\end{equation*}
 \begin{remark}\label{rem:cchedging2}
  As in Remark \ref{rem:equilibriumtrans2} again, by change of variable $\bo{z}'_i =\Delta \bo{q}_i$  and $R'_i  = R_i+\bo{q}_i^\top P $, $i\in \theD\setminus\{d\}$, the clearing condition \eqref{eq:cchedgingonEownbis} relative to  the post-default equilibrium $((\bo{q}'_i)_{i\in    \theD'}, \bo{p}'^\theD)$ can be converted to a zero clearing condition as per Definition \ref{defi:radner} on $\theD'$.\ \finenv
  \end{remark}
 
\subsection{The CCP fully replicates on another exchange}\label{rem:fullreplicationOtherExch}
Alternatively, the considered CCP of $d$ can replicate the portfolio $\bo{q}_d$ (if not liquidated) that it inherits from $d$ by mirroring positions $\bo{q}'_c=-\bo{q}_d$ on an external exchange  $\theE\notni d$, hence $\theE'=\theE \cup \{\thec\}$.
As in Section \ref{rem:fullreplication}, replication means that the only admissible trading strategy for the post-default trading participant $\thec$ is $-\bo{q}_d$.
In this case, $\MC_E=0, E\neq \theE,\theD$, and $ \sum_{\theE'= \theE \cup\{\thec\} }\Delta  {\bo{q}}_i  =0$ (the amount demanded must be equal to the amount supplied on the post-default exchange $\anE'$ where the hedge is implemented), $\sum_{i\in \theE } \bo{q}_i=0$. The ensuing post-default clearing condition on $\theE'$ is
\begin{equation*}
  \sum_{i\in  \theE' = \theE\cup \{\thec\}}\bo{q}'_i = \underbrace{\sum_{i\in  \theE}\bo{q}'_i}_{\bo{q}_d} + \underbrace{\bo{q}'_\thec}_{-\bo{q}_d} =0.
\end{equation*}
We also have $\Delta\brho_\thec =\rho_\thec \Big(\bo{q}_d^\top   (\bo{p}^\theD-P)- \bo{q}_d^\top   (\bo{p}'^\theE-P)\Big)= \bo{q}_d^\top \big(\bo{p}^\theD-\bo{p}'^\theE \big) $.

\noindent On the own exchange $\theD$ of the CCP, we have $\sum_{i\in \theD'=  \theD\setminus\{d\} }\Delta\bo{q}_i =0$, whence  the post-default clearing condition  
\begin{equation*}
    \sum_{i\in \theD' =\theD\setminus\{d\}} \bo{q}'_i  = -\bo{q}_d. 
\end{equation*}

\noindent Therefore $\MC=  \MC_\theE +\MC_\theD $, where
\begin{equation*}
    \MC_\theE  = \underbrace{0}_{\LC_\theE}+ \sum_{i\in \theE } \Delta\brho_i +\Delta\brho_\thec \quad \text{and} \quad \MC_\theD =  \underbrace{-\bo{q}_d^\top  (\bo{p}^{\theD}-\bo{p}'^{\theD})}_{\LC_ \theD}+ \sum_{i\in \theD' =\theD\setminus\{d\}} \Delta\brho_i.
\end{equation*}

\begin{remark}\label{rem:ccreplicateonE}
  As in Remark \ref{rem:equilibriumtrans2}, by change of variable $\bo{z}'_i =\Delta \bo{q}_i$  and $R'_i  = R_i+\bo{q}_i^\top P $, $i\in \theD\setminus\{d\})$,   we recover member optimally and zero clearing conditions as per Definition \ref{defi:radner} on $\theD'$.\ \finenv
  \end{remark}

\subsection{The CCP partially liquidates and hedges on another exchange}\label{rem:hybrid2}
The CCP can also liquidate a portion $\bo{q}_d^l$ of the defaulted position $\bo{q}_d$ and hedge the remaining $\bo{q}_d^h=\bo{q}_d-\bo{q}_d^l$ on another exchange $\theE\notni d$.
In this case, $\MC_E=0, E\neq \theE,\theD$.
Since the amount demanded should be equal to the amount supplied on each leg of the strategy, we have 
$\sum_{i\in \theE'} \Delta \bo{q}^l_i = \bo{q}^l_d$ and $\sum_{i\in \theE' } \Delta \bo{q}^h_i = 0 ,$ hence $\sum_{i\in \theE'} \Delta \bo{q}_i =\bo{q}^l_d.$
As $\sum_{i\in \theE}\bo{q}_i=0$ and $\bo{q}_c=0$, the ensuing post-default equilibrium clearing condition on $ \theE'$ is
\begin{equation}\label{eq:ccother}
    \sum_{i\in \theE'} \bo{q}'_i = \bo{q}^l_d.
\end{equation}
We assume that both liquidation and hedging happen simultaneously on the exchange $\theE$ with the same price $\bo{p}'^\theE$. 
Hence each trading participant on the post-default market $\theE'$ has a single member optimality condition \eqref{eq:mo}  (with, in particular,  
$R_\thec = (\bo{q}^h_d )^\top (P-\bo{p}^\theD)$). 

Regarding the own exchange of the CCP, the post-default equilibrium clearing condition on $\theD'$ is
\begin{equation}\label{eq:ccown}
    \sum_{i\in \theD'=  \theD\setminus\{d\} }\bo{q}'_i =-\bo{q}_d
\end{equation}
Hence $  \MC=\MC_{\theE} + \MC_{\theD} , $
where 
\begin{equation*}
     \MC_\theE = \underbrace{(\bo{q}^l_d)^\top(\bo{p}^\theE - \bo{p}'^\theE) }_{\LC_\theE}+ \sum_{\theE'= \theE \cup \{\thec\}}\Delta\brho_i \sp   \MC_\theD =  \underbrace{-\bo{q}_d ^\top  (\bo{p}^\theD-\bo{p}'^\theD)}_{\LC_\theD}+\sum_{\theD' = \theD\setminus\{d\}}\Delta\brho_i.
\end{equation*}
\begin{remark}\label{rem:equilibriumtrans22}
  Similarly to Remarks \ref{rem:equilibriumtrans} and \ref{rem:equilibriumtrans2}, by change of variables, the clearing conditions \eqref{eq:ccother}  and \eqref{eq:ccown} relative to  the post-default equilibria  $((\bo{q}'_i)_{i\in    \theE'}, \bo{p}'^\theE)$ and   $((\bo{q}'_i)_{i\in    \theD'}, \bo{p}'^\theD)$ can be converted to zero clearing conditions on the exchanges $\theE'$ and $\theD'$.\ \finenv
  \end{remark}

  \begin{table}[htp]  
        \begin{centering}
        \resizebox{\columnwidth}{!}{
           \begin{tabular}{@{}lcc@{}}
           \toprule
             & $\mathrm{LC}$ &   $\displaystyle  \sum_\anE \sum_{i\in  \anE' }  \Delta\brho_i$\\ 
                \midrule
                1.  &  $-\bo{q}_d ^\top  (\bo{p}^\theD-\bo{p}'^\theD) +\bo{q}_d ^\top  (\bo{p}^\theE-\bo{p}'^\theE)$  & $ \displaystyle\sum_{i\in\theD\setminus\{d\}}\Delta\brho_i+\displaystyle\sum_{i\in\theE}\Delta\brho_i$ \\ 
                2.  & $-\bo{q}_d ^\top  (\bo{p}^\theD-\bo{p}'^\theD)$    & $ \displaystyle\sum_{i\in \theD\setminus\{d\}}\Delta\brho_i + \displaystyle\sum_{i\in\theE\cup\{c\}}\Delta\brho_i$\\ 
                3.  & $-\bo{q}_d ^\top  (\bo{p}^\theD-\bo{p}'^\theD)$    & $ \displaystyle\sum_{i\in  \theD\setminus\{d\} }\Delta\brho_i + \displaystyle\sum_{i\in\theE\cup\{c\}}\Delta\brho_i$      with $\Delta\brho_\thec = \bo{q}_d^\top \big(\bo{p}^\theD-\bo{p}'^\theE \big) $\\ 
                4.  & $-\bo{q}_d ^\top  (\bo{p}^\theD-\bo{p}'^\theD) + (\bo{q}^h_d)^\top  (\bo{p}^\theE-\bo{p}'^\theE) $    &  $\displaystyle\sum_{i\in \theD\setminus\{d\}}\Delta\brho_i + \displaystyle\sum_{i\in\theE\cup\{c\}}\Delta\brho_i$ \\
                \bottomrule
            \end{tabular}
            }
            \caption{Decomposition of the  market costs in the four default resolution strategies of Section \ref{app:examples}.}
       \label{tab:SixCasesSummary2}
   \end{centering}
\end{table}

\section{XVA Gaussian Setup}\label{app:xva}

The purpose of this part is to provide a bridge from the equilibrium setup of Sections \ref{sec:re}--\ref{sec:es} to the XVA setup of \citet{BastideCrepeyDrapeauTadese21a}, 
so that we are able to provide an overall FTP \eqref{e:FTPexpr} 
quantifying the 
market but also credit
costs of a given default resolution strategy. 
We leave for future research the extension of the approach of this paper to a setup where not only the market costs, but also the credit costs, would be treated endogenously as part of a global (or perhaps two-stage\footnote{accounting for the segregation between the market and credit spheres in banks \citep[Article 92]{EU2013}.}) equilibrium, ideally in the setup of a dynamic model.

We endorse the structure of exchanges $E=A\cup B$ depicted in \hyperref[ss:StructExch]{Section \ref{ss:StructExch}}. Note that even for those clients clearing through a CCP member and also having OTC positions with that same member, the corresponding exposures must be treated separately, hence $\CLset \cap \OTCset=\varnothing$. For each participant $i$ to an exchange $E$, a comparison of the present setup with \citet[Eqns.~(15)-(16)]{BastideCrepeyDrapeauTadese21a} yields the mapping of Table 
\ref{tab:mapping}. 

\begin{table}[htbp]
    \begin{centering}
        \resizebox{0.95\columnwidth}{!}{
    \begin{tabular}{@{}p{3.5cm}cp{6.25cm}@{}}
    \toprule
         \bf{\cite[Eqns.~($\bm{15}$)-($\bm{16}$)]{BastideCrepeyDrapeauTadese21a}} & \bf{This paper} &  \bf{Description}\\
        \midrule 
       $\mathcal{P}^{ccp}_c- \mathrm{MtM}^{ccp}_c$ & $ (\bo{q}_b^0)^\top(\bo{p}^{E}-P)$  & Cash flows from a cleared client ($c$ in \cite{BastideCrepeyDrapeauTadese21a},  $b$ in this paper) of a CCP  ($ccp$ in \cite{BastideCrepeyDrapeauTadese21a}, the CCP of exchange $E$ in this paper) to the participant $0$.
       If the latter is not a clearing member of the CCP, these cash flows are simply zero. \\
       \cmidrule{1-3}
       $\mathcal{P}^{ccp}_i- \mathrm{MtM}^{ccp}_i$ &$\sum_{b\in\CLset}(\bo{q}_b^a)^\top(\bo{p}^\anE-P)$ & Client account cash flows from a clearing member ($i$ in \cite{BastideCrepeyDrapeauTadese21a}, $a$ in this paper)  to his CCP ($ccp$ in \cite{BastideCrepeyDrapeauTadese21a}, the CCP of exchange $E$ in this paper).  \\\cmidrule{1-3} 
       $\overline{\mathcal{P}}^{ccp}_i - \overline{\mathrm{MtM}}^{ccp}_i$  & $\bo{q}_a^\top (\bo{p}^{E}-P)$ & Proprietary account cash flows from a clearing member ($i$ in \cite{BastideCrepeyDrapeauTadese21a}, $a$ in this paper)  to his CCP. \\\cmidrule{1-3}
       $\mathcal{P}_b- \mathrm{VM}_b $ & $ R_0^o$ & Cash flows from an end-user  ($b$ in \cite{BastideCrepeyDrapeauTadese21a}, $o$ in this paper)  to participant $0$. \\
         \bottomrule
    \end{tabular}
    }
     \caption{Some notation adaptation for the cash flows of some market participants after variation margin is subtracted, in the setup of \cite{BastideCrepeyDrapeauTadese21a} and in this paper.}
        \label{tab:mapping}
    \end{centering}
\end{table}

Let $J_i$ be the survival indicator of participant $i$ to an exchange $E$, i.e.~$J_i=\mathds{1}_{\{\tau_i>T\}}$, where $\tau_i$ is the default time of participant $i$ over the period $[0,T]$, with probability $\gamma_i =\mathbb{P}(J_i=0)$ of default over $[0,T]$. We denote likewise $J_o=\mathds{1}_{\{\tau_o>T\}}$
for any end-user $o \in\OTCset$. 
Via the mapping of Table \ref{tab:mapping}, \cite[Eqns.~(15)-(16)]{BastideCrepeyDrapeauTadese21a} 
yield the  following (pre-default equilibrium) credit loss profile $    \dloss_0 $ of a participant $i=0$ to the exchanges, on which we focus in what follows: $   \dloss_0 
= \sum_{E=\CMset \cup \CLset}( \dloss_0^\CLset+ \dloss_0^\CMset)+\dloss_0^\OTCset$, where 
\begin{align}\label{e:AllocLossProfilePreEq}
\begin{split}
 &  \dloss_0^\CLset=\sum_{b\in\CLset} (1-J_b)\left(  (\bo{q}_b^0)^\top (\bo{p}^\anE-P)  -\IM_0^b\right)^+,\\
&  \dloss_0^\CMset
 =  w_0^\CMset \sum_{ a\in \CMset }(1-J_{a}) 
 \Big[ \Big(  \Big(\sum_{b\in\CLset}\bo{q}_b^a \Big)^\top  (\bo{p}^\anE-P ) -\IM_a^\CMset \Big)^+
+   \\
 & \qqq  \Big(  \bo{q}_a^\top  (\bo{p}^\anE-P\ )-\overline{\IM} ^\CMset _{a} \Big)^+ -\DF_{a}^\CMset \Big]^+, \\
 &  \dloss_0^\OTCset=\sum_{o\in O}(1-J_o) ( R_0^o -\IM_0^o )^+.
\end{split}
\end{align}
Here $\IM_0^b$  is the initial margin (IM) requested by the participant $0$ to the simple participant $b\in\CLset$ on the cleared position $
\bo{q}_b^0$ (equal to 0 if $0\notin A$); $w_0^\CMset$ (equal to 0 if $0\notin A$) is the loss allocation coefficient of the participant $0$ w.r.t. the CCP of the exchange $\anE=A\cup B$; $\IM_a^\CMset$,  $\overline{\IM}_a^\CMset$, and $\DF_a^\CMset$ are the initial margins for the cleared clients and proprietary accounts as well as the default fund contribution requested by (the CCP of) exchange $\anE$ to the clearing member $a$; $R_0^o$ is the exogenous receivable of the participant $0$ from its OTC bilateral counterparty $o$, with corresponding initial margin $\IM_0^o$ (which can be null under OTC agreement) requested by $0$ to $o$ (cf.~Figure \ref{fig:splitingclvsotcbilat}).

Likewise, the post-default equilibrium\footnote{post-default referring as usual in the paper to the instant default at time 0 of a clearing member $d$, here assumed $\neq  $ the reference clearing member 0.} 
default loss profile $\dloss'_0$ of the participant $0$ is
$\dloss'_0=    
\sum_{E'=\CMset ' \cup \CLset '}( \dloss_0'^{\CLset}+ \dloss_0'^{\CMset}) +\dloss_0^\OTCset
,$ for $\dloss_0^\OTCset$ as in \eqref{e:AllocLossProfilePreEq} and  (cf.~Table~\ref{tab:mapping} and \eqref{e:PostDefTdgLoss})
\begin{align}\label{e:AllocLossProfilePostEq}
\begin{split}
 &  \dloss_0'^{\CLset}=\sum_{b\in \CLset '} (1-J_b)\left( ( \bo{q}_b'^0)   ^\top(\bo{p}'^\anE-P)
+\big(\bo{q}_b^0 + (\Delta\bo{q}_b^0)^l\big)^\top  (\bo{p}^\anE -\bo{p}'^\anE)
 - \IM_0'^b  \right)^+ ,\\ 
 & \dloss_0'^{\CMset}\underset{\eqref{e:PostDefTdgLoss}}{=}  \,\, w'^{\CMset}_0\sum_{ a\in \CMset '}(1-J_a) \times\\&\qqq
 \Bigg( \Big(  \sum_{b\in \CLset '}(\bo{q}_b'^a)^\top (\bo{p}'^\anE-P)+
  \sum_{b\in\CLset '} \big(\bo{q}_b^a  + (\Delta\bo{q}_b^a )^l\big)^\top  (\bo{p}^\anE -\bo{p}'^\anE) -\IM_a'^{\CMset}\Big)^+ +\Bigg. \\
 & \qqq \Bigg.  \Big((\bo{q}'_a)^\top (\bo{p}'^\anE-P)+
(\bo{q}_a + \Delta\bo{q}^l_a)^\top  (\bo{p}^\anE -\bo{p}'^\anE) -\overline{\IM}'^{\CMset}_a\Big)^+   -\DF_{a}'^\CMset \Bigg)^+, 
\end{split}
\end{align}
where $w'^{\CMset}_0 
$, $\IM'^{\CMset}_a$, $\overline{\IM}'^{\CMset}_a$, $\DF'^{\CMset}_a$, and ${\IM'^b_0}$ are the post-default analogs of  $w^{\CMset}_0 
$, $\IM^{\CMset}_a$, $\overline{\IM}^{\CMset}_a$,  $\DF^{\CMset}_a$,  and ${\IM^b_0}$ in \eqref{e:AllocLossProfilePreEq}. 

By \citet[Theorem 3.7]{BastideCrepeyDrapeauTadese21a}, the pre-{ and post-}default CVA of the participant 0 are given by 
\begin{align}\label{e:cvaexpression}
    \CVA_0 =\E  \left[\dloss_0\big|J_0=1\right]
    =(1-\gamma_0)^{-1}\E  \left[J_0 \dloss_0 \right] \sp   \CVA'_0 = (1-\gamma_0)^{-1}\E     \left[ J_0\dloss'_0\right],
\end{align}
Denoting by $\overline{\IM}_0^o$ the initial margin from the participant $0$ to its OTC bilateral counterparty $o$, based on \hyperref[rem:Ri]{Remark \ref{rem:Ri}}, 
such margin remains constant in the post-default equilibrium, hence the pre- and post-default MVA of participant $0$ are given by 
\begin{align}\label{e:mvaexpression}
\begin{split}
    & \MVA_0=\widetilde{\gamma}_0\left(\sum_{o\in\OTCset} \overline{\IM}_0^o+\sum_\anE\left(\IM_0^\anE+\overline{\IM}_0^\anE+\DF_0^\anE\right)\right), \\
& \MVA'_0=\widetilde{\gamma}_0\left(\sum_{o\in\OTCset} \overline{\IM}_0^o+\sum_\anE\left(\IM_0^{\anE'}+\overline{\IM}_0^{\anE'}+\DF_0^{\anE'}\right)\right),
\end{split}
\end{align}
for some possibly blended funding rate  $\widetilde{\gamma}_0\le\gamma_0$ as detailed in \citet*[Section 5]{ArmentiCrepey16}. The pre-default KVA, defined for a hurdle rate $h$, is calculated based on an expected shortfall $\mathbb{ES}^0_{\tilde{\alpha}_0}$ of the participant $0$ under its own survival measure, $\mathbb{P}(\cdot J_0)/(1-\gamma_0)$ (with $\tilde{\alpha}_0>$ the confidence level $\alpha_0$ introduced for the market cost computation in \hyperref[sec:es]{Section \ref{sec:es}} when the risk measures used by the hedgers are expected shortfall risk measures\footnote{regulatory and economic capital aim at capturing extreme losses that can occur once every 1000 years \citep[paragraph 5.1]{BCBS05}, which leads to considering a much higher confidence level $\tilde{\alpha}_0$ for economic capital calculation, such as $0.9975$, from which the  KVA is defined, than the $\alpha_0$ used for market risk, set to $0.975$, in line with \cite[Section 1.4  (i)]{FRTB2013}.}), as
\beql{e:KVA}
 &  \KVA_0=\frac{h}{1+h}\mathbb{ES}^0_{\tilde{\alpha}_0}\left(\dloss_0-\CVA_0\right) \\
& =\frac{h}{1+h}\E  \left[J_0\left(\dloss_0-\CVA_0\right)\big| \dloss_0-\CVA_0\geq\mathbb{V}\mathrm{a}\mathbb{R}^0_{\tilde{\alpha}_0}\left(J_0(\dloss_0-\CVA_0)\right), J_0=1\right],
\eeql
by \citet[Theorem 3.7, last row of Table 2]{BastideCrepeyDrapeauTadese21a}, where $\mathbb{V}\mathrm{a}\mathbb{R}^0_{\tilde{\alpha}_0}$ denotes the value-at-risk at the confidence level $\tilde{\alpha}_0$ under the measure $(1-\gamma_0)^{-1}\mathbb{P}\left(\cdot J_0\right)$. The post-default KVA has a similar expression substituting $\dloss'_0$ to $\dloss_0$ and $\CVA'_0$ to $\CVA_0$ in \eqref{e:KVA}:
\beql{e:KVAprime}
 &  \KVA'_0=\frac{h}{1+h}\mathbb{ES}^0_{\tilde{\alpha}_0}\left(\dloss'_0-\CVA'_0\right) \\
& =\frac{h}{1+h}\E  \left[J_0\left(\dloss'_0-\CVA_0\right)\big| \dloss'_0-\CVA_0\geq\mathbb{V}\mathrm{a}\mathbb{R}^0_{\tilde{\alpha}_0}\left(J_0(\dloss'_0-\CVA'_0)\right), J_0=1\right],
\eeql
Finally, by \citet[Theorem 3.7, next to last row of Table 2]{BastideCrepeyDrapeauTadese21a}, 
the pre- and post- FVA of the participant $0$ is given by 
\begin{align}\label{e:FVA}
\begin{split}
    & \FVA_0=\displaystyle\frac{\gamma_0 }{1+\gamma_0}   \left(\displaystyle\sum_{o\in\OTCset} \ \E R_0^o  -(\CVA_0+\MVA_0)- \mathbb{ES}^0_{\tilde{\alpha}_0}\left(\dloss_0-\CVA_0\right)  \right)^+, \\
    & \FVA'_0=\displaystyle\frac{\gamma_0}{1+\gamma_0}   \left(\displaystyle\sum_{o\in\OTCset}  \E R_0^o  -(\CVA'_0+\MVA'_0)- \mathbb{ES}^0_{\tilde{\alpha}_0}\left(\dloss'_0-\CVA'_0\right)  \right)^+.
\end{split}
\end{align}

\subsection{XVA details in the setup of Section \ref{ss:econt}}\label{ss:xvad}

In the setup of Section \ref{ss:econt}, only clearing members $a$ participate to the only exchange of interest $\theD$ (so all participants $i$ are clearing members $a$ and there are no cleared clients $b$) and all the end-users (OTC bilateral counterparties) $o$ are assumed to be default risk-free,
Hence
the pre-default credit loss \eqref{e:AllocLossProfilePreEq} of member $0\in\theD$ reduces to
 \beql{e:SimpleTdgLossWithCr}
 \dloss_0= 
 w_ 0^\theD \sum_{ j\in \theD }(1-J_j)\left(\bo{q}_j(\bo{p}^\theD-P)-\overline{\IM}^\theD_j-\DF^\theD_j\right)^+,
 \eeql
where  $w_0^\theD=\frac{\DF^\theD_0 J_0}{\sum_{ j\in \theD }\DF^\theD_j J_j}$. Under the post-default equilibrium  when the CCP fully liquidates on $\theD$ (so that $\Delta \bo{q}_i^h=0$), the post-default credit loss \eqref{e:AllocLossProfilePostEq} of member $0\in\theD$ reduces to
\begin{align}\label{e:SimpleTdgLossWithCrPOst}
\begin{split}
\dloss'_0 
 & =
 w'^\theD_0\sum_{i\in \theD'}(1-J_i)\left(\bo{q}'_i (\bo{p}'^\theD - P )+{\bo{q}'_i}(\bo{p}^\theD- \bo{p}'^\theD)-\overline{\IM}'^\theD_i-\DF'^\theD_i\right)^+\\
 & =
 w'^\theD_0\sum_{i\in \theD'}(1-J_j)\left(\bo{q}'_i (\bo{p}^\theD - P ) -\overline{\IM}'^\theD_i-\DF'^\theD_i\right)^+,
 \end{split}
 \end{align}
 whereas, when the CCP fully hedges on $\theD$ (so that $\Delta \bo{q}_i^l=0$, $ \bo{q}_j+\Delta \bo{q}_j^h=\bo{q}'_j$), 
 \begin{align}\label{e:SimpleTdgLossWithCrPOstHdg}
\begin{split}
\dloss'_0  
 & =
 w'^\theD_0\sum_{i\in \theD'}(1-J_i)\left(\bo{q}'_i (\bo{p}'^\theD - P )+{\bo{q}_i}(\bo{p}^\theD- \bo{p}'^\theD)-\overline{\IM}'^\theD_i-\DF'^\theD_i\right)^+ .
 \end{split}
 \end{align}
In both cases $w'^\theD_0 $, $\overline{\IM}'^\theD_i$ and $\DF'^\theD_i$, $i\in\theD'$, are the post-default analogs of $w^\theD_0$, $\overline{\IM}^\theD_i$ and $\DF^\theD_i$, $i\in\theD$, based on the post-default updated portfolio positions.
The pre- and post-default CVA, MVA and KVA of member $0\in D\setminus \{d\}$ are given by
\beql{e:CVApostdef}  
 & \CVA_0 =(1-\gamma_0)^{-1}\E  \left[ J_0\dloss_0\right]\sp   \CVA'_0 = (1-\gamma_0)^{-1}\E     \left[ J_0\dloss'_0\right],\\
 &\MVA_0 =\widetilde{\gamma}_0\left(\overline{\IM}_0+\DF_0\right)\sp\MVA'_0 =\widetilde{\gamma}_0\left(\overline{\IM}'_0+\DF'_0\right),\\
& \KVA_0 = \frac{h}{1+h}\E  \left[J_0\left(\dloss_0-\CVA_0\right)\big| \dloss_0-\CVA_0\geq\mathbb{V}\mathrm{a}\mathbb{R}^0_{\tilde{\alpha}_0}\left(J_0(\dloss_0-\CVA_0)\right), J_0=1\right], \\
& \KVA'_0 =  \frac{h}{1+h}\E  \left[J_0\left(\dloss'_0-\CVA'_0\right)\big| \dloss'_0-\CVA'_0\geq\mathbb{V}\mathrm{a}\mathbb{R}^0_{\tilde{\alpha}_0}\left(J_0(\dloss'_0-\CVA'_0)\right), J_0=1\right].
\eeql
Moreover, in the setup of Section \ref{ss:econt},
$\E R_0^o$ in \eqref{e:FVA} corresponds to 
$\mathrm{MtM}_b-\mathrm{VM}_b$ in the setup of \cite{BastideCrepeyDrapeauTadese21a}, i.e. a difference of received ($\mathrm{VM}_b$) variation margin by the member $0$ and posted ($\mathrm{MtM}_b$) variation margin by the member $0$ for an OTC bilateral position between the clearing member $0$ and the end-user $o$.
We assume, as it is the case in practice, that there is only marginal, if any, difference between the two quantities. Hence we have $\E R_0^o \approx0$ and, in any case, dominated by $(\CVA_0+\MVA_0)- \mathbb{ES}^0_{\tilde{\alpha}_0}\left(\dloss_0-\CVA_0\right)$ and $(\CVA'_0+\MVA'_0)- \mathbb{ES}^0_{\tilde{\alpha}_0}\left(\dloss'_0-\CVA'_0\right)$ in \eqref{e:FVA}, leading to negligible $\FVA_0$ and $\FVA'_0$, which we therefore simply take as 0 (and do not report) in the numerics of Section \ref{ss:econt}.

\subsubsection*{Latent Factor Model} For default modeling purposes, we introduce for each member $i\in\theD$ a latent variable $X_i\sim\mathcal{N}_1(0,1)$ such that $\{J_i=0\}\Longleftrightarrow\left\{X_i\leq \Phi^{-1}(\gamma_i)\right\}$. 
These default latent variables are correlated as per $ X_i=\sqrt{\varrho^{cr}} \varepsilon +
\sqrt{1-\varrho^{cr}}\varepsilon_i$, where $\varepsilon$ and $\varepsilon_i$ are i.i.d. $\mathcal{N}_1(0,1)$, while $\varrho^{cr}$ is a positive credit/credit correlation coefficient. Writing $P=\mu + \sigma Y$ with $Y\sim\mathcal{N}_1(0,1)$,  
the IM posted to the CCP by member $i$, based on the idea of a $\bo{q}_i(\bo{p}^\theD - P)$ VM call not fulfilled over its corresponding time period T (versus $\Delta_s$ in \cite{BastideCrepeyDrapeauTadese21a}), is computed by the VaR metric\footnote{under the member survival measure.} at a confidence level $\alpha_{im}\in(1/2,1)$ as
\begin{equation}\label{e:IMCaldDef}
\overline{\IM}^\theD_i={\mathbb{V}\mathrm{a}\mathbb{R}}_{\alpha_{im}} \left(  \bo{q}_i(\bo{p}^\theD-P)\right)=\bo{q}_i(\bo{p}^\theD-\mu) +|\bo{q}_i|\,\sigma {\Phi}^{-1}(\alpha_{im}).
\end{equation}
The liquidation time period $\Delta_l$ in \cite{BastideCrepeyDrapeauTadese21a} is also taken as the one-period of time considered in the Radner equilibrium setup of the present paper, so that $\Delta_s=\Delta_l=T$. The default fund is calculated at the level of the considered CCP of $d$ as the sum of the two highest stress loss over IM (SLOIM), where SLOIM is given for each member $i$ as $$\mathrm{SLOIM}^\theD_i={\mathbb{V}\mathrm{a}\mathbb{R}}_{{\alpha}_{df}}\left(\bo{q}_i(\bo{p}^\theD-P)-\overline{\IM}^\theD_i\right)= |\bo{q}_i|\,\sigma\left(\Phi^{-1}({\alpha}_{df})-\Phi^{-1}({\alpha}_{df})\right),$$ 
for some confidence level ${\alpha}_{df}>\alpha_{im}$. The default fund contribution of member $i$ is given as 
\begin{equation}\label{e:DFCaldDef}
\DF^\theD_i = \frac{\mathrm{SLOIM}^\theD_i}{\sum_{j\in\theD}\mathrm{SLOIM}^\theD_j}\left({\rm SLOIM}^\theD_{(0)} + {\rm SLOIM}^\theD_{(1)}\right),
\end{equation}
based on the cover-2 amount given as the sum of two largest stressed losses over IM (${\rm SLOIM}_i$) among its members, identified with subscripts $(0)$ and $(1)$.

\begin{table}[htp]  
        \begin{centering}
            \begin{tabular}{@{}lr@{}}
                \toprule
                Portfolios maturity $T$ & $5$ years \\ 
                Liquidation period at default $\Delta_l$ & 5 years \\
                Credit factors correlation $\varrho^{cr}$  & 20\% \\ 
                IM covering period (margin period of risk) $\Delta_s$ & 5 years\\ 
                Default probabilities $\gamma_i$ & $39.3\%$ \\ 
                MVA funding blending ratio $\widetilde{\gamma}_i/\gamma_i$  & 25\% \\ 
                Quantile levels $\tilde{\alpha}_i$ used for clearing members $\KVA$ & $99.75\%$ \\
                Hurdle rate $h$ used for $\KVA$ computations & 10.0\%\\ 
                Number of Monte-Carlo simulations (for $\CVA$ and $\KVA$ computations) & $10\mathrm{M}$ \\ 
                Number of batches (for $\KVA$ computations) & 100 \\
                \bottomrule
            \end{tabular}
             \caption{XVAs calculation configuration.}
        \label{tab:XVAconfig1CCP20Mbs}
    \end{centering}
\end{table}

\end{document}